\documentclass[a4paper,11pt]{article}
\usepackage[utf8]{inputenc}
\usepackage{xcolor}
\usepackage{varioref}
\usepackage{amsmath,amsfonts,amsthm,mathrsfs}
\usepackage{amssymb}
\usepackage{amstext}
\usepackage{enumerate}
\usepackage{float}
\usepackage{fancyvrb}
\usepackage{verbatim}
\usepackage{wrapfig}
\usepackage{appendix}
\usepackage{braket}
\usepackage{graphicx}
\usepackage{color}
\usepackage{varioref}
\usepackage{multirow,graphics}

\usepackage{bbm}
\usepackage{slashed}
\usepackage{relsize}
\usepackage{comment}

\usepackage{hyperref} 
\usepackage[english]{babel}
\usepackage{mathtools} 
\usepackage{mathtext} 
\usepackage{bm}
\usepackage{dsfont}
\usepackage{tikz}
\usetikzlibrary{patterns, calc, decorations.markings, arrows.meta}

\tikzset{->-/.style={line width=0.25mm,decoration={
			markings,
			mark=at position 0.53 with {\arrow{Stealth}}},postaction={decorate}}}
\tikzset{-<-/.style={line width=0.25mm,decoration={
			markings,
			mark=at position 0.47 with {\arrowreversed{Stealth}}},postaction={decorate}}}

\hypersetup{colorlinks=true, linktocpage=false, linkcolor=blue!80!black, citecolor=red!85!black, urlcolor=blue} % цвет гиперссылок

\renewcommand{\imath}{\mathrm{i}}
\renewcommand{\Re}{\operatorname{Re}}
\renewcommand{\Im}{\operatorname{Im}}
\renewcommand{\phi}{\varphi}

\newcommand{\tr}{\mathop{\rm tr}\nolimits}

\def\II{\hbox{{1}\kern-.25em\hbox{l}}}

\usepackage{cite}

\usepackage{tocloft}

\numberwithin{equation}{section}

\begin{document}
\thispagestyle{empty}

\begin{center}
{\bf \large Baxter $\mathbb{Q}$-operator and Separation of Variables for BC-type open $SL(2,\mathbb{C})$ spin chain}
\vspace{0.7cm}

{I. Kotikov$^{\diamond}$, P. Antonenko$^{\dagger\ast}$}
\vspace{0.7cm}

{\small \it
	$^\diamond$Moscow State University, Moscow, Russia\vspace{0.3cm}\\
	$^\dagger$Steklov Mathematical Institute, Fontanka 27, \\St.~Petersburg, 191023, Russia\vspace{0.3cm}\\
	$^\ast$Leonhard Euler International Mathematical Institute,\\ Pesochnaya nab. 10, St.~Petersburg, 197022, Russia
}
	
\end{center}

\vspace{0.1cm}

\begin{abstract} \noindent
	We construct two-parameter Baxter $\mathbb{Q}$-operators for the transfer matrices of noncompact open $SL(2,\mathbb{C})$ BC-type spin chains with general boundary interactions, treating both one-boundary and two-boundary models. The $\mathbb{Q}$-operators are defined as traces of general monodromy matrices over infinite-dimensional auxiliary spaces carrying principal series representations of $SL(2,\mathbb{C})$. 
	
	At specific values of parameters the two-parameter $\mathbb{Q}$-operators degenerate into simpler one-parameter Baxter operators. We prove that the general transfer matrices factorize into products of these one-parameter $Q$-operators. We provide both an algebraic derivation based on local operator identities and a diagrammatic proof using Feynman diagrammatic techniques.
	
	Finally, we apply the quantum Separation of Variables (SoV) method to the spectral problem of the one-boundary transfer matrix. We derive the system of separated difference equations for the transfer matrix eigenfunctions and show that, under the assumption of kernel factorization, it reduces to a one-dimensional difference equation coinciding precisely with the Baxter equation for eigenvalues of appropriately normalized one-parameter Baxter $Q$-operator.
	\end{abstract}

\newpage

 \tableofcontents

 \newpage

\section{Introduction}

We study a noncompact open $SL(2,\mathbb{C})$ spin magnet with boundary interaction~\cite{Skl:Boundary,Ch} of the most general form at both ends of the spin chain.
Noncompact $SL(2,\mathbb{C})$ spin chains are infinite-dimensional counterparts of the well-known spin-$\frac{1}{2}$ XXX Heisenberg magnet~\cite{TaFa,KulSk,Fad:BetheAns}, which corresponds to the two-dimensional representation of the group.
Beyond their mathematical interest, noncompact $SL(2,\mathbb{C})$ spin magnets are central to the analysis of high-energy quantum chromodynamics~\cite{DKM,L0,L1,L2,FK}. Moreover, they provide an integrable framework for two-dimensional and higher-dimensional conformal fishnet field theories~\cite{KO,DKO,DO1,DO2,DO3,Oliv,O1,O2,GKK,KG,GGKK,GKKNS,CK,BCF,BD,BD1,DFO,Duhr2,AO,AFKO,KO1}.

The model is treated within the framework of the quantum inverse scattering method~\cite{Skl91,Fad:BetheAns,FST,KulSk,TaFa}.
To the $k$-th site of the chain we associate the copy $V_k$ of space $\mathrm{L}^2(\mathbb{C})$ and the $L$-operator
\begin{align*}
	L(u) =
	\left (\begin{array}{cc}
		u + S & S_{-} \\
		S_{+} & u - S \end{array} \right )
\end{align*}
constructed from generators $S,\,S_\pm$ of unitary principal series representation of $SL(2,\mathbb{C})$ group acting in~$V_k$ and satisfying the standard commutation relations $[S_+,S_-]=2S$, $[S,S_\pm]=\pm S_\pm$.
The $L$-operator depends on the additional numerical parameter $u$ called spectral parameter.
The Hilbert space $H$ of the chain consisting of $n$ sites is given by the tensor product of $n$ copies of $\mathrm{L}^2(\mathbb{C})$: $H=V_1\otimes\ldots\otimes V_n$.

A systematic approach to constructing exactly solvable quantum chains with boundary conditions differing from (quasi)periodic ones was developed by Sklyanin~\cite{Skl:Boundary} on the basis of Cherednik's work~\cite{Ch} devoted to factorized scattering on a half-line.
In accordance with Sklyanin's method we consider the matrices
\begin{equation} \nonumber 
	K_-(u) = \begin{pmatrix}
		\gamma\bigl(g - \frac{1}{2} \bigr) & u - \frac{1}{2} \\[6pt]
		\gamma^2 \bigl(u - \frac{1}{2} \bigr) & \gamma\bigl(g - \frac{1}{2} \bigr)
	\end{pmatrix} \,,
	\qquad
	K_+(u) = \begin{pmatrix}
		\tilde{\gamma}\bigl(\tilde{g} - \frac{1}{2} \bigr) & -u - \frac{1}{2} \\[6pt]
		\tilde{\gamma}^2\bigl(-u - \frac{1}{2} \bigr) & \tilde{\gamma}\bigl(\tilde{g} - \frac{1}{2} \bigr)
	\end{pmatrix} ,
\end{equation}
which define the boundary interaction at different ends of the chain.
They are general scalar solutions of the so-called reflection equations (boundary Yang-Baxter equations), with Yang's $R$-matrix acting in $\mathbb{C}^2\otimes\mathbb{C}^2$. The latter satisfies the Yang-Baxter equation.
Reflection equations yield the integrability of the model.
A complete set of observables is constructed with the help of the monodromy matrix.
We consider two cases: with boundary interaction at one of the ends and at both ends of the chain, and refer to the corresponding models as one-boundary and two-boundary spin chain.
These models are defined by monodromy matrices of the form~\cite{Skl:Boundary, KS, PP}
\begin{align}
	& \label{Tint}
	T_n(u) = L_n(u) \cdots L_1(u) K_-(u) L_1(u) \cdots L_n(u) = \begin{pmatrix}
		A_n(u) & B_n(u) \\
		C_n(u) & D_n(u)
	\end{pmatrix} ,
	\\
	& \nonumber
	\hat{T}_n(u) =K_+(u)\,T_n(u) =K_+(u) L_n(u) \cdots L_1(u) K_-(u) L_1(u) \cdots L_n(u) = \begin{pmatrix}
		\hat{A}_n(u) & \hat{B}_n(u) \\
		\hat{C}_n(u) & \hat{D}_n(u)
	\end{pmatrix} ,
\end{align}
where each $L$-operator $L_k$ acts nontrivially in the space $V_k$.
The generating function of the integrals of motion is the trace of the monodromy matrix, which is called the transfer matrix
\begin{equation} \nonumber
	t_n(u) = \tr_{\mathbb{C}^2}T_n(u)=A_n(u)+D_n(u) \,, \quad
	\hat{t}_n(u) = \tr_{\mathbb{C}^2}\hat{T}_n(u) \,.
\end{equation}

By construction, $t_n(u)$ and $\hat{t}_n(u)$ are operator-valued polynomials in spectral parameter~$u$.
Similarly to~\cite[Section~4]{Skl:Boundary} one can show that in the present case of reflection algebras determined by Yang's $R$-matrix these transfer matrices are even functions of~$u$.
As it is demonstrated in Section~\ref{sect:Monodromy}, $t_n(u)$ and $\hat{t}_n(u)$ have the form
\begin{align}
	& \nonumber
	t_n(u) \sim
	2\gamma\bigl(u^2-\frac{1}{4}\bigr)\biggl(u^{2n-2}H_1+u^{2n-4}H_2+\ldots+H_n\biggr) \,,\\
	& \label{t_expl_int}
	\hat{t}_n(u) \sim
	-\bigl(\gamma^2+\tilde{\gamma}^2\bigr)\bigl(u^2-\frac{1}{4}\bigr)\biggl(u^{2n}+u^{2n-2}\hat{H}_1+\ldots+\hat{H}_n\biggr)
\end{align}
up to constant terms (independent of $u$) proportional to identity operator.
The operators $H_k$ and $\hat{H}_k$ form the sets of $n$ independent integrals of motion of the corresponding models.
Commutativity of these sets follows from commutativity of transfer-matrices at different values of spectral parameter
\begin{equation} \nonumber
	[t_n(u), t_n(v)] = 0 \,, \qquad [\hat{t}_n(u),\hat{t}_n(v)] = 0 \,.
\end{equation}
The latter properties are, in turn, consequences of reflection equations for $K_+$ and monodromy matrix $T_n$ (the latter generalizes the reflection equation for $K_-$).
This way, the problem of determining eigenstates of the model (the common eigenstates of commuting integrals of motion) is equivalent to the spectral problem for transfer matrix at arbitrary values of spectral parameter.

From~\eqref{t_expl_int} it follows that eigenvalues of transfer matrices $t_n(u)$ and $\hat{t}_n(u)$ are also even polynomial functions of $u$ proportional to $u^2-\frac{1}{4}$ (again, up to constant terms), the roots of these polynomials are of the form $\pm x_1,\dots, \pm x_{n-1}$ and $\pm y_1, \dots, \pm y_n$, correspondingly, not counting the obvious roots $\pm\frac{1}{2}$.
This way, the eigenvalues and eigenfunctions of $t_n(u)$ and $\hat{t}_n(u)$ are parametrized by the numbers $x_1,\dots, x_{n-1}$ and $y_1, \dots, y_n$, respectively.\footnote{In the case of $t_n(u)$ one also needs to introduce an additional parameter $\lambda$ (see~\eqref{eigenvalue}), because the leading coefficient of the polynomial is not the identity.}
We call these numbers spectral variables.
Obviously, the eigenvalues of $t_n(u)$ are invariant under permutations of $x_1,\dots,x_{n-1}$ and reflections $x_k\to -x_k$. In other words, the group of eigenvalues' symmetries is $\mathbb{Z}_2\ltimes\mathfrak{S}_{n-1}$ -- the Weyl group of B and C root systems.
If the spectrum is simple, the eigenfunctions also possess this symmetry.
A similar situation occurs for the transfer matrix $\hat{t}_n(u)$: its eigenvalues (and eigenfunctions in the case of simple spectrum) are symmetric with respect to the same transformations of spectral variables~$y_1, \dots, y_n$.
Therefore, we call the considered model \textit{the spin chain of BC-type}, by analogy with BC-type Toda lattice~\cite{KuzSk,IS,BDK1,BDK2}.
Note that the spectrum and eigenfunctions of the relative model determined by the operator $B_n(u)$~\eqref{Tint} enjoy the same symmetry~\cite{ADV2}, as well as the Gustafson integral of BC-type (generalized to the complex field)~\cite[eq.~(4.2)]{DM:Gust3}, \cite[eq.~(2.3b)]{DM:Gust4} used to prove the completeness of those eigenfunctions. In the present work we use the latter eigenfunctions to construct the representation of separated variables in the case of spectral problem for $t_n(u)$.

In~\cite{Skl:Boundary} the transfer matrix of open spin-$\frac{1}{2}$ XXZ Heisenberg magnet was diagonalized in the case of diagonal boundary $K$-matrices by means of Algebraic Bethe Ansatz (ABA)~\cite{Fad:BetheAns,FST,KulSk}.
It is much harder to implement this approach when $K_\pm$ are not diagonal.
Modifications of ABA for spin chains with general open boundary conditions were developed in the case of XXZ chain of spin~$\frac{1}{2}$ in~\cite{Belliard2,Belliard3,Belliard4} and \cite{Cao,YangZhang}.

A more powerful tool for studying open spin magnets is the quantum Separation of Variables (SoV)~\cite{Skl:VarSep,Skl91,KK}.
It was successfully used for constructing a complete set of eigenvectors of the transfer matrix for XXX, XXZ and XYZ chains of spin $\frac{1}{2}$ and $GL(3)$ chain (in the fundamental representation) with $K$-matrices of general form \cite{Nicc12,FKN14,FN14,KMN14,KMNT17,KMNT18,MNopen}.
The work~\cite{MNopen} also contains some results for higher-rank $GL(N)$ chains.

All the mentioned results were obtained for compact spin chains, that is, those determined by the finite-dimensional representations of corresponding symmetry groups (or quantum groups).
In the case of a noncompact chain with an infinite-dimensional principal series representation of $SL(2,\mathbb{C})$ at each site, there is no reference state for the algebraic Bethe ansatz, and therefore this method cannot be applied.
Instead, one can use the aforementioned quantum Separation of Variables invented by Sklyanin in the context of determining the spectrum of (quasi)periodic 
Toda chain~\cite{Skl:Toda} (not amenable to solution by means of ABA for the same reason) and further developed by Kharchev and Lebedev~\cite{KL1,KL2,KL3} for constructing the wavefunctions of this model.
The combination of SoV and Baxter $Q$-operator method~\cite{Bax82} proved to be effective in transfer matrix diagonalization for various noncompact quantum chains with periodic boundary conditions~\cite{KSS00,DKM3,DKM,KMS03,BT06,FMM17,DV19,O1} and with boundary interaction~\cite{DKM2,IS,Belits19}.

In the present work we construct the two-parameter $Q$-operator for the considered model and derive the eigenvalue equations for wavefunctions of one-boundary chain in the SoV representation.
We prove the characteristic properties of $Q$-operator: commutation with the transfer matrix, commutativity of $Q$-operators at different values of parameters and the Baxter equation ($tQ$-relation) with respect to each of the parameters.
It is shown that at two special points it reduces to one-parameter $Q$-operators of simpler form.
We generalized to the case of open spin chain the idea of $Q$-operator construction from the paper~\cite{Derkachov:2005fg} devoted to the periodic $SL(2,\mathbb{C})$ chain in the same principal series representations of the group.
The $Q$-operator coincides with the general transfer matrix (up to a linear change of parametrization), which is defined as the trace of a general monodromy matrix over an infinite-dimensional auxiliary space carrying a principal series representation.
The latter is given by product of $\mathbb{R}$-operators~\cite{DKM,DM09} and $\mathbb{K}_\pm$-operators~\cite{ADV2,ABDV} -- the general infinite-dimensional solutions of the Yang-Baxter and reflection equations which generalize the notions of $L$-operator and $K_\pm$-matrices.

A similar idea that the $Q$-operator can be expressed as a trace of some general monodromy matrix associated with an infinite-dimensional representation of the corresponding quantum group was first formulated in~\cite{BLZ} for CFT models and in~\cite{AF} for quantum-mechanical models described by the $U_q(\widehat{sl(2)})$-invariant $R$-matrix.
In the papers~\cite{Frassek_2015,VlaarWest,CoopVlaarWest} these results were generalized to the case of XXX and XXZ chains of spin $\frac{1}{2}$ with boundary interaction.

The key ingredients of the proof of Baxter equations for noncompact $SL(2,\mathbb{C})$ spin chains are the so-called triangularity relations for $\mathbb{R}$ and $L$ operators~\cite{Derkachov:2005fg}: at some special points their product, conjugated by a coordinate-dependent matrix, becomes an upper-triangular matrix.
These relations follow from factorization of $\mathbb{R}$-operator~\cite{Derkachov:2005hw,DM09} into the product of operators $\mathcal{R}^{(1)}$ and $\mathcal{R}^{(2)}$ having simpler structure.
The latter solve the Yang-Baxter $\mathcal{R}LL$-relations with $L$-operator.
We prove similar triangularity relations for $\mathbb{K}_\pm$-operators.
Note that the factorization property of the $\mathbb{R}$-operator also plays an important role in the one-parameter reductions of the $Q$-operator: in the first case, the $\mathbb{R}$-operator reduces to~$\mathcal{R}^{(1)}$, whereas in the second case, it reduces to~$\mathcal{R}^{(2)}$.

The general transfer matrix factorizes into the product of two one-parameter $Q$-operators, a similar property holds for various integrable models~\cite{BLZ,AF,DKM}.
$Q$-operators take the form of integral operators, and we give a convenient diagrammatic representation of their kernels.
The diagram technique formulated in Appendix~\ref{App-DiagTech} simplifies calculations with integral operators reducing them to elementary transformation rules for Feynman diagrams~\cite{DKM,DM14,DM:Gust3,DKM3,DKM2}.
We use it to derive the transfer matrix factorization.

The Separation of Variables in our case consists in decomposing the eigenfunctions of one-boundary transfer matrix $t_n(u)$ over the basis of eigenfunctions of the $B$-element of monodromy matrix $T_n(u)$~\eqref{Tint}.
The kernel of the corresponding integral expansion plays the role of transfer matrix eigenfunction in the representation of separated variables.
The complete orthogonal set of $B$-element's eigenfunctions was constructed by induction in~\cite{ADV2}.
The function for the chain consisting of $n$ sites is obtained by applying the raising integral $\Lambda_n$-operator to the function for $n-1$ sites.

The paper is organized as follows. Section~\ref{sect:Rep} contains all the needed information about principal series representations of $SL(2,\mathbb{C})$.
In Section~\ref{sect:ReflEq} we give explicit formulas for $\mathbb{R}$- and $\mathbb{K}_\pm$-operators and describe the hierarchy of Yang-Baxter and reflection equations in the case of $SL(2,\mathbb{C})$ symmetry group.
The most general are the reflection equations for $\mathbb{R}$ and $\mathbb{K}_\pm$ proven in~\cite{ADV2} and the Yang-Baxter equation for $\mathbb{R}$-operator~\cite{DKM,DM09}.
Other relations, which involve the $L$-operator, $K_\pm$-matrices and Yang's $R$-matrix, are finite-dimensional reductions of these general identities.
All mentioned relations yield the commutativity of transfer matrices and $Q$-operators.

In Section~\ref{sect:Monodromy} we define the model in the language of quantum inverse scattering method and specify the form of $t_n(u)$ and $\hat{t}_n(u)$. At the very end we formulate the spectral problem for the transfer matrix $t_n(u)$ of one-boundary chain.
The general monodromy and transfer matrices are introduced in Section~\ref{sect:generalT}.
We prove that general transfer matrices form the commutative family and commute with the ordinary transfer matrix (the trace over $\mathbb{C}^2$).
These commutativity relations are in fact the first two characteristic properties of the $Q$-operator.
The proofs are given for the two-boundary chain and derivations in the one-boundary case are similar.

The section~\ref{sect:Q} is devoted to Baxter $Q$-operator. The two-parameter $Q$-operator is defined as the general transfer matrix with a linear change of parameters.
We establish the equivalence of the two triangularity relations for the $\mathbb{R}$-operator derived in~\cite{Derkachov:2005fg}, derive the triangularity relations for the $\mathbb{K}_\pm$-operators, and prove the Baxter equations with respect to both parameters of the $Q$-operator in the two-boundary case. The corresponding calculations for the one-boundary chain are similar but somewhat shorter.
Finally, we describe the reductions to one-parameter $Q$-operators.

The factorization formulas for general transfer matrices of one- and two-boundary chains are proven in Section~\ref{sect:Qfact}.
First, we present the derivations using an algebraic approach similar to those in~\cite{ADV1},~\cite{ADV2}. It is based on local operator relations like the Yang-Baxter equations for $\mathcal{R}$-operators or formulas for traces of $\mathcal{R}$-operator products over~$\mathrm{L}^2(\mathbb{C})$ space.
Then we give the diagrammatic proof.

In Section~\ref{sect:SoV} we apply the quantum Separation of Variables method to the spectral problem for transfer-matrix $t_n(u)=A_n(u)+D_n(u)$ of one-boundary chain.
Our approach is inspired by the work~\cite{IS}, where the same method was used to construct the wavefunctions of B-type Toda chain.
We calculate the action of elements $A_n(u)$ and $D_n(u)$ of monodromy matrix $T_n(u)$ on the eigenfunctions of $B_n(u)$ using a certain relation for $T_n(u)$ and the $\Lambda_n$-operator used in the inductive construction of these eigenfunctions.
With the help of the obtained key relations we derive the system of separated difference equations for the eigenfunctions of transfer matrix in SoV representation.
At the end of the Section it is demonstrated that these equations coincide with the Baxter equation for eigenvalues of properly normalized one-parameter $Q$-operator, provided that the eigenfunctions in SoV representation can be written in the factorized form with each factor depending on one of the separated variables.

In Section~\ref{sect:conclusion} we collect and discuss the main results of the present work. Some auxiliary formulas can be found in Appendices.

\section{Principal series representations of $SL(2,\mathbb{C})$} \label{sect:Rep}

To define the model, we first recall the standard formulas for the \textit{principal series representations} of the group $SL(2,\mathbb{C})$~\cite[Ch.~III]{GGV}.
These representations are defined on the Hilbert space $\mathrm{L}^2(\mathbb{C})$ of square-integrable functions on $\mathbb{C}$, equipped with the inner product
\begin{equation} \label{sp}
    \langle \Phi|\Psi\rangle = \int \mathrm{d}^2z \; \overline{\Phi(z,\bar{z})}\,\Psi(z, \bar{z}) \,.
\end{equation}
The integration measure is the standard Lebesgue measure on $\mathbb{C}$: $\mathrm{d}^2 z = \mathrm{d}\Re z \,\mathrm{d}\Im z$. Throughout what follows, the integration symbol $\int \mathrm{d}^2z$ denotes integration over the entire complex plane.

We denote complex conjugation by $*: a \mapsto a^*$, so that in general $\bar{a}$ is distinct from the complex conjugate of $a$.
The only exceptions to this rule are the coordinate variable $z$, for which $\bar{z} \equiv z^*$, and the complex conjugation of functions, written as $\overline{\Psi(z, \bar{z})}$. 

Each representation is parametrized by a pair of ``spins'' $(s, \bar{s})$ satisfying $2(s - \bar{s}) \in \mathbb{Z}$.
An element $g$ of the group $SL(2,\mathbb{C})$,
\begin{equation} \nonumber
    g =
    \begin{pmatrix}
        a & b \\
        c & d
    \end{pmatrix}, \qquad ad - bc = 1,
\end{equation}
is represented by the operator $T^{(s,\bar{s})}(g)$ acting on functions $\Psi(z, \bar{z})$ according to the formula
\begin{equation} \nonumber
    [T^{(s,\bar{s})}(g) \, \Psi] (z) =
    [d-bz]^{-2s} \, \Psi \biggl(\frac{-c+az}{d-bz}\biggr) \,.
\end{equation}  
For conciseness, we display only the holomorphic argument $z$ implying that $\Psi(z) \equiv \Psi(z, \bar{z})$.
We also introduce here the notation for complex powers
\begin{align} \label{power}
    [z]^a \equiv z^a \bar{z}^{\bar{a}} = |z|^{a+\bar{a}} \, {e}^{\imath(a-\bar{a}) \arg z} \,,
\end{align}
where the pair $(a,\bar{a})$ must satisfy the condition $a-\bar{a} \in\mathbb{Z}$ so that the power function is single-valued.
The condition $2(s - \bar{s}) \in \mathbb{Z}$ follows exactly from this restriction.

This representation is unitary provided that the following condition holds
\begin{equation} \label{scond}
    s^\ast + \bar{s} = 1 \,,
\end{equation}
where $s^\ast$ denotes the complex conjugate of $s$. Combining it with the restriction $2(s - \bar{s}) \in \mathbb{Z}$, we obtain the following parametrization of the spins
\begin{equation}\label{sparam}
    s = \frac{1+n_s}{2} + \imath\nu_s, \qquad
    \bar{s} = \frac{1-n_s}{2} + \imath\nu_s \,,
    \qquad
    n_s \in \mathbb{Z}+\sigma, \quad \nu_s \in \mathbb{R},
\end{equation}
where the parameter
\begin{equation} \label{sigma}
    \sigma \in \left\{0, \tfrac{1}{2}\right\}
\end{equation}
is fixed for the rest of the paper.

We also introduce the generators of the representation
\begin{align} \label{gen}
    \begin{aligned}
        &S = z\partial_z + s\,, \qquad S_- = -\partial_z\,, \qquad
        S_+ = z^2\partial_z+2sz\,, \\[2pt]
        &
        \bar{S} = \bar{z}\partial_{\bar{z}} + \bar{s}\,, \qquad \bar{S}_- = -\partial_{\bar{z}}\,, \qquad
        \bar{S}_+ = \bar{z}^2\partial_{\bar{z}}+2\bar{s}\bar{z} \,.
    \end{aligned}
\end{align}
The generators in the set $\{S, S_{\pm}\}$ are called holomorphic, they commute with the antiholomorphic generators $\{\bar{S}, \bar{S}_\pm\}$. Each set obeys the commutation relations of the Lie algebra~$sl(2,\mathbb{C})$,
\begin{align} \nonumber
    [S_+, S_-] = 2S, \qquad [S,S_\pm] = \pm S_\pm,
\end{align}
and similarly for the antiholomorphic generators. Under the inner product~\eqref{sp} and subject to condition~\eqref{scond}, the two sets of generators are Hermitian conjugates of each other:
\begin{equation} \nonumber
    S^\dagger = -\bar{S}, \qquad S_-^\dagger = -\bar{S}_-, \qquad
    S_+^\dagger = -\bar{S}_+ \,.
\end{equation}

The representations characterized by the parameters $(s, \bar s)$ and $(1-s, 1-\bar{s})$ are isomorphic \cite{Gelfand-Naimark}.
For generic complex $s$ and $\bar{s}$ there exists an intertwining isomorphism $W(s,\bar{s})$ between the principal series representations $T^{(s,\bar{s})}$ and $T^{(1-s,1-\bar{s})}$, satisfying
\begin{equation} \nonumber
    W(s,\bar{s})\,T^{(s,\bar{s})}(g) =
    T^{(1-s,1-\bar{s})}(g)\,W(s,\bar{s})\,.
\end{equation}
The operator $W$ is an integral operator given by \cite{GGV}
\begin{align} \nonumber
    W(s,\bar{s}) := \left[\hat{p}\right]^{1-2s} \,.
\end{align}
Here, for any pair $(\alpha,\bar{\alpha})$ such that $\alpha-\bar{\alpha}\in\mathbb{Z}$, the operator $[\hat{p}]^{\alpha}$ is defined as
\begin{align} \label{d} 
    \left[\hat{p}\right]^{\alpha}\Phi(z,\bar z) :=
    c(\alpha)\,
    \int \mathrm{d}^2 w\,
    \frac{\Phi(w,\bar w)}{[z-w]^{1+\alpha}} \,.
\end{align}
As before, for the sake of brevity we display only the holomorphic parameter $\alpha$.
The normalization coefficient $c(\alpha)$ is given by
\begin{align}\label{c}
    c(\alpha) = \frac{[\imath]^{\alpha}\,\bm{\Gamma}(\alpha+1)}{\pi} \,,
\end{align}
where $[\imath]^{\alpha} = \imath^{\alpha-\bar{\alpha}}$ and $\bm{\Gamma}(\alpha)$ denotes the gamma function of the complex field~\cite{GGR,N}:
\begin{equation} \nonumber
    \bm{\Gamma}(\alpha):=\frac{\Gamma(\alpha)}{\Gamma(1-\bar\alpha)} \,.
\end{equation}

We adopt the notation $[\hat{p}]^\alpha$ for the following reason: via the Fourier transform, operator \eqref{d} can be interpreted as the double power $\hat{p}^\alpha\,\hat{\bar{p}}^{\bar{\alpha}}$ of the momentum operators $\hat{p}=-\imath\partial_z$ and $\hat{\bar{p}}=-\imath\partial_{\bar{z}}$ \cite[Section~2.1]{ADV1}, \cite[Section~2.3]{DM09}. For the motivation behind the choice of normalization \eqref{c}, see \cite[Section~2.1]{ADV1}.

Operators of type \eqref{d} are used extensively throughout the text. Using the chain relation \eqref{Chain} and its limiting case \eqref{delta2}, one can verify that with the normalization \eqref{c}, these operators satisfy
\begin{equation} \nonumber
    [\hat{p}]^\alpha\,[\hat{p}]^\beta=[\hat{p}]^{\alpha+\beta} \,, \qquad
    [\hat{p}]^\alpha\,[\hat{p}]^{-\alpha} = \II \,,
\end{equation}
where $\II$ is the identity operator.
From definition \eqref{d} and the expression for the inner product \eqref{sp}, it is straightforward to derive the conjugation rule
\begin{align} \nonumber
    \left(\left[\hat{p}\right]^{\alpha}\right)^{\dagger} =  
    \left[\hat{p}\right]^{\bar{\alpha}^*} \,.
\end{align}
We also require the conjugation rule for the operator of multiplication by a power function:
\begin{equation} \nonumber
    ([z]^\alpha)^\dagger=([z]^\alpha)^\ast=[z]^{\bar{\alpha}^\ast}=z^{\bar{\alpha}^\ast}\bar{z}^{\alpha^\ast} \,.
\end{equation}

\section{$SL(2,\mathbb{C})$ spin chain of BC type}

We are going to introduce the model in section~\ref{sect:Monodromy}.
But first let us consider in section~\ref{sect:ReflEq} the objects which are used to define our model in the language of quantum inverse scattering method, including $L$-operator, Yang's $R$-matrix and Sklyanin's $K$-matrix.

The first object is the general $\mathbb{R}$-operator acting in the tensor product of two principal series representations. It is the general solution of the Yang-Baxter equation corresponding to the group $SL(2,\mathbb{C})$. The $L$-operator and $R$-matrix can be obtained as certain reductions of this universal $\mathbb{R}$-operator.
In addition, we consider two reflection operators $\mathbb{K}_-$ and $\mathbb{K}_+$ that also act in the space of principal series representation. Together with the $\mathbb{R}$-operator they form two isomorphic reflection algebras. Their finite-dimensional reductions are the Sklyanin's $K_{+}$ and $K_-$ matrices which define the boundary interaction in the model.

\subsection{Reflection equations} \label{sect:ReflEq}

The general reflection equations have the following form \cite{Skl:Boundary,PP,Oliv}
\begin{align}
	& \nonumber
	\mathbb{R}_{12}(u-v)\,\overset{1}{\mathbb{K}}_-(u,s_1)\,\mathbb{R}_{12}(u+v)\,\overset{2}{\mathbb{K}}_-(v,s_2) \\
	& \label{Refl}
	=
	\overset{2}{\mathbb{K}}_-(v,s_2)\,\mathbb{R}_{12}(u+v)\,\overset{1}{\mathbb{K}}_-(u,s_1)\,\mathbb{R}_{12}(u-v) \,.
\end{align}
and 
\begin{align}
	& \nonumber
	\mathbb{R}_{12}(v-u)\,\overset{1}{\mathbb{K}}_+(u,s_1)\,\mathbb{R}_{12}(-u-v-2)\,\overset{2}{\mathbb{K}}_+(v,s_2) \\
	& \label{Refl+}
	=
	\overset{2}{\mathbb{K}}_+(v,s_2)\,\mathbb{R}_{12}(-u-v-2)\,\overset{1}{\mathbb{K}}_+(u,s_1)\,\mathbb{R}_{12}(v-u) \,.
\end{align}
Operator $\mathbb{R}_{12}(u)$ acts in the tensor product $V_1\otimes V_2$ of two principal series representations of $SL(2,\mathbb{C})$ with spins $s_1$ and $s_2$.
It depends on spectral parameter $u$ and the spins which determine the representations.
It is also a function of antiholomorphic counterparts of these parameters, which are not shown for brevity.
Generally speaking, we omit the dependence on the antiholomorphic parameters for all objects throughout the text and show them only if it is necessary.
Reflection operators $\overset{1}{\mathbb{K}}_{\pm}(u,s_1)$ act nontrivially in the space $V_1$, and the corresponding operators $\overset{2}{\mathbb{K}}_{\pm}(v,s_2)$ act in the space $V_2$.
The $\mathbb{R}$-operator is the general solution of the Yang-Baxter equation with symmetry group $SL(2,\mathbb{C})$~\cite{DM09,DKM,Derkachov:2005hw}
\begin{align} \label{YBgen}
	\mathbb{R}_{12}(u-v)\,\mathbb{R}_{13}(u)\,\mathbb{R}_{23}(v) = 
	\mathbb{R}_{23}(v)\,\mathbb{R}_{13}(u)\,\mathbb{R}_{12}(u-v) \,.
\end{align}
The last identity is the operator relation in the tensor product $V_1\otimes V_2\otimes V_3$ of three representations, each operator $\mathbb{R}_{ij}$ acts nontrivially in spaces $V_i$ and $V_j$ and as the identity operator in the third space.

In the case of two-dimensional representations $V_1=V_2=\mathbb{C}^2$ the operator $\mathbb{R}_{12}(u)$ reduces \cite{DM09,CDS} to the Yang's $R$-matrix in $\mathbb{C}^2\otimes \mathbb{C}^2$ -- a finite-dimensional solution to the Yang-Baxter equation
\begin{equation} \label{Rmat1}
	R_{12}(u) = R(u) = u  + P \,, \qquad P\,a \otimes b = b \otimes a \,.
\end{equation}
The reflection equations \eqref{Refl},~\eqref{Refl+} in this case take the form
\begin{equation} \nonumber
    \begin{aligned}
        &R(u-v)\, \bigl(k_-(u)\otimes \bm{1}\bigr)\,R(u+v)\, \bigl(\bm{1} \otimes k_-(v)\bigr) \\
        &\quad = \bigl(\bm{1} \otimes k_-(v)\bigr)\,R(u+v)\,\bigl(k_-(u)\otimes \bm{1}\bigr)\, R(u-v) \,, \\[0.6em]
        &R(v-u)\, \bigl(k_+(u)\otimes \bm{1}\bigr)\,R(-u-v-2)\, \bigl(\bm{1} \otimes k_+(v)\bigr) \\
        &\quad = \bigl(\bm{1} \otimes k_+(v)\bigr)\,R(-u-v-2)\,\bigl(k_+(u)\otimes \bm{1}\bigr)\, R(v-u) \,.
    \end{aligned}
\end{equation}
where $k_-(u)$, $k_+(u)$ are $2\times 2$ matrices and $\bm{1}$ is a $2\times 2$ identity matrix. 
It appears to be convenient to shift the spectral parameter $u\to u-\frac{1}{2}$ and define
\begin{align*}
	K_-(u) = k_-(u-\frac{1}{2}) \,, \qquad K_+(u) = k_+(u-\frac{1}{2}) \,,
\end{align*}
Then the reflection equations read
\begin{align} 
    & \nonumber
    R(u-v)\, \bigl(K_-(u)\otimes \bm{1}\bigr)\,R(u+v-1)\, \bigl(\bm{1} \otimes K_-(v)\bigr) \\
    \label{Reflshifted}
    &\quad = \bigl(\bm{1} \otimes K_-(v)\bigr)\,R(u+v-1)\,\bigl(K_-(u)\otimes \bm{1}\bigr)\, R(u-v) \,, \\[0.6em]
    & \nonumber
    R(v-u)\, \bigl(K_+(u)\otimes \bm{1}\bigr)\,R(-u-v-1)\, \bigl(\bm{1} \otimes K_+(v)\bigr) \\
    \label{Reflshifted+}
    &\quad = \bigl(\bm{1} \otimes K_+(v)\bigr)\,R(-u-v-1)\,\bigl(K_+(u)\otimes \bm{1}\bigr)\, R(v-u) \,.
\end{align}

If we take $V_1 = V_2 = \mathbb{C}^2$ and $V_3$ to be
the space of arbitrary representation  
$T^{(s,\bar{s})}$ with generators~\eqref{gen}
\begin{align*}
	S = z\partial_z + s \,, \qquad S_- = -\partial_z \,, \qquad 
	S_+ = z^2\partial_z+2sz \,,
\end{align*}
then we get a more complicated solution to the Yang-Baxter equation -- the $L$-operator
\begin{align} \label{L}
	L(u) =
	\left( \begin{array}{cc}
		u + S & S_{-} \\
		S_{+} & u - S \end{array} \right) .
\end{align}
The operator $\mathbb{R}_{12}$ reduces in this case to the Yang's $R$-matrix, and the operators $\mathbb{R}_{13}$, $\mathbb{R}_{23}$ degenerate into $L$-operators~\cite{DM09,CDS} (up to the shift of spectral parameters), so one obtains
\begin{align*}
	\textstyle R(u-v)\, \bigl(L(u+\frac{1}{2})\otimes \bm{1}\bigr)\, 
	\bigl(\bm{1} \otimes L(v+\frac{1}{2})\bigr)
	= \bigl(\bm{1} \otimes L(v+\frac{1}{2})\bigr)\,
	\bigl(L(u+\frac{1}{2})\otimes \bm{1}\bigr)\, R(u-v) \,.
\end{align*}
Substituting $u\to u-\frac{1}{2}$ and $v\to v-\frac{1}{2}$ we rewrite this Yang-Baxter equation in the more common form
\begin{align} \label{RLL}
	\textstyle R(u-v)\, \bigl(L(u)\otimes \bm{1}\bigr)\, 
	\bigl(\bm{1} \otimes L(v)\bigr)
	= \bigl(\bm{1} \otimes L(v)\bigr)\,
	\bigl(L(u)\otimes \bm{1}\bigr)\, R(u-v) \,.
\end{align}

Concerning reflection equations, in the case when $V_1 = \mathbb{C}^2$ and $V_2$ 
is the space of representation $T^{(s,\bar{s})}$ we arrive at the defining relations for reflection operators
\begin{align*} 
	&\textstyle L(u-v+\frac{1}{2})\,k_-(u)\,L(u+v+\frac{1}{2})\,\mathbb{K}_-(v\,,s) = 
	\mathbb{K}_-(v\,,s)\,L(u+v+\frac{1}{2})\,k_-(u)\,L(u-v+\frac{1}{2}) \,,\\
	&\textstyle L(v-u+\frac{1}{2})\,k_+(u)\,L(-u-v-\frac{3}{2})\,\mathbb{K}_+(v\,,s) = 
	\mathbb{K}_+(v\,,s)\,L(-u-v-\frac{3}{2})\,k_+(u)\,L(v-u+\frac{1}{2}) \,,
\end{align*}
or after the shift $u\to u-\frac{1}{2}$
\begin{align} \label{ReflL1}
	&\textstyle L(u-v)\,K_-(u)\,L(u+v)\,\mathbb{K}_-(v\,,s) = 
	\mathbb{K}_-(v\,,s)\,L(u+v)\,K_-(u)\,L(u-v) \,,\\
    \label{ReflL2}
	&\textstyle L(v-u+1)\,K_+(u)\,L(-u-v-1)\,\mathbb{K}_+(v\,,s) = 
	\mathbb{K}_+(v\,,s)\,L(-u-v-1)\,K_+(u)\,L(v-u+1) \,.
\end{align}

If we choose~$V_1$ and~$V_2$ to be the spaces of arbitrary representations~$T^{(s_1,\bar{s}_1)}$, $T^{(s_2,\bar{s}_2)}$ and take~$V_3=\mathbb{C}^2$, then the Yang-Baxter equation reduces to the defining relation for $\mathbb{R}$-operator
\begin{align} \label{YBRLL}
	\mathbb{R}_{12}(u-v)\,L_1(u)\,L_2(v) = 
	L_2(v)\,L_1(u)\,\mathbb{R}_{12}(u-v) \,,
\end{align}
where $L_k$ is the $L$-operator corresponding to representation $T^{(s_k,\bar{s}_k)}$
\begin{align*}
	L_1(u) = 
	\left(\begin{array}{cc} u+s_1+z_1 \partial_1 & -\partial_1\\
		z_1^2 \partial_1 + 2s_1 z_1 & u-s_1 -z_1
		\partial_1
	\end{array}\right) , \quad
	L_2(v) = 
	\left(\begin{array}{cc} v+s_2+z_2 \partial_2 & -\partial_2\\
		z_2^2 \partial_2 + 2s_2 z_2 & v-s_2 -z_2
		\partial_1
	\end{array}\right) .
\end{align*}
The general solution of equation~\eqref{YBRLL} was constructed in~\cite{DKM}, \cite[\S3]{DM09}, it reads
\begin{align}\label{RYB}
	\mathbb{R}_{12}(u) = \mathbb{P}_{12}\,\check{R}_{12}(u) = \mathbb{P}_{12}\,
	[z_{12}]^{u+1-s_1-s_2}\,
	[\hat{p}_1]^{u+s_2-s_1}\,[\hat{p}_2]^{u+s_1-s_2}\,
	[z_{12}]^{u+s_1+s_2-1} \,,
\end{align}
where $\mathbb{P}_{12}$ is the operator of permutation 
\begin{align} \label{perm}
	\mathbb{P}_{12}\,\Psi(z_1\,,z_2) = \Psi(z_2\,,z_1) \,.
\end{align}

We consider the general solution of the reflection equation~\eqref{Reflshifted} with Yang's $R$-matrix~\eqref{Rmat1}~\cite{Skl:Boundary,PP,KulSk}
\begin{align*}
	K_-(u) = \begin{pmatrix}
		\gamma\bigl(g - \frac{1}{2} \bigr) & u - \frac{1}{2} \\
		\gamma^2 \bigl(u - \frac{1}{2} \bigr) & \gamma\bigl(g - \frac{1}{2} \bigr)
	\end{pmatrix} \,,
\end{align*}
The corresponding operator $\mathbb{K}_-(u\,,s)$ was constructed in~\cite{ABDV}
\begin{align}\label{Kvs}
	\mathbb{K}_-(v\,,s) =
	[z+\gamma]^{v-s+g} \,
	[z-\gamma]^{v-s+1-g} \,
	[\hat{p}]^{2v} \,
	[z+\gamma]^{v+s-g} \,
	[z-\gamma]^{v+s+g-1} \,,
\end{align}
the reflection equation for $\mathbb{R}$- and $\mathbb{K}_-$-operators \eqref{RYB}, \eqref{Kvs} in the form~\eqref{Refl} was formulated in~\cite{ABDV,Oliv} and proven in~\cite{ADV2}.

Denote by $\tilde{K}_-(u)$ and $\tilde{\mathbb{K}}_-(u,s)$ the operators obtained from $K_-(u)$ and $\mathbb{K}_-(u,s)$ by replacing the parameters $\gamma$ and $(g,\bar{g})$ with corresponding parameters $\tilde{\gamma}$ and $(\tilde{g},\bar{\tilde{g}})$, which do not coincide with the previous ones in general
\begin{align}
	& \nonumber
	\tilde{K}_-(u) = \begin{pmatrix}
		\tilde{\gamma}\bigl(\tilde{g} - \frac{1}{2} \bigr) & u - \frac{1}{2} \\
		\tilde{\gamma}^2 \bigl(u - \frac{1}{2} \bigr) & \tilde{\gamma}\bigl(\tilde{g} - \frac{1}{2} \bigr)
	\end{pmatrix} , \\
	& \nonumber
	\tilde{\mathbb{K}}_-(v\,,s) =
	[z+\tilde{\gamma}]^{v-s+\tilde{g}} \,
	[z-\tilde{\gamma}]^{v-s+1-\tilde{g}} \,
	[\hat{p}]^{2v} \,
	[z+\tilde{\gamma}]^{v+s-\tilde{g}} \,
	[z-\tilde{\gamma}]^{v+s+\tilde{g}-1} \,.
\end{align}
In the following we define
\begin{align}
	\label{Kmat+}
	K_+(u) & := \tilde{K}_-(-u) = 
	\begin{pmatrix}
		\tilde{\gamma}\bigl(\tilde{g} - \frac{1}{2} \bigr) & -u - \frac{1}{2} \\
		\tilde{\gamma}^2 \bigl(-u - \frac{1}{2} \bigr) & \tilde{\gamma}\bigl(\tilde{g} - \frac{1}{2} \bigr)
	\end{pmatrix}
	, \\
	\nonumber
	\mathbb{K}_+(u\,,s) & :=\tilde{\mathbb{K}}_-(-1-u\,,s) \\
	\label{Kvs+}
	& \; =
	[z+\tilde{\gamma}]^{\tilde{g}-u-s-1} \,
	[z-\tilde{\gamma}]^{-\tilde{g}-u-s} \,
	[\hat{p}]^{-2u-2} \,
	[z+\tilde{\gamma}]^{-\tilde{g}-u+s-1} \,
	[z-\tilde{\gamma}]^{\tilde{g}-u+s-2} \,.
\end{align}	
Then it is easy to see that reflection equations~\eqref{Refl+}, \eqref{Reflshifted+} and \eqref{ReflL2} follow from~\eqref{Refl}, \eqref{Reflshifted+} and \eqref{ReflL1}, correspondingly.

In the paper~\cite{ABDV} the defining relation for the $\mathbb{K}_-$-operator \eqref{ReflL1} was considered in the following form
\begin{multline} \label{Kdef}
	\mathcal{K}(s,x) \, L(u +x - 1, u - s) \, K_-(u) \, L(u  + s - 1, u - x) \\
	= L(u + s - 1, u - x) \, K_-(u) \, L(u + x - 1, u - s) \, 
	\mathcal{K}(s,x) \,,
\end{multline}
where $L(u_1,u_2)$ is the alternative parametrization of $L$-operator $L(u)$~\cite{Derkachov:2005hw}
\begin{align} \label{Lfact}
	L(u_1, u_2) = 
	\left(\begin{array}{cc} u_1+1+z \partial_z & -\partial_z\\
		z^2 \partial_z + (u_1-u_2+1) z & u_2-z\partial_z
	\end{array}\right) 
	= 
	\left(%
	\begin{array}{cc}
		1 & 0 \\
		z & 1 \\
	\end{array}%
	\right)\,\left(%
	\begin{array}{cc}
		u_1 & -\partial_{z} \\
		0 & u_2 \\
	\end{array}%
	\right)\, \left(%
	\begin{array}{cc}
		1 & 0 \\
		-z & 1 \\
	\end{array}%
	\right)
\end{align}
corresponding to the change of variables $(u,s) \to (u_1, u_2)$,
\begin{align*}
	u_1 = u+s-1 \,, \quad  u_2 = u-s \,.
\end{align*}
Its solution was expressed as
\begin{align}\label{Kop-d}
	\mathcal{K}(s,x) =
	[z+\gamma]^{g-s} \,
	[z-\gamma]^{1-s-g} \,
	[\hat{p}]^{x-s} \,
	[z+\gamma]^{x-g} \,
	[z-\gamma]^{x+g-1}.
\end{align}
Note that the operator $\mathcal{K}(s,x)$ possesses the following properties 
\begin{equation}\nonumber
	\mathcal{K}(s,x)\,\mathcal{K}(x,y) = \mathcal{K}(s,y) \,, \quad \mathcal{K}(x,x) = \II \,,
\end{equation}
where by $\II$ we denote the identity operator.
These properties are direct consequences of the expression~\eqref{Kop-d}.

The equation~\eqref{Kdef} transforms into~\eqref{ReflL1} after the substitution
$s\to s-v, \, x\to s+v$.
Correspondingly, we have
\begin{equation} \nonumber
	\mathbb{K}_-(v,s) = \mathcal{K}(s-v,s+v) \,,
\end{equation}
and, taking into account~\eqref{Kvs+}, one obtains
\begin{align}\nonumber
	\mathbb{K}_+(v,s) = \tilde{\mathcal{K}}(s+v+1,s-v-1) \,,
\end{align}	
where by $\tilde{\mathcal{K}}(x,y)$ we denote the operator $\mathcal{K}(x,y)$ with $\gamma$ and $(g,\bar{g})$ replaced by $\tilde{\gamma}$ and $(\tilde{g},\bar{\tilde{g}})$
\begin{align} \label{Kop-d-tilde}
	\tilde{\mathcal{K}}(x,y) =
	[z+\tilde{\gamma}]^{\tilde{g}-s} \,
	[z-\tilde{\gamma}]^{1-s-\tilde{g}} \,
	[\hat{p}]^{x-s} \,
	[z+\tilde{\gamma}]^{x-\tilde{g}} \,
	[z-\tilde{\gamma}]^{x+\tilde{g}-1} \,.
\end{align}

\subsection{Monodromy matrix} \label{sect:Monodromy}

The Hilbert space of the model is given by the tensor product of the $\mathrm{L}^2(\mathbb{C})$ spaces,
\begin{equation}\label{HN}
	H=V_1\otimes V_2\otimes\cdots\otimes V_n,  \qquad V_k=\mathrm{L}^2(\mathbb{C})\,,\qquad
	k=1,\ldots,n \,.
\end{equation}
To each site $k$ we associate the quantum $L$-operators \eqref{L}
with subscript $k$, acting nontrivially in the $k$-th component of the tensor product~(\ref{HN})
\begin{equation} \nonumber
	L_k(u) =
	\left (\begin{array}{cc}
		u + S^{(k)} & S_{-}^{(k)} \\
		S_{+}^{(k)} & u - S^{(k)} \end{array} \right ) , \quad 
	\bar L(\bar u) = \left(\begin{array}{cc}
		\bar u + \bar S^{(k)} & \bar S_{-}^{(k)}\\
		\bar S_{+}^{(k)}& \bar u - \bar S^{(k)}
	\end{array}\right) .
\end{equation}
Functions in the space $V_k$ depend on variable $z_k$, and copies of generators \eqref{gen} acting in this space have the form
\begin{equation} \nonumber
	S^{(k)} = z_k\partial_k + s, \qquad  S^{(k)}_- = -\partial_k, \qquad
	S^{(k)}_+ = z_k^2\partial_k+2sz_k \,,
\end{equation}
expressions for the antiholomorphic generators are similar. That is, we consider only homogeneous chains with spin parameters $(s,\bar{s})$ being equal at all sites.\\
The $L$-operator possesses the property of unitarity 
\begin{equation} \label{unitarity}
	L(u)\,L(1-u) = \rho_s(u)\,\bm{1}, \qquad \rho_s(u) = (u-s)(u+s-1) \,.
\end{equation} 

The monodromy matrices are defined as follows \cite{Skl:Boundary,KS,PP}
\begin{align}
	& \label{Top}
	T_n(u) = L_n(u) \cdots L_1(u)\,K_-(u)\,L_1(u) \cdots L_n(u) = \begin{pmatrix}
		A_n(u) & B_n(u) \\
		C_n(u) & D_n(u)
	\end{pmatrix} , 
	\\
	& \label{Top1}
	\hat{T}_n(u) = K_+(u)\,L_n(u) \cdots L_1(u)\,K_-(u)\,L_1(u) \cdots L_n(u) = \begin{pmatrix}
		\hat{A}_n(u) & \hat{B}_n(u) \\
		\hat{C}_n(u) & \hat{D}_n(u)
	\end{pmatrix} .
\end{align}
Their antiholomorphic counterparts $\bar{T}_n(\bar{u})$, $\bar{\hat{T}}_n(\bar{u})$ are defined by 
means of $\bar{L}_k(\bar{u})$ and $\bar{K}_\pm(\bar{u})$ in the same way.
The $K$-matrix $K_-$ is the general solution of the reflection equation~\eqref{Reflshifted} with Yang's $R$-matrix~\cite{Skl:Boundary,PP,KulSk}
\begin{align} \label{K-}
	K_-(u) = \begin{pmatrix}
		\gamma\bigl(g - \frac{1}{2} \bigr) & u - \frac{1}{2} \\[6pt]
		\gamma^2 \bigl(u - \frac{1}{2} \bigr) & \gamma\bigl(g - \frac{1}{2} \bigr)
	\end{pmatrix},
	\qquad
	\bar{K}_-(\bar{u}) = \begin{pmatrix}
		\bar{\gamma}\bigl(\bar{g} - \frac{1}{2} \bigr) & \bar{u} - \frac{1}{2} \\[6pt]
		\bar{\gamma}^2 \bigl(\bar{u} - \frac{1}{2} \bigr) & \bar{\gamma}\bigl(\bar{g} - \frac{1}{2} \bigr)
	\end{pmatrix},
\end{align}
and the matrix $K_+$ is obtained from $K_-$ by means of~\eqref{Kmat+}
\begin{align} \label{K+}
	K_+(u) = \begin{pmatrix}
		\tilde{\gamma}\bigl(\tilde{g} - \frac{1}{2} \bigr) & -u - \frac{1}{2} \\[6pt]
		\tilde{\gamma}^2 \bigl(-u - \frac{1}{2} \bigr) & \tilde{\gamma}\bigl(\tilde{g} - \frac{1}{2} \bigr)
	\end{pmatrix},
	\qquad
	\bar{K}_+(\bar{u}) = \begin{pmatrix}
		\bar{\tilde{\gamma}}\bigl(\bar{\tilde{g}} - \frac{1}{2} \bigr) & -\bar{u} - \frac{1}{2} \\[6pt]
		\bar{\tilde{\gamma}}^2 \bigl(-\bar{u} - \frac{1}{2} \bigr) & \bar{\tilde{\gamma}}\bigl(\bar{\tilde{g}} - \frac{1}{2} \bigr)
	\end{pmatrix},
\end{align}
it solves the reflection equation~\eqref{Reflshifted+}.
The transfer matrix $T_n(u)$ satisfies the reflection equation \cite{Ch,Skl:Boundary,KS,PP}
\begin{multline}\label{refl_eq}
	R(u-v)\, \bigl(T_n(u)\otimes \bm{1}\bigr)\,R(u+v-1)\, \bigl(\bm{1} \otimes T_n(v)\bigr) \\[6pt]
	= \bigl(\bm{1} \otimes T_n(v)\bigr)\,R(u+v-1)\,\bigl(T_n(u)\otimes \bm{1}\bigr)\, R(u-v)
\end{multline}
(as well as its antiholomorphic analog $\bar{T}(\bar{u})$),
where we remind that Yang's $R$-matrix $R(u)$ acts in the tensor product $\mathbb{C}^2\otimes \mathbb{C}^2$
\begin{equation} \label{Rmat}
	R(u) = u  + P, \qquad P \, a \otimes b = b \otimes a.
\end{equation}
Using the \eqref{unitarity} we see that the $T_n$-matrix can be written in the following form
\begin{align*}
	T_n(u) = \rho_s^n(1-u) \, M_n(1-u)^{-1} \, K_-(u) \, M_n(u) \,.
\end{align*}
where $M_n(u)$ is the monodromy matrix of A-type open spin chain~\cite{DKM,ADV1}
\begin{equation} \nonumber
	M_n(u) = L_1(u) \cdots L_n(u) = \begin{pmatrix}
		a_n(u) & b_n(u) \\
		c_n(u) & d_n(u)
	\end{pmatrix} \,.
\end{equation}
Using the explicit expression for $K(u)$ we see that
\begin{equation} \label{T_n(1/2)}
	T_n\left(\frac{1}{2}\right) = \gamma \, \Bigl(g-\frac{1}{2}\Bigr) \, \rho_s^n\Bigl(\frac{1}{2}\Bigr) \, \bm{1} \,.
\end{equation}	
The transfer matrices $t_n(u)$ and $\hat{t}_n(u)$ are defined as traces of $T_n(u)$ and $\hat{T}_n(u)$ over the matrix space
\begin{align}
	& \nonumber
	t_n(u) = \tr_{\mathbb{C}^2} T_n(u)
	= \tr_{\mathbb{C}^2}\Bigl[L_n(u) \cdots L_1(u)\,K_-(u)\,L_1(u) \cdots L_n(u)\Bigr]
	= A_n(u)+D_n(u) \,, \\
	& \nonumber
	\hat{t}_n(u) = \tr_{\mathbb{C}^2} \hat{T}_n(u)
	= \tr_{\mathbb{C}^2}\Bigl[K_+(u)\,L_n(u) \cdots L_1(u)\,K_-(u)\,L_1(u) \cdots L_n(u)\Bigr] \,.
\end{align}
From \eqref{refl_eq}, \eqref{Reflshifted+} and the antiholomorphic analogues of these identities follows the commutativity~\cite[Section~3]{Skl:Boundary}
\begin{align} \nonumber
	[t_n(u), t_n(v)] = 0 \,, \qquad [\bar{t}_n(\bar{u}), \bar{t}_n(\bar{v})] = 0 \,, \\
	[\hat{t}_n(u), \hat{t}_n(v)] = 0 \,, \qquad [\bar{\hat{t}}_n(\bar{u}), \bar{\hat{t}}_n(\bar{v})] = 0 \,.
\end{align}
The holomorphic and antiholomorphic operators also commute with each other
\begin{equation} \nonumber
	[t_n(u), \bar{t}_n(\bar{v})] = 0 \,, \quad [\hat{t}_n(u), \bar{\hat{t}}_n(\bar{v})] = 0 \,.
\end{equation}

Now, denote by $u$ and $\bar{u}$ the pair of complex conjugate variables
\begin{equation} \nonumber
	\bar{u} = u^\ast \,.
\end{equation}
By analogy with the paper \cite{ADV2} we choose the same parametrization for constants $(g,\bar{g})$ and $(\tilde{g},\bar{\tilde{g}})$ as for spins
\begin{equation} \label{gparam}
	(g, \bar{g}) = \Bigl(\frac{1+n_g}{2} + \imath\nu_g \,, \frac{1-n_g}{2} + \imath\nu_g \Bigr) \,, \quad
	(\tilde{g}, \bar{\tilde{g}}) = \Bigl(\frac{1+\tilde{n}_g}{2} + \imath\tilde{\nu}_g \,, \frac{1-\tilde{n}_g}{2} + \imath\tilde{\nu}_g \Bigr)
\end{equation}
where $n_g,\,\tilde{n}_g \in \mathbb{Z}+\sigma $, $\nu_g,\,\tilde{\nu}_g \in \mathbb{R}$ and the parameter $\sigma$ was introduced in \eqref{sigma}.
We also assume that the constants $\gamma,\bar{\gamma}$ are complex conjugate to each other, and similarly for $\tilde{\gamma},\bar{\tilde{\gamma}}$
\begin{equation} \nonumber
	\gamma^\ast = \bar{\gamma} \,, \qquad
	\tilde{\gamma}^\ast = \bar{\tilde{\gamma}} \,.
\end{equation}
First, the pairs in \eqref{gparam} satisfy the same requirement as spins $g-\bar{g} \in \mathbb{Z}+\sigma$, $\tilde{g}-\bar{\tilde{g}} \in \mathbb{Z}+\sigma$ so that the double powers in formulas like \eqref{Kop-d} are well defined.
Second, for such values of $K$-matrices' parameters the operators $t_n(u)$ and $\hat{t}_n(u)$ possess the following rule of hermitian conjugation
\begin{equation} \label{tconj}
	t_n^\dagger(u) = - \bar{t}_n(\bar{u}) \,, \quad \hat{t}_n^\dagger(u) = \bar{\hat{t}}_n(\bar{u}) \,.
\end{equation}
We prove these relations in the end of this section.

We also prove that the operators $t_n(u)$ and $\hat{t}_n(u)$ have the following form 
\begin{align}
	& \nonumber
	t_n(u) = 2\gamma\bigl(g-\frac{1}{2}\bigr)\bigl(s-\frac{1}{2}\bigr)^{2n}+2\gamma\bigl(u^2-\frac{1}{4}\bigr)\Bigl(u^{2n-2}H_1+u^{2n-4}H_2+\ldots+H_n\Bigr) \,,\\
	& \label{xn}
	\hat{t}_n(u) = 2\gamma\tilde{\gamma}\bigl(g-\frac{1}{2}\bigr)\bigl(\tilde{g}-\frac{1}{2}\bigr)\bigl(s-\frac{1}{2}\bigr)^{2n}-\bigl(\gamma^2+\tilde{\gamma}^2\bigr)\bigl(u^2-\frac{1}{4}\bigr)\Bigl(u^{2n}+u^{2n-2}\hat{H}_1+\ldots+\hat{H}_n\Bigr).
\end{align}
(and similarly for the antiholomorphic counterparts),
where $H_k$ and $\hat{H}_k$ are the integrals of motion of the one-boundary and two-boundary open spin chain, respectively. 

To derive this explicit expressions we express the monodromy matrix
\begin{equation} \nonumber
	T_n(u) = L_n(u) \cdots L_1(u) K(u) L_1(u) \cdots L_n(u)
\end{equation}
in terms of the monodromy matrix $M_n(u)$ of A-type chain
\begin{equation} \nonumber
	M_n(u) = L_1(u) \cdots L_n(u) = \begin{pmatrix}
		a_n(u) & b_n(u) \\
		c_n(u) & d_n(u)
	\end{pmatrix} \,.
\end{equation}
We have
\begin{align}
	\nonumber
	T_n(u) & = (-1)^n\,\sigma_2\,M_n^t(-u)\,\sigma_2\,K(u)\,M_n(u)
	\\ 
	\nonumber
	& =
	(-1)^n \begin{pmatrix}
		d_n(-u) & -b_n(-u) \\
		-c_n(-u) & a_n(-u)
	\end{pmatrix} \begin{pmatrix}
		\gamma\bigl(g - \frac{1}{2} \bigr) & u - \frac{1}{2} \\[6pt]
		\gamma^2 \bigl(u - \frac{1}{2} \bigr) & \gamma\bigl(g - \frac{1}{2} \bigr)
	\end{pmatrix} \begin{pmatrix}
		a_n(u) & b_n(u) \\
		c_n(u) & d_n(u)
	\end{pmatrix} \,,
\end{align}
where~$M_n^t$ is transposed matrix $M_n$ and $\sigma_2$ is the Pauli matrix.
This expression is a simple consequence of the relation for the $L$-operator
\begin{align} \nonumber
	L_k(u) = -\sigma_2\,L^t_k(-u)\,\sigma_2 \,.
\end{align}
In the paper \cite{ADV2} it was shown that the operator $B_n(u)$ has the following form
\begin{equation} \label{B_n}
	B_n(u) = (u-\frac{1}{2})\,P_{2n}(u)
\end{equation}	
where $P_{2n}(u)$ is an even polynomial of degree $2n$ in $u$ with the leading coefficient equal to $1$.
Using the same arguments it's easy to show that the operator $C_n(u)$ has the form
\begin{equation} \label{C_n}
	C_n(u) = (u-\frac{1}{2})\,F_{2n}(u)
\end{equation}
where $F_{2n}(u)$ is some even polynomial in $u$ of degree $2n$ with the leading coefficient~$\gamma^2$.

Now we express the transfer matrices $t_n(u)$ and $\hat{t}_n(u)$ in terms of entries of~$M_n(u)$
\begin{align}
	& \nonumber
	t_n(u) = (-1)^n \biggl\{ \gamma \bigl(g-\frac{1}{2}\bigr) \Bigl[ d_n(-u)a_n(u) + a_n(-u)d_n(u) - b_n(-u)c_n(u) - c_n(-u)b_n(u) \Bigr] \\
	& \label{t_n_abcd} 
	\quad + \bigl(u-\frac{1}{2}\bigr)
	\Bigl[ d_n(-u)c_n(u) - c_n(-u)d_n(u) \Bigr] + \gamma^2 \bigl(u-\frac{1}{2}\bigr)\Bigl[ a_n(-u)b_n(u) - b_n(-u)a_n(u) \Bigr] \biggr\} ,
	\\ 	
	& \nonumber
	\hat{t}_n(u) = \tilde{\gamma} \bigl(\tilde{g} - \frac{1}{2} \bigr) \, t_n(u) - \tilde{\gamma}^2 \bigl(u + \frac{1}{2} \bigr) \,B_n(u) - \bigl(u + \frac{1}{2} \bigr) \,C_n(u) \,.
\end{align}	
The Yang-Baxter equation for $M_n(u)$ with Yang's $R$-matrix \eqref{Rmat} \cite{Fad:BetheAns,TaFa,Skl:VarSep,Skl91,KulSk}
\begin{align} \nonumber
	R(u-v)\, \bigl(M_n(u)\otimes \bm{1}\bigr)\,\bigl(\bm{1} \otimes M_n(v)\bigr) = 
	\bigl(\bm{1} \otimes M_n(v)\bigr)\,\bigl(M_n(u)\otimes \bm{1}\bigr)\, R(u-v) \,,
\end{align}
(which is the direct consequence of the $RLL$ relation~\eqref{RLL})
implies the following commutation relations for the entries of $M_n$
\begin{gather*}
	(u-v)\,a_n(v)\,b_n(u) = (u-v+1)\,b_n(u)\,a_n(v) + b_n(v)\,a_n(u) \,, \\
	(u-v)\,d_n(v)\,c_n(u) = (u-v+1)\,c_n(u)\,d_n(v) + c_n(v)\,d_n(u) \,.
\end{gather*}
Substituting $v=-u$ in the last two equations we obtain
\begin{gather*}
	\frac{a_n(-u)\,b_n(u) - b_n(-u)\,a_n(u)}{-u-\frac{1}{2}} = \frac{a_n(u)\,b_n(-u) - b_n(u)\,a_n(-u)}{u-\frac{1}{2}} \,, \\
	\frac{d_n(-u)\,c_n(u) - c_n(-u)\,d_n(u)}{-u-\frac{1}{2}} = \frac{d_n(u)\,c_n(-u) - c_n(u)\,d_n(-u)}{u-\frac{1}{2}} \,.
\end{gather*}
From these identities it follows that 
\begin{align*}
	& a_n(-u)\,b_n(u) - b_n(-u)\,a_n(u) = \bigl(u+\frac{1}{2}\bigr)\,\tilde{P}_{2n-2}(u)\,, \\
	& d_n(-u)\,c_n(u) - c_n(-u)\,d_n(u) = \bigl(u+\frac{1}{2}\bigr)\,\tilde{F}_{2n-2}(u) \,,
\end{align*}	
where $\tilde{P}_{2n-2}(u)$ and $\tilde{F}_{2n-2}(u)$ are even polynomials of degree $2n-2$.
Next we use the following relation which is a cosequence of the Yang-Baxter equation for~$M_n(u)$
\begin{multline} \nonumber
	(u-v-1)\left[\sigma_2\,M_n(u)^t\,\sigma_2\,M_n(v)\right]_{ij} \\
	= 
	(u-v)\left[M_n(v)^t\,\sigma_2\,M_n(u)\,\sigma_2\right]_{ji} - \left[\sigma_2\,M_n(v)^t\,\sigma_2\,M_n(u)\right]_{ij} \,.
\end{multline}
Substituting $v=-u$ and taking trace we obtain
\begin{align*} \nonumber
	&
	a_n(-u)\,d_n(u) + d_n(-u)\,a_n(u) - b_n(-u)\,c_n(u) - c_n(-u)\,b_n(u) \\
	&
	= a_n(u)\,d_n(-u) + d_n(u)\,a_n(-u) - b_n(u)\,c_n(-u) - c_n(u)\,b_n(-u)\,.
\end{align*}
This shows that the expression in the first square brackets in the formula for~$t_n(u)$~\eqref{t_n_abcd} is an even polynomial of degree~$2n$ in~$u$.
Combining the obtained results together we see that~$t_n(u)$ and~$\hat{t}_n(u)$ are even polynomials in~$u$. 
Next, using~\eqref{T_n(1/2)} and~\eqref{B_n}, \eqref{C_n} we see that~$t_n(u)$ and~$\hat{t}_n(u)$ have the form~\eqref{xn}.

It remains to prove the hermitian conjugation formula \eqref{tconj} for $t_n(u)$ and $\hat{t}_n(u)$. In the paper \cite{ADV2} it was demonstrated that the operator $B_n(u)$ satisfies the hermitian conjugation formula
\begin{equation} \nonumber
	B_n^\dagger(u) = \bar{B}_n(\bar{u}) \,,
\end{equation}
Using the same arguments one can show that a similar relation holds for~$C_n(u)$
\begin{equation} \nonumber
	C_n^\dagger(u) = \bar{C}_n(\bar{u}) \,.
\end{equation}
According to \eqref{t_n_abcd}, for proving \eqref{tconj} one needs to know the conjugation rules for entries of matrix $M_n(u)$.
Recall that the holomorphic and antiholomorphic generators of principal series representations are related via conjugation
\begin{equation} \nonumber
	S^\dagger = -\bar{S}, \qquad S_-^\dagger = -\bar{S}_-, \qquad
	S_+^\dagger = -\bar{S}_+ \,,
\end{equation}
so we obtain the following property for matrix elements of every $L$-operator
\begin{equation} \nonumber
	(L_k)_{ij}^\dagger(u) = -(\bar{L}_k)_{ij}(-\bar{u}) \,.
\end{equation}
By virtue of the last formula and commutativity of matrix entries of $L$-operators corresponding to different $L^2(\mathbb{C})$ spaces one gets the conjugation rule for the elements of~$M_n(u)$
\begin{equation} \label{bdDagger}
	(M_n)_{ij}^\dagger(u) = (-1)^n\,(\bar{M}_n)_{ij}(-\bar{u}) \,.
\end{equation}
Using \eqref{bdDagger} and the relations $\gamma^\ast=\bar{\gamma}$, $u^\ast=\bar{u}$, $g^\ast=1-\bar{g}$, where the last one follows from the parametrization of $g$ and $\bar{g}$ given by \eqref{gparam}, one can easily show that the hermitian conjugate of the expression \eqref{t_n_abcd} for $t_n(u)$ is equal to $ - \bar{t}_n(\bar{u})$. 
The conjugation rule  $\hat{t}_n^{\dagger}(u) = \bar{\hat{t}}_n(\bar{u})$ is then a consequence of the conjugation rules for $t_n$, $B_n$ and $C_n$.

Taking to account the explicit expression \eqref{xn} for $t_n(u)$ it's convenient to consider the matrix 
\begin{equation} \label{t_ren}
	\tilde{t}_n(u) = \frac{t_n(u) - 2\gamma\bigl(g-\frac{1}{2}\bigr)\bigl(s-\frac{1}{2}\bigr)^{2n}}{2\gamma (u^{2} - \frac{1}{4})} = u^{2n-2}H_1+u^{2n-4}H_2+\ldots+H_n \,.
\end{equation}
Formula \eqref{xn} implies also that the leading coefficient takes the form
\begin{align*}
	H_1 = g-\frac{1}{2} + \gamma^{-1}(S_+ + \gamma^2 S_-) = g-\frac{1}{2} + \gamma^{-1}\sum\limits_{k = 1}^n \bigl[(z_k^2-\gamma^2)\partial_k + 2sz_k\bigr] \,.
\end{align*}
The complete orthogonal set of common eigenfunctions $f_{\mu_1,\dots,\mu_n}$ of $H_1$ and of its antiholomorphic counterpart $\bar{H}_1$ (the corresponding coefficient in the transfer matrix $\bar{t}_n(\bar{u})$) can be found by separating variables
\begin{equation} \nonumber
	f_{\mu_1,\dots,\mu_n}(z_1,\dots,z_n) = \prod_{k = 1}^n f_{\mu_k}(z_k) \,, \quad
	(\mu_k,\bar{\mu}_k)=\bigl(\tfrac{r_k}{2}+\imath\omega_k, -\tfrac{r_k}{2}+\imath\omega_k\bigr), \; r_k\in\mathbb{Z}+\sigma,\,\omega_k\in\mathbb{R},
\end{equation}
where $f_\mu(z)$ form the complete orthogonal set of common eigenfunctions of operator $(z^2-\gamma^2)\partial_z + 2sz$ and the corresponding antiholomorphic operator
\begin{equation} \nonumber
	f_{\mu}(z) = [z+\gamma]^{-\mu-s}\,[z-\gamma]^{\mu-s} \,, \quad
	(\mu,\bar{\mu})=\bigl(\tfrac{r}{2}+\imath\omega, -\tfrac{r}{2}+\imath\omega\bigr)\,, \;
	r\in\mathbb{Z}+\sigma,\,\omega\in\mathbb{R} \,.
\end{equation}
The eigenvalues $(\lambda,\bar{\lambda})$ of $H_1$ and $\bar{H}_1$ corresponding to $f_{\mu_1,\dots,\mu_n}$ read
\begin{align*}
	\lambda = g-\frac{1}{2} + \sum \limits_{k = 0}^n 2\mu_k \,, \quad
	\bar{\lambda} = \bar{g}-\frac{1}{2} + \sum \limits_{k = 0}^n 2\bar{\mu}_k \,.
\end{align*}
This way, each pair of eigenvalues of $H_1$ and $\bar{H}_1$ corresponding to their common eigenfunction has the form
\begin{equation} \label{lambdaPar}
	(\lambda,\bar{\lambda}) = \bigl(\tfrac{m}{2}+\imath\rho, -\tfrac{m}{2}+\imath\rho\bigr)\,, \quad
	m\in\mathbb{Z}+\sigma,\,\rho\in\mathbb{R} \,.
\end{equation}

The eigenfunctions of the one-boundary chain are defined by the following spectral problem for the renormalized transfer matrix~\eqref{t_ren}
\begin{equation} \label{eigenvalue}
\begin{aligned}
\tilde{t}_n(u)\Phi_{\bm{x}_{n-1}}^{\lambda}(\bm{z_{n-1}}) 
&= \lambda \bigl(u^2-x_1^2\bigr) \ldots \bigl(u^2-x_{n-1}^2\bigr) \Phi_{\bm{x}_{n-1}}^{\lambda} \,, \\
\bar{\tilde{t}}_n(\bar{u})\Phi_{\bm{x}_{n-1}}^{\lambda}(\bm{z_{n-1}}) 
&= \bar{\lambda} \bigl(\bar{u}^2-\bar{x}_1^2\bigr) \ldots \bigl(\bar{u}^2-\bar{x}_{n-1}^2\bigr) \Phi_{\bm{x}_{n-1}}^{\lambda} \,,
\end{aligned}
\end{equation}
where we use the following compact notations for tuples of variables
\begin{align}\nonumber
	\bm z_n=(z_1,\bar{z}_1, \ldots, z_n, \bar{z}_n) \ \,, \  
	\bm x_n=(x_1,\bar{x}_1, \ldots, x_n, \bar{x}_n)  \,.
\end{align}
On parameters $x_k,\bar{x}_k$ we impose a similar restriction $x_k-\bar{x}_k\in\mathbb{Z}+\sigma$ as on $s,\bar{s}$ and $g,\bar{g}$. In addition, due to \eqref{tconj} and \eqref{lambdaPar} one has $(x_k^\ast)^2=\bar{x}_k^2$. We choose $x_k^\ast=-x_k$.
Thus we obtain the following parametrization for spectral variables
\begin{equation} \nonumber
	(x_k, \bar{x}_k) = \Bigl(\frac{n_k}{2}+\imath\nu_k \,, -\frac{n_k}{2}+\imath\nu_k\Bigr)\,, \qquad
	(\lambda,\bar{\lambda}) = \Bigl(\frac{m}{2}+\imath\rho, -\frac{m}{2}+\imath\rho\Bigr) \,,
\end{equation}
where $n_k, m\in\mathbb{Z}+\sigma$ and $\nu_k$, $\rho$ are real.

\subsection{General Transfer matrix}
\label{sect:generalT}

Now we are going to define general monodromy matrices $\mathbb{T}_n(u,s_0)$ and $\hat{\mathbb{T}}_n(u,s_0)$ for the one- and two-boundary cases respectively, and the corresponding general transfer matrices $\mathbbm{t}_n(u,s_0)$, $\hat{\mathbbm{t}}_n(u,s_0)$.
In contrast to monodromy matrices $T_n(u)$ and $\hat{T}_n(u)$~\eqref{Top}, \eqref{Top1}, the auxiliary space for the general ones is infinite-dimensional -- it is the space of principal series representation of spin $(s_0,\bar{s}_0)$.
In correspondence with the hierarchy of Yang-Baxter and reflection relations discussed in Section~\ref{sect:ReflEq}, when the auxiliary space becomes infinite-dimensional the $L$-operator turns into $\mathbb{R}$-operator~\eqref{RYB} and $K_\pm$-matrices turn into $\mathbb{K}_\pm$-operators~\eqref{Kvs}, \eqref{Kvs+}.
This way, we define the general monodromy matrices as follows
\begin{align}
	\label{generalT}
	&\mathbb{T}_n(u,s_0) = \mathbb{R}_{n0}(u)\cdots \mathbb{R}_{10}(u)\overset{0}{\mathbb{K}}_-(u,s_0)\mathbb{R}_{10}(u)\cdots \mathbb{R}_{n0}(u) \,,\\
	\label{generalhatT}
	&\hat{\mathbb{T}}_n(u,s_0) = \overset{0}{\mathbb{K}}_+(u,s_0)\mathbb{T}_n(u) \,.
\end{align}

Similarly to transfer matrices $t_n(u)$ and $\hat{t}_n(u)$, the general ones $\mathbbm{t}_n(u,s_0)$ and $\hat{\mathbbm{t}}_n(u,s_0)$ are defined as traces of $\mathbb{T}_n(u,s_0)$ and $\hat{\mathbb{T}}_n(u,s_0)$, respectively, over the auxiliary space
\begin{align} \label{general-t}
	\mathbbm{t}_n(u,s_0) = \tr_0[\mathbb{T}_n(u,s_0)] \,, \qquad
	\hat{\mathbbm{t}}_n(u,s_0) = \tr_0[\hat{\mathbb{T}}_n(u,s_0)] \,.
\end{align}
Let us explain how the partial trace of an operator in the infinite-dimensional space is defined.
Consider an integral operator $\mathcal{O}$ acting in the tensor product of $\mathrm{L}^2(\mathbb{C})$ spaces $V_0\otimes V_1\otimes\dots\otimes V_n$
\begin{equation} \nonumber
	[\mathcal{O}\,\Psi](z_0,\bm{z}_n) = \int \mathrm{d}^2z_0\,\mathrm{d}^{2n}\bm{z}_n \,
	\mathcal{O}(z_0,\bm{z}_n|w_0,\bm{w}_n) \,
	\Psi(w_0,\bm{w}_n) \,,
\end{equation}
where $\bm{z}_n = (z_1,\dots,z_n)$, $\mathrm{d}^{2n}\bm{z}_n = \prod_{k=1}^n \mathrm{d}^2z_k$, and $\mathcal{O}(z_0,\bm{z}_n|w_0,\bm{w}_n)$ is the kernel of~$\mathcal{O}$.
The trace of $\mathcal{O}$ over the space $V_0$ is an integral operator $\tilde{\mathcal{O}}$ in~$V_1\otimes\dots\otimes V_n$ whose kernel is given by the integral over $w_0$ of the kernel of $\mathcal{O}$ with $z_0=w_0$
\begin{align}
	& \label{trInt}
	\tr_0[\mathcal{O}] = \tilde{\mathcal{O}} \,, \\
	& \nonumber
	\text{where} \quad
	[\tilde{\mathcal{O}}\,\Phi](\bm{z}_n) =
	\int \mathrm{d}^{2n}\bm{z}_n \, \tilde{\mathcal{O}}(\bm{z}_n|\bm{w}_n)\,\Phi(\bm{w}_n) \,, \quad
	\tilde{\mathcal{O}}(\bm{z}_n|\bm{w}_n) = \int \mathrm{d}^2w_0\,\mathcal{O}(w_0,\bm{z}_n|w_0,\bm{w}_n) \,.
\end{align}
In Appendix~\ref{app:TrTransp} we collect and derive the basic relations for the trace~\eqref{trInt} which are used throughout the paper.

Using the general Yang-Baxter equation~\eqref{YBgen} and reflection equations~\eqref{Refl},~\eqref{ReflL1} one can show that the general monodromy matrix $\mathbb{T}_n(u,s_0)$ satisfies the following reflection equations
\begin{align}
	\label{RTRTgen}
	&\mathbb{R}_{00'}(u-v)\,\overset{0}{\mathbb{T}}_n(u,s_0)\,\mathbb{R}_{00'}(u+v)\,\overset{0'}{\mathbb{T}}_n(v,s_0') = \overset{0'}{\mathbb{T}}_n(v,s_0')\,\mathbb{R}_{00'}(u+v)\,\overset{0'}{\mathbb{T}}_n(u,s_0)\,\mathbb{R}_{00'}(u-v) \,,\\
	\label{TLTL}
	&\mathbb{T}_n(v,s_0)\,L_0(u+v)\,T_n(u)\,L_0(u-v) = L_0(u-v)\,T_n(u)\,L_0(u+v)\,\mathbb{T}_n(v,s_0) \,,
\end{align}	
where the $L$-operator~$L_0(u)$ corresponds to spin~$s_0$.
From these relations it follows that general transfer matrices $\mathbbm{t}_n(u,s_0)$ and $\hat{\mathbbm{t}}_n(u,s_0)$ form commutative families and also commute with the transfer matrices $t_n(u)$ and $\hat{t}_n(u)$, respectively,
\begin{align}
	& \nonumber
	[\mathbbm{t}_n(u,s_0), \mathbbm{t}_n(v,s_0')] = 0 \,, \quad \quad [\mathbbm{t}_n(u,s_0), t_n(v)] = 0 \,, \\
	& \label{Gcomm}
	[\hat{\mathbbm{t}}_n(u,s_0), \hat{\mathbbm{t}}_n(v,s_0')] = 0 \,, \quad \quad [\hat{\mathbbm{t}}_n(u,s_0), \hat{t}_n(v)] = 0 \,.
\end{align}
	
Let us check the commutativity~\eqref{Gcomm} for the two boundary case. The derivation for one boundary case is similar and just a bit shorter.
All the derivations follow the ideas of Sklyanin's proof~\cite[Section~3]{Skl:Boundary} of transfer matrix commutativity of the type~$[\hat{t}_n(u),\hat{t}_n(v)]=0$.
For the sake of brevity, till the end of the Section we will omit the index $n$ (the number of sites of the spin chain) for transfer and monodromy matrices.

We need four important relations. First two are unitarity and crossing unitarity of $L$-operator
\begin{align} \label{Lunitar}
	L(u)\,L(1-u) = \rho_s(u)\,\bm{1} \,, \qquad L^t(u)\,L^t(-u-1) = \tilde{\rho}_s(u)\,\bm{1} \,,
\end{align}
where $\rho_s(u) = -(u-s)(u+s-1)$ and $\tilde{\rho}_s(u) = -(u+s)(u-s+1)$.
They can be proven by a direct computation.
Second two are the unitarity and the crossing unitarity of the general $\mathbb{R}$-matrix proven in Appendix~\ref{app:ucu}
\begin{align}
	& \label{Runi1}
	\mathbb{R}_{12}(u)\mathbb{R}_{12}(-u) = \II \,, \\
	& \label{Rcuni1}
	\mathbb{R}_{12}^{t_1}(u)\mathbb{R}_{12}^{t_1}(-u-2)= \mathbb{R}_{12}^{t_2}(u)\mathbb{R}_{12}^{t_2}(-u-2) = \rho_{s_1,s_2}(u)\,\II \,,
\end{align}
where by $\mathbb{R}_{12}^{t_1}(u)$ and $\mathbb{R}_{12}^{t_2}(u)$ we denote the transposition of $\mathbb{R}_{12}$ in the $1$-st and $2$-nd space, respectively, and
\begin{align}
	\nonumber
	\rho_{s_1,s_2}(u) & = 
	(-1-u\pm(s_1-s_2))^{-1}(1+u\pm(1-s_1-s_2))^{-1} \\
	& \nonumber
	\quad\times
	(-1-\bar{u}\pm(\bar{s}_1-\bar{s}_2))^{-1}(1+\bar{u}\pm(1-\bar{s}_1-\bar{s}_2))^{-1} \,,
\end{align}
where $(x\pm y)$ denotes $(x+y)(x-y)$.
We define the partial transposition of an integral operator by transposing the corresponding variables in its kernel.
Namely, if $\mathcal{O}$ is an operator in~$V_1\otimes V_2, \, V_k = \mathrm{L}^2(\mathbb{C})$,
\begin{equation} \label{transpDef1}
	[\mathcal{O}\,\Psi](z_1,z_2) = \int \mathrm{d}^2z_1\,\mathrm{d}^2z_2 \,
	\mathcal{O}(z_1,z_2|\,w_1,w_2) \, \Psi(w_1,w_2) \,,
\end{equation}
then the kernels of~$\mathcal{O}^{t_1}$ and~$\mathcal{O}^{t_2}$ (where $\mathcal{O}^{t_k}$ denotes the transposition in the space~$V_k$) are equal to
\begin{equation} \label{transpDef2}
	\mathcal{O}^{t_1}(z_1, z_2|\,w_1, w_2) = \mathcal{O}(w_1, z_2|\,z_1, w_2) \,, \quad
	\mathcal{O}^{t_2}(z_1, z_2|\,w_1, w_2) = \mathcal{O}(z_1, w_2|\,w_1, z_2) \,.
\end{equation}
This definition is obviously generalized to the tensor product of arbitrary number of $\mathrm{L}^2(\mathbb{C})$ copies, that is, to the case of an operator acting on functions of any number of variables.
Some useful relations involving this transposition operation and their proofs can be found in Appendix~\ref{app:TrTransp}.

Let us start by deriving the commutativity of $\hat{\mathbbm{t}}(x,s_0)$ and $\hat{t}(u)$. 
We recall that
\begin {align*}
	\hat{\mathbbm{t}}(u,s_0) = \tr_0\Bigl[\overset{0}{\mathbb{K}}_+(u,s_0)\,\mathbb{T}(u,s_0)\Bigr], \qquad
	\hat{t}(v) = \tr_{\mathbb{C}^2}[K_+(v)\,T(v)]
\end{align*}
Using the cyclic property of the trace over~$\mathbb{C}^2$ and its invariance under matrix transposition we can write
\begin{multline} \label{bbmtt1}
	\hat{\mathbbm{t}}(u,s_0)\,\hat{t}(v) = \tr_0\Bigl[\overset{0}{\mathbb{K}}_+(u,s_0)\,\mathbb{T}(u,s_0)\Bigr]\,\tr_{\mathbb{C}^2}[K_+^t(v)\,T^t(v)] \\
	=
	\tr_{0,\mathbb{C}^2}[\overset{0}{\mathbb{K}}_+(u,s_0)\,\mathbb{T}(u,s_0)\,K_+^t(v)\,T^t(v)]
	= \tr_{0,\mathbb{C}^2}[K_+^t(v)\,\overset{0}{\mathbb{K}}_+(u,s_0)\,\mathbb{T}(u,s_0)\,T^t(v)] \,,
\end{multline}
where we also used the commutativity of numerical matrix~$K_+^t(v)$ with~$\overset{0}{\mathbb{K}}_+(u,s_0)$ and $\mathbb{T}(u,s_0)$.
The idea is to permute the factors $\mathbb{T}(u,s_0)$ and $T(v)$ in the argument of the trace~\eqref{bbmtt1} with the help of reflection equation~\eqref{TLTL} which involves this pair of monodromy matrices.
In order to apply~\eqref{TLTL}, we add the $L_0$-operators into the expression~\eqref{bbmtt1} by means of~\eqref{Lunitar}.
First, we insert the identity
\begin{equation} \label{ident1}
	\bm{1} = \tilde{\rho}_{s_0}^{-1}(u+v)\,L_0^t(u+v)\,L_0^t(-u-v-1)
\end{equation}
and use the cyclicity of traces over $0$-th space~\eqref{tr-cycle} and~$\mathbb{C}^2$ arriving at
\begin{align*}
	\hat{\mathbbm{t}}(u,s_0)\,\hat{t}(v) = \tilde{\rho}_{s_0}^{-1}(u+v)\,
	\tr_{0,\mathbb{C}^2}\bigl[L_0^t(-u-v-1) K_+^t(v)\,\overset{0}{\mathbb{K}}_+(u,s_0)\,\mathbb{T}(u,s_0)\,T^t(v)\, L_0^t(u+v)\bigr] .
\end{align*}
Next, using the commutativity of entries of matrices $L_0$ and $K_+$ (and similarly for $T$ and $L_0$) we rewrite
\begin{align*}
	& \nonumber
	L_0^t(-u-v-1)\,K_+^t(v)\,\overset{0}{\mathbb{K}}_+(u,s_0) =
	\bigl(K_+(v)\,L_0(-u-v-1)\,\overset{0}{\mathbb{K}}_+(u,s_0)\bigr)^t \,, \\
	& \nonumber
	\mathbb{T}(u,s_0)\,T^t(v)\,L_0^t(u+v) =
	\bigl(\mathbb{T}(u,s_0)\,L_0(u+v)\,T(v)\bigr)^t
\end{align*}
and apply the property
\begin{equation} \label{trAtBt}
	\tr_{\mathbb{C}^2}\bigl[\mathcal{A}^t\,\mathcal{B}^t\bigr] =
	\tr_{\mathbb{C}^2}\bigl[\mathcal{A}\,\mathcal{B}\bigr]
\end{equation}
which can be easily checked for any $2\times 2$ matrices $\mathcal{A}$ and $\mathcal{B}$ even with non-commutative entries.
As the result we have
\begin{align} \label{bbmtt2}
	\hat{\mathbbm{t}}(u,s_0)\,\hat{t}(v) = \tilde{\rho}_{s_0}^{-1}(u+v)\,
	\tr_{0,\mathbb{C}^2}\bigl[K_+(v)\,L_0(-u-v-1)\,\overset{0}{\mathbb{K}}_+(u,s_0) \, \mathbb{T}(u,s_0)\,L_0(u+v)\,T(v)\bigr].
\end{align}
Inserting the identity
\begin{equation} \label{ident2}
	\bm{1} = \rho_{s_0}^{-1}(v-u) \, L_0(v-u) \, L_0(1-v+u)
\end{equation}
and employing the cyclic property of the trace~$\tr_{0,\mathbb{C}^2}$ again we rearrange the last expression in the form
\begin{multline*}
	\hat{\mathbbm{t}}(u,s_0)\,\hat{t}(v) 
	= \tilde{\rho}_{s_0}^{-1}(u+v)\, \rho_{s_0}^{-1}(v-u) \\
	\quad \times
	\tr_{0,\mathbb{C}^2}\bigl[L_0(1-v+u)\,K_+(v)\,L_0(-u-v-1)\,\overset{0}{\mathbb{K}}_+(u,s_0) \,\mathbb{T}(u,s_0)\,L_0(u+v)\,T(v)\,L_0(v-u)\bigr].
\end{multline*}
This way, we obtain the product of the left hand sides of reflection equations~\eqref{ReflL2} and~\eqref{TLTL} in the argument of the trace (up to the transposition $u\leftrightarrow v$), and using these relations we rewrite everything as
\begin{multline*}
	\hat{\mathbbm{t}}(u,s_0)\,\hat{t}_n(v) 
	= \tilde{\rho}_{s_0}^{-1}(u+v)\, \rho_{s_0}^{-1}(v-u) \\
	\quad \times
	\tr_{0,\mathbb{C}^2}\bigl[\overset{0}{\mathbb{K}}_+(u,s_0)\,L_0(-u-v-1)\,K_+(v)\,L_0(1-v+u) \, L_0(v-u)\,T(v)\,L_0(u+v)\, \mathbb{T}(u,s_0)\bigr].
\end{multline*}
Thus, we have changed the order of factors $T(v)$ and $\mathbb{T}(u,s_0)$ in the operator product, and this is the key part of the proof of transfer matrix commutativity.
Now, cancelling the pair of mutually inverse $L$-operators~\eqref{ident2} $L_0(1-v+u)$ and $L_0(v-u)$, which are not needed anymore, and doing the cyclic permutation of $\mathbb{K}_+$-operator we get the following
\begin{multline} \label{bbmtt}
	\hat{\mathbbm{t}}(u,s_0)\,\hat{t}(v)
	= \tilde{\rho}_{s_0}^{-1}(u+v) \\
	\times
	\tr_{0}\Bigl[ \, 
	\tr_{\mathbb{C}^2}\bigl[L_0(-u-v-1)\,K_+(v)\,T(v)\,L_0(u+v)\bigr] \, \mathbb{T}(u,s_0)\,\overset{0}{\mathbb{K}}_+(u,s_0) \Bigr].
\end{multline}
The last step is the cancellation of the remaining pair of mutually inverse $L_0$-operators~\eqref{ident1} with the help of~\eqref{trAtBt}
\begin{align*}
	&
	\tr_{\mathbb{C}^2}\bigl[L_0(-u-v-1)\,K_+(v)\,T(v)\,L_0(u+v)\bigr] \\
	& =
	\tr_{\mathbb{C}^2}\bigl[\bigl(L_0(-u-v-1)\,K_+(v)\,T(v)\bigr)^t \, L_0^t(u+v)\bigr] \\
	& =
	\tr_{\mathbb{C}^2}\bigl[T^t(v)\,K_+^t(v)\,L_0^t(-u-v-1)\, L_0^t(u+v)\bigr] \\
	& =
	\tilde{\rho}_{s_0}(u+v)\,\tr_{\mathbb{C}^2}\bigl[T^t(v)\,K_+^t(v)\bigr]
	= \tilde{\rho}(u+v)\,\hat{t}(v) \,.
\end{align*}
Substituting this result into~\eqref{bbmtt} and removing $\hat{t}$ from the trace over the $0$-th copy of~$\mathrm{L}^2(\mathbb{C})$ (since it acts trivially in this space), we finish the proof of commutativity of~$\hat{\mathbbm{t}}(u,s_0)$ and~$\hat{t}(v)$
\begin{align} \nonumber
	\hat{\mathbbm{t}}(u,s_0)\,\hat{t}(v) =
	\tr_0\bigl[\hat{t}(v)\,\mathbb{T}(u,s_0)\,\overset{0}{\mathbb{K}}_+(u,s_0)\bigr]
	=
	\hat{t}(v) \, \tr_0\bigl[\mathbb{T}(u,s_0)\,\overset{0}{\mathbb{K}}_+(u,s_0)\bigr] =
	\hat{t}(v) \, \hat{\mathbbm{t}}(u,s_0) \,.
\end{align}

The proof of commutativity of $\hat{\mathbbm{t}}(u,s_0)$ and $\hat{\mathbbm{t}}(v,s_0')$ goes along the same lines.
One just replaces $L_0$-operators with $\mathbb{R}$-operators and uses the unitarity~\eqref{Runi1} and crossing unitarity~\eqref{Rcuni1} of $\mathbb{R}$-operator instead of the similar relations for $L$-operator~\eqref{Lunitar}.
First we use the cyclic property of the trace over an infinite-dimensional space~\eqref{tr-cycle} and its invariance under transposition of its argument~\eqref{tr-transp} to write
\begin{equation}
	\hat{\mathbbm{t}}(v,s_0') =
	\tr_{0'}\bigl[\overset{0'}{\mathbb{K}}_+(v,s_0')\,\overset{0'}{\mathbb{T}}(v,s_0')\bigr] =
	\tr_{0'}\bigl[\overset{0'}{\mathbb{K}}{}_+^{t_{0'}}(v,s_0')\,\overset{0'}{\mathbb{T}}{}^{t_{0'}}(v,s_0')\bigr] \,.
\end{equation}
Thus we have
\begin{align*}
	\hat{\mathbbm{t}}(u,s_0)\,\hat{\mathbbm{t}}(v,s_0') & =
	\tr_{0,0'}\bigl[\overset{0}{\mathbb{K}}_+(u,s_0)\,\overset{0}{\mathbb{T}}(u,s_0)\,\overset{0'}{\mathbb{K}}{}_+^{t_{0'}}(v,s_0')\,\overset{0'}{\mathbb{T}}{}^{t_{0'}}(v,s_0')\bigr] \\
	& =
	\tr_{0,0'}\bigl[\overset{0'}{\mathbb{K}}{}_+^{t_{0'}}(v,s_0')\,\overset{0}{\mathbb{K}}_+(u,s_0)\,\overset{0}{\mathbb{T}}(u,s_0)\,\overset{0'}{\mathbb{T}}{}^{t_{0'}}(v,s_0')\bigr] \,,
\end{align*}
where we moved the operator $\overset{0'}{\mathbb{K}}{}_+^{t_{0'}}$ leftwards owing to its commutativity with $\overset{0}{\mathbb{K}}_+$ and $\overset{0}{\mathbb{T}}$ which act trivially in the $0'$-th space.

Now, in order to apply the reflection equation~\eqref{RTRTgen} for permuting the monodromy matrices~$\mathbb{T}(u,s_0)$ and~$\mathbb{T}(v,s_0')$, we insert the identity by virtue of~\eqref{Rcuni1}
\begin{equation}
	\II = \rho_{s_0,s_0'}^{-1}(u+v)\,\mathbb{R}_{00'}^{t_{0'}}(u+v)\,\mathbb{R}_{00'}^{t_{0'}}(-u-v-2)
\end{equation}
and employ the cyclicity of double trace~$\tr_{0,0'}$, getting the following result
\begin{multline*}
	\hat{\mathbbm{t}}(u,s_0)\,\hat{\mathbbm{t}}(v,s_0') \\
	\quad =
	\rho_{s_0,s_0'}^{-1}(u+v) \,
	\tr_{0,0'}\bigl[\mathbb{R}_{00'}^{t_{0'}}(-u-v-2)\,\overset{0'}{\mathbb{K}}{}_+^{t_{0'}}(v,s_0')\,\overset{0}{\mathbb{K}}_+(u,s_0)\,\overset{0}{\mathbb{T}}(u,s_0)\,\overset{0'}{\mathbb{T}}{}^{t_{0'}}(v,s_0')\,\mathbb{R}_{00'}^{t_{0'}}(u+v)\bigr] .
\end{multline*}
Next we rewrite by means of~\eqref{transp} and~\eqref{transp-triv}
\begin{align}
	& \nonumber 
	\mathbb{R}_{00'}^{t_{0'}}(-u-v-2)\,\overset{0'}{\mathbb{K}}{}_+^{t_{0'}}(v,s_0')\,\overset{0}{\mathbb{K}}_+(u,s_0)
	= \bigl(\overset{0'}{\mathbb{K}}{}_+(v,s_0')\,\mathbb{R}_{00'}(-u-v-2)\bigr)^{t_{0'}} \, \overset{0}{\mathbb{K}}_+(u,s_0) \\
	& \nonumber 
	\qquad =
	\bigl(\overset{0'}{\mathbb{K}}{}_+(v,s_0')\,\mathbb{R}_{00'}(-u-v-2)\,\overset{0}{\mathbb{K}}_+(u,s_0)\bigr)^{t_{0'}} \,, \\
	& \nonumber
	\overset{0}{\mathbb{T}}(u,s_0)\,\overset{0'}{\mathbb{T}}{}^{t_{0'}}(v,s_0')\,\mathbb{R}_{00'}^{t_{0'}}(u+v) \\
	& \nonumber
	\qquad
	= \overset{0}{\mathbb{T}}(u,s_0) \, \bigl(\mathbb{R}_{00'}(u+v)\,\overset{0'}{\mathbb{T}}(v,s_0')\bigr)^{t_{0'}}
	= \bigl(\overset{0}{\mathbb{T}}(u,s_0)\,\mathbb{R}_{00'}(u+v)\,\overset{0'}{\mathbb{T}}(v,s_0')\bigr)^{t_{0'}}
\end{align}
and using  the formula~\eqref{tr-tt} (the infinite-dimensional version of~\eqref{trAtBt}) express the trace as
\begin{multline} \nonumber
	\hat{\mathbbm{t}}(u,s_0)\,\hat{\mathbbm{t}}(v,s_0') \\
	\quad =
	\rho_{s_0,s_0'}^{-1}(u+v) \,
	\tr_{0,0'}\bigl[
	\overset{0'}{\mathbb{K}}{}_+(v,s_0')\,\mathbb{R}_{00'}(-u-v-2)\,\overset{0}{\mathbb{K}}_+(u,s_0) \,
	\overset{0}{\mathbb{T}}(u,s_0)\,\mathbb{R}_{00'}(u+v)\,\overset{0'}{\mathbb{T}}(v,s_0')
	\bigr] .
\end{multline}
Inserting the identity again with the help of~\eqref{Runi1}
\begin{equation}
	\II = 
	\mathbb{R}_{00'}(v-u)\,\mathbb{R}_{00'}(u-v)
\end{equation}
we eventually obtain in the argument of the trace the product of left hand side and right hand side of reflection equations~\eqref{RTRTgen} and~\eqref{Refl+}, correspondingly,
\begin{multline} \nonumber
	\hat{\mathbbm{t}}(u,s_0)\,\hat{\mathbbm{t}}(v,s_0') = \rho_{s_0,s_0'}^{-1}(u+v)  \\
	\times
	\tr_{0,0'}\bigl[
	\overset{0'}{\mathbb{K}}{}_+(v,s_0')\,\mathbb{R}_{00'}(-u-v-2)\,\overset{0}{\mathbb{K}}_+(u,s_0) \,
	\mathbb{R}_{00'}(v-u)\,\mathbb{R}_{00'}(u-v)\,
	\overset{0}{\mathbb{T}}(u,s_0)\,\mathbb{R}_{00'}(u+v)\,\overset{0'}{\mathbb{T}}(v,s_0')
	\bigr] .
\end{multline}
Applying the reflection equations, performing the cyclic permutation of $\mathbb{R}_{00'}(v-u)\,\overset{0}{\mathbb{K}}_+(u,s_0)$ and cancelling the mutually inverse operators $\mathbb{R}_{00'}(v-u)$ and $\mathbb{R}_{00'}(u-v)$ we arrive at
\begin{multline} \nonumber
	\hat{\mathbbm{t}}(u,s_0)\,\hat{\mathbbm{t}}(v,s_0') \\
	= \rho_{s_0,s_0'}^{-1}(u+v)
	\tr_{0}\bigl[ \,
	\tr_{0'}\bigl[
	\mathbb{R}_{00'}(-u-v-2)\,\overset{0'}{\mathbb{K}}{}_+(v,s_0') \,	\overset{0'}{\mathbb{T}}(v,s_0')  \, \mathbb{R}_{00'}(u+v) \bigr] \,
	\overset{0}{\mathbb{T}}(u,s_0)\,\overset{0}{\mathbb{K}}_+(u,s_0)
	\bigr] .
\end{multline}
The final step is the cancellation of the second pair of mutually inverse $\mathbb{R}$-operators
\begin{align*}
	&
	\tr_{0'}\bigl[
	\mathbb{R}_{00'}(-u-v-2)\,\overset{0'}{\mathbb{K}}{}_+(v,s_0') \,	\overset{0'}{\mathbb{T}}(v,s_0')  \, \mathbb{R}_{00'}(u+v) \bigr] \\
	& =
	\tr_{0'}\Bigl[
	\bigl(\mathbb{R}_{00'}(-u-v-2)\,\overset{0'}{\mathbb{K}}{}_+(v,s_0') \,	\overset{0'}{\mathbb{T}}(v,s_0')\bigr)^{t_{0'}}  \, \mathbb{R}_{00'}^{t_{0'}}(u+v) \Bigr] \\
	& =
	\tr_{0'}\bigl[
	\overset{0'}{\mathbb{T}}{}^{t_{0'}}(v,s_0') \,
	\overset{0'}{\mathbb{K}}{}_+^{t_{0'}}(v,s_0') \, \mathbb{R}_{00'}^{t_{0'}}(-u-v-2) \, \mathbb{R}_{00'}^{t_{0'}}(u+v) \bigr] \\
	& =
	\rho_{s_0,s_0'}(u+v) \,
	\tr_{0'}\bigl[
	\overset{0'}{\mathbb{T}}{}^{t_{0'}}(v,s_0') \,
	\overset{0'}{\mathbb{K}}{}_+^{t_{0'}}(v,s_0') \bigr]
	= \rho_{s_0,s_0'}(u+v)\,\hat{\mathbbm{t}}(v,s_0') \,,
\end{align*}
where we have used~\eqref{tr-tt} and~\eqref{transp} again.
As the result we get the desired general transfer matrix commutativity
\begin{multline} \nonumber
	\hat{\mathbbm{t}}(u,s_0)\,\hat{\mathbbm{t}}(v,s_0')
	= \tr_{0}\bigl[ \hat{\mathbbm{t}}(v,s_0') \,
	\overset{0}{\mathbb{T}}(u,s_0)\,\overset{0}{\mathbb{K}}_+(u,s_0)
	\bigr] \\
	= \hat{\mathbbm{t}}(v,s_0') \,
	\tr_{0}\bigl[  \overset{0}{\mathbb{T}}(u,s_0)\,\overset{0}{\mathbb{K}}_+(u,s_0) \bigr]
	= \hat{\mathbbm{t}}(v,s_0') \, \hat{\mathbbm{t}}(u,s_0) \,.
\end{multline}

\section{Baxter $Q$-operator} \label{sect:Q}

In the previous section we introduced the general transfer matrices~\eqref{general-t} --
the two-parameter families of operators which act in the quantum space of our models~\eqref{HN}: $\mathbbm{t}_n(x,s_0)$ for one-boundary spin chain and $\hat{\mathbbm{t}}_n(x,s_0)$ for the two-boundary chain.
They are traces of monodromy matrices~\eqref{generalT},~\eqref{generalhatT} with infinite-dimensional auxiliary space, which are obtained from ordinary monodromy matrices~\eqref{Top},~\eqref{Top1} (with auxiliary space $\mathbb{C}^2$) by replacing the $L$-operator and $K_-$-, $K_+$-matrices with their infinite-dimensional analogs -- the $\mathbb{R}$-operator~\eqref{RYB} and $\mathbb{K}_-$-, $\mathbb{K}_+$-operators~\eqref{Kvs},~\eqref{Kvs+}.
We proved that general transfer matrices form commutative sets of operators
\begin{equation} \nonumber %\label{bmtbmt-comm}
	[\mathbbm{t}_n(x,s_0) \,, \mathbbm{t}_n(y,s_0')] = 0 \,, \qquad
	[\hat{\mathbbm{t}}_n(x,s_0) \,, \hat{\mathbbm{t}}_n(y,s_0')] = 0
\end{equation}
and commute with transfer matrices of the corresponding spin chains
\begin{equation} \nonumber %\label{bmtt-comm}
	[\mathbbm{t}_n(x,s_0) \,, t_n(y)] = 0 \,, \qquad
	[\hat{\mathbbm{t}}_n(x,s_0) \,, \hat{t}_n(y)] = 0 \,.
\end{equation}
The last two properties are attributes of Baxter's $Q$-operator,
and the general transfer matrix indeed plays the role of two-parameter $Q$-operator (up to a linear change of parameters).
In the current section we prove Baxter's $tQ$-relations (Baxter equations) for this $Q$-operator with respect to each of the parameters and reduce it to simpler one-parameter operators.

Let us define the following operators for the one-boundary and two-boundary chain
\begin{align} \label{Qdef}
	\mathbb{Q}_n(x_+;x_-) := \mathbbm{t}_n(x(x_+,x_-), s_0(x_+,x_-)) \,, \quad
	\hat{\mathbb{Q}}_n(x_+;x_-) := \hat{\mathbbm{t}}_n(x(x_+,x_-), s_0(x_+,x_-)) \,,
\end{align}
where we passed from the $(x,s_0)$-parametrization to the equivalent pair of spectral parameters
\begin{equation} \label{x+-}
	x_+ = x + s_0 \,, \quad x_- = x - s_0 + 1 \,.
\end{equation}
As well as the general transfer matrix, these operators also depend on the antiholomirphic counterparts of the parameters $\bar{x}_+ = \bar{x} + \bar{s}_0$, $\bar{x}_- = \bar{x} - \bar{s}_0 + 1$.

Operators~\eqref{Qdef} possess the characteristic properties of two-parameter Baxter $Q$-operator:
\begin{itemize}
	\item each of them commutes with the corresponding transfer matrices: $t_n(u)$,~$\bar{t}_n(\bar{u})$ or $\hat{t}_n(u)$,~$\bar{\hat{t}}_n(\bar{u})$
		\begin{equation} \label{qtcom}
			\begin{aligned}
				\mathbb{Q}_n(x_+;x_-)\,t_n(u) = t_n(u)\,\mathbb{Q}_n(x_+;x_-) \,, \quad  \mathbb{Q}_n(x_+;x_-)\,\bar{t}_n(\bar{u}) = \bar{t}_n(\bar{u})\,\mathbb{Q}_n(x_+;x_-) \,, \\
				\hat{\mathbb{Q}}_n(x_+;x_-)\,\hat{t}_n(u) = \hat{t}_n(u)\,\hat{\mathbb{Q}}_n(x_+;x_-) \,, \quad  \hat{\mathbb{Q}}_n(x_+;x_-)\,\bar{\hat{t}}_n(\bar{u}) = \bar{\hat{t}}_n(\bar{u})\,\hat{\mathbb{Q}}_n(x_+;x_-) \,,
			\end{aligned}	
		\end{equation}	
	\item they commute at different values of spectral parameters
		\begin{align}
			& \nonumber
			\mathbb{Q}_n(x_+;x_-)\,\mathbb{Q}_n(y_+;y_-) = \mathbb{Q}_n(y_+;y_-)\,\mathbb{Q}_n(x_+;x_-) \,, \\
			& \label{qqcom}
			\hat{\mathbb{Q}}_n(x_+;x_-)\,\hat{\mathbb{Q}}_n(y_+;y_-) = \hat{\mathbb{Q}}_n(y_+;y_-)\,\hat{\mathbb{Q}}_n(x_+;x_-) \,,
		\end{align}	
	\item they satisfy the Baxter equations with respect to each parameter
		\begin{align}
		\nonumber
		& \mathbb{Q}_n(u,\bar{u};x_-)\,t_n(u) = \Delta_1(u,x_-)\,\mathbb{Q}_n(u + 1,\bar{u};x_-) + \Delta_2(u;x_-)\,\mathbb{Q}_n(u - 1,\bar{u};x_-) \,, \\
		\label{baxterqeq1}
		& t_n(u)\,\mathbb{Q}_n(x_+;u,\bar{u}) = \Delta_1(u,x_+)\,\mathbb{Q}_n(x_+,u + 1,\bar{u}) + \Delta_2(u,x_+)\,\mathbb{Q}_n(x_+;u - 1,\bar{u}) \,,
		\end{align}
		\begin{align}
		\nonumber
		& \hat{\mathbb{Q}}_n(u,\bar{u};x_-)\,\hat{t}_n(u) = \hat{\Delta}_1(u,x_-)\,\hat{\mathbb{Q}}_n(u + 1,\bar{u};x_-) + \hat{\Delta}_2(u,x_-)\,\hat{\mathbb{Q}}_n(u - 1,\bar{u};x_-) \,, \\
		\label{baxterqeq2}
		& \hat{t}_n(u)\,\hat{\mathbb{Q}}_n(x_+;u,\bar{u}) = \hat{\Delta}_1(u,x_+)\,\hat{\mathbb{Q}}_n(x_+;u + 1,\bar{u}) + \hat{\Delta}_2(u,x_+)\,\hat{\mathbb{Q}}_n(x_+;u - 1,\bar{u}) \,.
		\end{align}
\end{itemize}
Here we denote 
\begin{align}
	\nonumber
	& \Delta_1(u,y) := \frac{(-1)^n\,\imath\,(u-\frac{1}{2})(u+y+1)}{2u(u+y)}\,, \\ 
	\nonumber
	&\Delta_2(u,y) := \frac{(-1)^n\,\imath\gamma^2}{u(u+y)}\,(2u+1)(u+y-1)(u-g)(u+g-1)\Bigl((u-s)(u+s-1)\Bigr)^{2n} , \\
	\nonumber
	& \hat{\Delta}_1(u,y) := \frac{(-1)^{n+1}\,\tilde{\gamma}^2}{2u(u+y)} \, (2u-1)(u+y+1)(u-\tilde{g}+1)(u+\tilde{g}) \,, \\
	\label{DeltaCoeff}
	& \hat{\Delta}_2(u,y) := \frac{(-1)^{n+1}\,\gamma^2}{2u(u+y)} \, (2u+1)(u+y-1)(u-g)(u+g-1)\Bigl((u-s)(u+s-1)\Bigr)^{2n} .
\end{align}
Note that in every Baxter equation we show the dependence of the $\mathbb{Q}$-operator on the antiholomorphic counterpart of varying parameter in order to emphasize that this antiholomorphic counterpart remains unchanged.
For example, in the first equation in~\eqref{baxterqeq1} we have the operator $\mathbb{Q}(x_+;x_-)$ at three points $(x_+,\bar{x}_+)=(u,\bar{u})$, $(x_+,\bar{x}_+)=(u+1,\bar{u})$ and $(x_+,\bar{x}_+)=(u-1,\bar{u})$ (parameters $(x_-;\bar{x}_-)$ do not change).
At all these three points parameter $\bar{x}_+$ has one and the same value: $\bar{x}_+=\bar{u}$.
Throughout the current section we will use this notation for operators $\mathbb{Q}$, $\mathbb{R}$~\eqref{Rx+-} and $\mathbb{K}_{\pm}$~\eqref{K+x+-}, \eqref{K-x+-} in the $(x_+;x_-)$-parametrization where necessary.
In addition to~\eqref{baxterqeq1}, \eqref{baxterqeq2}, there are similar finite-difference Baxter equations for $\mathbb{Q}_n$ and $\hat{\mathbb{Q}}_n$ involving antiholomorphic transfer matrices $\bar{t}_n(\bar{u})$ and $\bar{\hat{t}}_n(\bar{u})$, respectively.
In those equations, the antiholomorphic parameter~$\bar{u}$
is shifted, whereas the holomorphic parameter~$u$ remains unchanged.

The properties \eqref{qtcom} and \eqref{qqcom} are direct consequences of the commutativity of general transfer matrices \eqref{Gcomm}.
We are going to prove the Baxter equations for the two-boundary case~\eqref{baxterqeq2}. The proof for the one-boundary case is similar.

Taking into account the definition of $\hat{\mathbb{Q}}_n(x_+,x_-)$ and the expression for $\hat{\mathbbm{t}}_n(x,s_0)$~\eqref{general-t} we can write
\begin{align}
	\nonumber
	\mathbb{Q}_n(x_+;x_-) = \, & \tr_0\bigl[\mathbb{R}_{n0}(x_+;x_-)\dots\mathbb{R}_{10}(x_+;x_-)\,\overset{0}{\mathbb{K}}_-(x_+;x_-)\,\mathbb{R}_{10}(x_+;x_-)\dots\mathbb{R}_{n0}(x_+;x_-)\bigr] \,, \\
	\nonumber
	\hat{\mathbb{Q}}_n(x_+;x_-) = \, & \tr_0\bigl[\overset{0}{\mathbb{K}}_+(x_+;x_-)\,\mathbb{R}_{n0}(x_+,x_-)\dots\mathbb{R}_{10}(x_+,x_-) \\
	& \nonumber
	\quad\;\;\times \overset{0}{\mathbb{K}}_-(x_+;x_-)\,\mathbb{R}_{10}(x_+;x_-)\dots\mathbb{R}_{n0}(x_+;x_-)\bigr] \,,
\end{align}
where by $\mathbb{R}_{k0}(x_+;x_-)$, $\mathbb{K}_+(x_+;x_-)$ and $\mathbb{K}_-(x_+;x_-)$ we denote the operators $\mathbb{R}_{k0}(x)$, $\mathbb{K}_+(x,s_0)$ and $\mathbb{K}_-(x,s_0)$ in new parametrization~\eqref{x+-}.
Recall that the operator $\mathbb{R}_{k0}$ acts in the tensor product of $k$-th $\mathrm{L}^2(\mathbb{C})$ space corresponding to representation with spins $(s,\bar{s})$ and $0$-th space of representation with spins $(s_0,\bar{s}_0)$.
In accordance with the formula for $\mathbb{R}$-operator~\eqref{RYB} we have
\begin{align}
	\nonumber
	\mathbb{R}_{k0}(x_+;x_-)
	& =
	\mathbb{P}_{k0}\,[z_{k0}]^{x+1-s-s_0}\,
	[\hat{p}_k]^{x+s_0-s}\,[\hat{p}_0]^{x+s-s_0}\,
	[z_{k0}]^{x+s+s_0-1} \\
	\label{Rx+-}
	& =
	\mathbb{P}_{k0}\,[z_{k0}]^{x_--s}\,
	[\hat{p}_k]^{x_+-s}\,[\hat{p}_0]^{x_-+s-1}\,
	[z_{k0}]^{x_++s-1} \,.
\end{align}
Recalling the formulas for $\mathbb{K}_+$, $\mathbb{K}_-$ \eqref{Kvs+}, \eqref{Kvs} we get the following expressions
\begin{align}
	& \label{K+x+-}
	\mathbb{K}_+(x_+;x_-) =
	[z+\tilde{\gamma}]^{-x_++\tilde{g}-1} \,
	[z-\tilde{\gamma}]^{-x_+-\tilde{g}} \,
	[\hat{p}]^{-x_+-x_--1} \,
	[z+\tilde{\gamma}]^{-x_--\tilde{g}} \,
	[z-\tilde{\gamma}]^{-x_-+\tilde{g}-1} \,, \\
	& \label{K-x+-}
	\mathbb{K}_-(x_+;x_-) =
	[z+\gamma]^{x_-+g-1} \,
	[z-\gamma]^{x_--g} \,
	[\hat{p}]^{x_++x_--1} \,
	[z+\gamma]^{x_+-g} \,
	[z-\gamma]^{x_++g-1} \,.
\end{align}

\subsection{Triangularity relations and $\mathbb{R}$-operator factorization}

The crucial role in the proof of Baxter equations play the following operator-valued matrices
\begin{align}
	\label{Mk01def}
	\mathcal{M}^{(1)}_{k0}(u,x_+,x_-) & :=
	Z_0^{-1}\,L_k(u)\,\mathbb{R}_{k0}(x_+;x_-)\,Z_0 \,, \\
	\label{Mk02def}
	\mathcal{M}^{(2)}_{k0}(u,x_+,x_-) & :=
	Z_0^{-1}\,\mathbb{R}_{k0}(x_+;x_-)\,L_k(u)\,Z_0 \,, \\
	\label{M-1def}
	\mathcal{M}^{(1)}_-(u,x_+,x_-) & :=
	Z^{-1}\,K_-(u)\,L(x_++u-1,x_-+u-1)\,\mathbb{K}_-(x_+;x_-)\,Z \,, \\
	\label{M-2def}
	\mathcal{M}^{(2)}_-(u,x_+,x_-) & :=
	Z^{-1}\,\mathbb{K}_-(x_+;x_-)\,L(x_++u-1,x_-+u-1)\,K_-(u)\,Z \,, \\
	\label{M+1def}
	\mathcal{M}^{(1)}_+(u,x_+,x_-) & := Z^{-1}\,\mathbb{K}_+(x_+;x_-)\,L(-x_--u-1,-x_+-u-1)\,K_+(u)\,Z \,, \\
	\label{M+2def}
	\mathcal{M}^{(2)}_+(u,x_+,x_-) & :=
	Z^{-1}\,K_+(u)\,L(-x_--u-1,-x_+-u-1)\,\mathbb{K}_+(x_+;x_-)\,Z \,,
\end{align}
where
\begin{equation} \nonumber
	Z =
	\begin{pmatrix}
		1 & 0 \\
		z & 1
	\end{pmatrix} ,
\end{equation}
by $Z_j$ we denote the matrix $Z$ with $z=z_j$ and
$L(u_1,u_2)$ is the parametrization of $L$-operator~\eqref{Lfact}.
The key property of these matrices is that they become upper-triangular at special points $u=x_-$
\begin{align}
	\label{Mk01}
	\mathcal{M}^{(1)}_{k0}(u,x_+,u) & =
	\begin{pmatrix}
		-\imath\,(u-s)(u+s-1)\,\mathbb{R}_{k0}(x_+;u-1,\bar{u}) &
		-\partial_k\,\mathbb{R}_{k0}(x_+;u,\bar{u}) \\
		0 & \imath\,\mathbb{R}_{k0}(x_+;u+1,\bar{u})
	\end{pmatrix} , \\
	\label{M-1}
	\mathcal{M}^{(1)}_-(u,x_+,u) & =
	\left(
	\begin{array}{c|c}
		\begin{array}{l}
			2\imath\gamma^2(u-g)(u+g-1) \\
			\times (u+x_+-1)\mathbb{K}_-(x_+;u-1,\bar{u})
		\end{array} &
		\begin{array}{l}
			\frac{-\imath[z(u-\frac{1}{2})+\gamma(g-\frac{1}{2})]}{z^2-\gamma^2}\,\mathbb{K}_-(x_+;u+1,\bar{u}) \\
			-\frac{2\gamma^2(u-g)(u+g-1)}{z^2-\gamma^2}\,\mathbb{K}_-(x_+;u,\bar{u})
		\end{array} \\[12pt]
		\hline
		0 & \imath\,\bigl(u-\frac{1}{2}\bigr)\,\mathbb{K}_-(x_+;u+1,\bar{u})
	\end{array}
	\right) \!, \\
	\label{M+1}
	\mathcal{M}^{(1)}_+(u,x_+,u) & =
	\left(
	\begin{array}{c|c}
		-\imath\bigl(u+\frac{1}{2}\bigr)\mathbb{K}_+(x_+;u-1,\bar{u}) &
		\begin{array}{l}
			-\mathbb{K}_+(x_+;u-1,\bar{u})\frac{\imath[z(u+\frac{1}{2})+\tilde{\gamma}(\tilde{g}-\frac{1}{2})]}{z^2-\tilde{\gamma}^2} \\
			-\mathbb{K}_+(x_+;u,\bar{u})\,\frac{2\tilde{\gamma}^2(u+\tilde{g})(u-\tilde{g}+1)}{z^2-\tilde{\gamma}^2} 
		\end{array} \\[12pt]
		\hline
		0 &
		\begin{array}{l}
			-2\imath\tilde{\gamma}^2(u+\tilde{g})(u-\tilde{g}+1) \\
			\times (u+x_++1) \mathbb{K}_+(x_+;u+1,\bar{u})
		\end{array}
	\end{array}
	\right)
\end{align}
and $u=x_+$
\begin{align}
	\label{Mk02}
	\mathcal{M}^{(2)}_{k0}(u,u,x_-) & =
	\begin{pmatrix}
		\imath\,\mathbb{R}_{k0}(u+1,\bar{u};x_-) &
		-\mathbb{R}_{k0}(u,\bar{u};x_-)\,\partial_k \\
		0 &
		-\imath\,(u-s)(u+s-1)\,\mathbb{R}_{k0}(u-1,\bar{u};x_-)
	\end{pmatrix} , \\
	\label{M-2}
	\mathcal{M}^{(2)}_-(u,u,x_-) & =
	\left(
	\begin{array}{c|c}
		\imath\bigl(u-\frac{1}{2}\bigr)\,\mathbb{K}_-(u+1,\bar{u};x_-) &
		\begin{array}{l}
			\mathbb{K}_-(u+1,\bar{u};x_-)\,\frac{\imath[z(u-\frac{1}{2})-\gamma(g-\frac{1}{2})]}{z^2-\gamma^2} \\
			-\mathbb{K}_-(u,\bar{u};x_-)\,\frac{2\gamma^2(u-g)(u+g-1)}{z^2-\gamma^2}
		\end{array} \\[12pt]
		\hline
		0 &
		\begin{array}{l}
			2\imath\gamma^2(u-g)(u+g-1) \\
			\times (u+x_--1)\mathbb{K}_-(u-1,\bar{u};x_-)
		\end{array}
	\end{array}
	\right) \!, \\
	\label{M+2}
	\mathcal{M}^{(2)}_+(u,u,x_-) & =
	\left(
	\begin{array}{c|c}
		\begin{array}{l}
			-2\imath\tilde{\gamma}^2(u+\tilde{g})(u-\tilde{g}+1) \\
			\times (u+x_-+1) \mathbb{K}_+(u+1,\bar{u};x_-)
		\end{array}
		&
		\begin{array}{l}
			\frac{\imath[z(u+\frac{1}{2})-\tilde{\gamma}(\tilde{g}-\frac{1}{2})]}{z^2-\tilde{\gamma}^2}\mathbb{K}_+(u-1,\bar{u};x_-) \\
			-\frac{2\tilde{\gamma}^2(u+\tilde{g})(u-\tilde{g}+1)}{z^2-\tilde{\gamma}^2}\mathbb{K}_+(u,\bar{u};x_-) 
		\end{array} \\[12pt]
		\hline
		0 &
		-\imath\bigl(u+\frac{1}{2}\bigr)\mathbb{K}_+(u-1,\bar{u};x_-)
	\end{array}
	\right) \!.
\end{align}
Relations \eqref{Mk01} -- \eqref{M+1} are used in the derivation of the second Baxter equation, and \eqref{Mk02} -- \eqref{M+2} -- in the derivation of the first one.

Further in this section we demonstrate that identities \eqref{Mk01} and \eqref{Mk02} are equivalent and can be obtained from each other by matrix inversion.
In the same manner we show the equivalence of~\eqref{M-1} and \eqref{M-2}. The relation~\eqref{M-2} will be derived by direct calculation and the other one will be deduced from it.
Equalities~\eqref{M+1} and~\eqref{M+2} are, in fact, identical to~\eqref{M-2} and~\eqref{M-1}, respectively.
To verify it one just needs to express $\mathcal{M}^{(1)}_+(u,x_+,u)$ and $\mathcal{M}^{(2)}_+(u,u,x_-)$ in terms of $\mathcal{M}^{(2)}_-(-u,-u,x_+)$ and $\mathcal{M}^{(1)}_+(-u,-x_-,-u)$, correspondingly, using definitions \eqref{M+1def}, \eqref{M+2def} and formulas
\begin{equation} \nonumber
	\mathbb{K}_+(x_+;x_-) = \tilde{\mathbb{K}}_-(-x_-;-x_+) \,, \qquad
	K_+(u) = \tilde{K}_-(-u) \,,
\end{equation}
where by $\tilde{\mathbb{K}}_-$ and $\tilde{K}_-$ we denote the operator $\mathbb{K}_-$ and matrix $K_-$ with parameters $\gamma$ and $(g,\bar{g})$ replaced by $\tilde{\gamma}$ and $(\tilde{g},\bar{\tilde{g}})$.

Expressions \eqref{Mk01} and \eqref{Mk02} were obtained in~\cite[Section~3]{Derkachov:2005fg}
with the help of the following factorization formula for general $\mathbb{R}$-operator \cite[Section~2.3]{Derkachov:2005hw}, \cite[Section~3.3]{DM09}
\begin{equation}\label{Rfactor}
	\mathbb{R}_{k0}(x_+;x_-) = \mathbb{P}_{k0}\,\mathcal{R}^{(1)}_{k0}(x_-)\, \mathcal{R}^{(2)}_{k0}(x_+)
\end{equation}		
where
\begin{align}
	\nonumber
	\mathcal{R}^{(1)}_{ij}(y,\bar{y}) & :=
	[z_{ij}]^{y-s}\,[\hat{p}_j]^{y+s-1}\,[z_{ij}]^{2s-1} = 
	[\hat{p}_j]^{2s-1}\,[z_{ij}]^{y+s-1}
	\,[\hat{p}_j]^{y-s} \,, \\
	\label{R12def}
	\mathcal{R}^{(2)}_{ij}(y,\bar{y}) & := 
	[z_{ij}]^{1-2s}\,[\hat{p}_i]^{y-s}\,[z_{ij}]^{y+s-1} = 
	[\hat{p}_i]^{y+s-1}\,[z_{ij}]^{y-s}
	\,[\hat{p}_i]^{1-2s} \,.
\end{align}
In what follows, for brevity we omit almost everywhere the dependence of $\mathcal{R}^{(1)}$ and $\mathcal{R}^{(2)}$ on the antiholomorphic parameter, as well as we do for other operators: $\mathcal{R}^{(m)}(y) := \mathcal{R}^{(m)}(y,\bar{y})$.
These operators are integral operators which act on functions $\Psi(z_i,z_j)$ in the following way
\begin{align}
	\nonumber
	\left[\mathcal{R}^{(1)}_{ij}(y,\bar{y}) \, \Psi\right](z_i, z_j) & = 
	c(y+s-1)\,
	\int \mathrm{d}^2 w \,
	\frac{[z_i-z_j]^{y-s}}{[z_j-w]^{y+s} \, [z_i-w]^{1-2s}} \,
	\Psi(z_i, w) \,,\\
	\label{R_kernel}
	\left[\mathcal{R}^{(2)}_{ij}(y,\bar{y}) \, \Psi\right](z_i, z_j) & = 
	c(y-s)\,
	\int \mathrm{d}^2 w \,
	\frac{[z_i-z_j]^{1-2s}}{[z_i-w]^{1-s+y} \, [w-z_j]^{1-s-y}} \,
	\Psi(w, z_j) \,.
\end{align}
They are the unique solutions to equations
\begin{align}
	& \nonumber
	\mathcal{R}^{(1)}_{1 2}(y) \, L_1(u_1, u_2) \, L_2(u-y, u_2) = 
	L_1(u-y, u_2) \, L_2(u_1, u_2) \, \mathcal{R}^{(1)}_{1 2}(y) \,, \\
	& \label{defR}
	\mathcal{R}^{(2)}_{1 2}(y) \, L_1(u_1, u_2) \, L_2(u_1, u-y) = 
	L_1(u_1, u-y) \, L_2(u_1, u_2) \, \mathcal{R}^{(2)}_{1 2}(y) \,.
\end{align}

Formulas \eqref{Mk01} and \eqref{Mk02} are particular cases of the more general identities
\begin{align}
	& \nonumber
	\mathcal{M}^{(1)}_{k0}(u,x_+,x_-) = Z_0^{-1}
	\,L_k(u)\,\mathbb{R}_{k0}(x_+;x_-)\,Z_0 \\
	& \label{tr1}
	=
	\left(
	\begin{array}{c|c}
		\begin{array}{l}
			(u-x_-)\mathbb{R}_{k0}(x_+;x_-,\bar{x}_-) \\
			-\imath(x_--s)(x_-+s-1)\mathbb{R}_{k0}(x_+;x_--1,\bar{x}_-)
		\end{array}
		& -\partial_k\,\mathbb{R}_{k0}(x_+;x_-,\bar{x}_-) \\[6pt]
		\hline
		\begin{array}{l}
			(u-y)z_{k0}\bigl[\mathbb{R}_{k0}(x_+;x_-,\bar{x}_-) \\
			+\imath(x_-+s-1)\mathbb{R}_{k0}(x_+;x_--1,\bar{x}_-)\bigr]
		\end{array} &
		\begin{array}{l}
			\imath\,\mathbb{R}_{k0}(x_+;x_-+1,\bar{x}_-) \\
			+ (u-x_-)\mathbb{R}_{k0}(x_+;x_-,\bar{x}_-)
		\end{array}
	\end{array}
	\right), \\
	& \nonumber
	\mathcal{M}^{(2)}_{k0}(u,x_+,x_-) = Z_0^{-1}
	\,\mathbb{R}_{k0}(x_+;x_-)\,L_k(u)\,Z_0 \\
	& \label{tr2}
	=
	\left(
	\begin{array}{c|c}
		\begin{array}{l}
			\imath\,\mathbb{R}_{k0}(x_++1,\bar{x}_+;x_-) \\
			+ (u-x_+)\mathbb{R}_{k0}(x_+,\bar{x}_+;x_-)
		\end{array}
		& -\mathbb{R}_{k0}(x_+,\bar{x}_+;x_-) \, \partial_k \\
		\hline
		(u-x_+)\,z_{k0}\,\mathbb{R}_{k0}(x_+,\bar{x}_+;x_-) &
		\begin{array}{l}
			(u-x_+)\mathbb{R}_{k0}(x_+,\bar{x}_+;x_-) \\
			-\imath(x_+-s)(x_++s-1)\mathbb{R}_{k0}(x_+-1,\bar{x}_+;x_-)
		\end{array}
	\end{array}
	\right) \!.
\end{align}
at the points $u=x_-$ and $u=x_+$, respectively.
The last two identities are, in turn, consequences of the following relations for $\mathcal{R}$-operators
\begin{align}
& \nonumber
Z_k^{-1}\,L_0(u)\,\mathcal{R}^{(1)}_{k0}(y)\,Z_0 \\
& \label{ZLRZ1}
\quad =
\left(
\begin{array}{c|c}
\begin{array}{l}
	(u-y)\mathcal{R}^{(1)}_{k0}(y,\bar{y}) \\
	-\imath(y-s)(y+s-1)\mathcal{R}^{(1)}_{k0}(y-1,\bar{y})
\end{array}
& -\partial_0\,\mathcal{R}^{(1)}_{k0}(y,\bar{y}) \\[10pt]
\hline
\begin{array}{l}
	-(u-y)z_{k0}\bigl[\mathcal{R}^{(1)}_{k0}(y,\bar{y}) \\
	+\imath(y+s-1)\mathcal{R}^{(1)}_{k0}(y-1,\bar{y})\bigr]
\end{array} &
\,\imath\,\mathcal{R}^{(1)}_{k0}(y+1,\bar{y}) +
(u-y)\mathcal{R}^{(1)}_{k0}(y,\bar{y})
\end{array}
\right),
\end{align}
\begin{align}
	& \nonumber
	Z_k^{-1}\,\mathcal{R}^{(2)}_{k0}(y)\,L_k(u)\,Z_0 \\
	& \label{ZLRZ2}
	\quad =
	\left(
	\begin{array}{c|c}
		\imath\mathcal{R}^{(2)}_{k0}(y+1,\bar{y})
		+ (u-y)\mathcal{R}^{(2)}_{k0}(y,\bar{y})
		& -\mathcal{R}^{(2)}_{k0}(y,\bar{y}) \, \partial_k \\[1pt]
		\hline
		-(u-y)\,z_{k0}\,\mathcal{R}^{(2)}_{k0}(y,\bar{y}) &
		\begin{array}{l}
			(u-y)\mathcal{R}^{(2)}_{k0}(y,\bar{y}) \\
			-\imath(y-s)(y+s-1)\mathcal{R}^{(2)}_{k0}(y-1,\bar{y})
		\end{array}
	\end{array}
	\right).
\end{align}
Namely, if we multiply~\eqref{ZLRZ1} with $y=x_-$ by the permutation operator $\mathbb{P}_{k0}$~\eqref{perm} from the left and by operator~$\mathcal{R}^{(2)}_{k0}(x_+)$ from the right, then we obtain~\eqref{tr1}.
And~\eqref{tr2} can be deduced from~\eqref{ZLRZ2} by taking~$y=x_+$ and multiplying this identity with the operator~$\mathbb{P}_{ik}\,\mathcal{R}^{(1)}_{ik}(x_-)$ from the left.

Let us show that~\eqref{Mk01} can be deduced from~\eqref{Mk02} by inversion.
On the one hand, using the definition of $\mathcal{M}^{(2)}_{k0}$, the unitarity of $L$-operator~\eqref{Lunitar} and inversion formula for $\mathbb{R}$-operator (which follows directly from the explicit expression~\eqref{Rx+-})
\begin{equation} \nonumber
	\mathbb{R}_{k0}^{-1}(x_+;x_-) = [-1]^{x_++x_-}\,\mathbb{R}_{k0}(1-x_-;1-x_+)
\end{equation}
we obtain
\begin{align}
	\nonumber
	\mathcal{M}^{(2)}_{k0}{}^{-1}(u,u,x_-) & =
	-\frac{[-1]^{u+x_-}}{(u-s)(u+s-1)}\,Z_0^{-1}\,L_k(1-u)\,\mathbb{R}_{k0}(1-x_-;1-u)\,Z_0 \\
	\label{Minv1}
	& = -\frac{[-1]^{u+x_-}}{(u-s)(u+s-1)}\,\mathcal{M}^{(1)}_{k0}(1-u,1-x_-,1-u) \,.
\end{align}
On the other hand, the matrix $\mathcal{M}^{(2)}_{k0}{}^{-1}(u,u,x_-)$ can be calculated by means of~\eqref{Mk02} and the inversion formula for an upper-triangular matrix with noncommutative entries
\begin{equation} \label{trinv}
	\mathcal{A} =
	\begin{pmatrix}
		\mathcal{A}_{11} & \mathcal{A}_{12} \\
		0 & \mathcal{A}_{22}
	\end{pmatrix}
	\Rightarrow
	\mathcal{A}^{-1} =
	\begin{pmatrix}
		\mathcal{A}_{11}^{-1} & -\mathcal{A}_{11}^{-1}\mathcal{A}_{12}\,\mathcal{A}_{22}^{-1} \\
		0 & \mathcal{A}_{22}^{-1}
	\end{pmatrix} .
\end{equation}
The corresponding expression reads
\begin{align}
	& \nonumber
	\mathcal{M}^{(2)}_{k0}{}^{-1}(u,u,x_-) \\
	& \label{Minv2}
	= -\frac{[-1]^{u+x_-}}{(u-s)(u+s-1)}
	\left(
	\begin{array}{c|c}
		\begin{array}{l}
			-\imath(u-s)(u+s-1) \\ \times\mathbb{R}_{k0}(1-x_-;-u,1-\bar{u})
		\end{array} &
		\begin{array}{l}
			\mathbb{R}_{k0}^{-1}(u+1,\bar{u};x_-)\,\mathbb{R}_{k0}(u,\bar{u};x_-) \\
			\times \partial_k\,\mathbb{R}_{k0}(1-x_-;2-u,1-\bar{u})
		\end{array}
		\\
		\hline
		0 &
		\imath\,\mathbb{R}_{k0}(1-x_-;2-u,1-\bar{u})
	\end{array}
	\right),
\end{align}
and it remains to calculate the element $(1,2)$ of the last matrix.
To do this we use two basic identities
\begin{align}
	\nonumber
	\imath(z_{k0}\partial_k-x_-+s)\,\mathbb{R}_{k0}(x_+;x_-) & =
	\mathbb{R}_{k0}(x_+;x_-+1,\bar{x}_-) \,, \\
	\nonumber
	\mathbb{R}_{k0}(x_+;x_-) \, \imath(z_{k0}\partial_k+x_++s) & =
	-\mathbb{R}_{k0}(x_++1,\bar{x}_+;x_-)
\end{align}
which can be easily derived with the help of relations
\begin{align}
	& \nonumber
	\bigl[z_{k0}\partial_0,[z_{k0}]^\alpha\bigr]=-\alpha\,[z_{k0}]^\alpha \,,
	\quad
	\bigl[z_{k0}\partial_k,[z_{k0}]^\alpha\bigr]=\alpha\,[z_{k0}]^\alpha \,, \\
	& \label{dp}
	(-i\partial_k)\,[\hat{p}_k]^\alpha=[\hat{p}_k]^\alpha\,(-i\partial_k)=[\hat{p}_k]^{\alpha+\Delta} \,,
\end{align}
where
\begin{equation} \label{Del}
	(\Delta,\bar{\Delta})=(1,0) \,,
\end{equation}
the last formula is the direct consequence of the definition~\eqref{d}.
We show the derivation of the first identity
\begin{align*}
	&
	\imath(z_{k0}\partial_k-x_-+s)\,\mathbb{R}_{k0}(x_+;x_-) \\
	&
	=
	\imath(z_{k0}\partial_k-x_-+s) \,
	\mathbb{P}_{k0}\,[z_{k0}]^{x_--s}\,[\hat{p}_0]^{x_-+s-1}\, [\hat{p}_k]^{x_+-s}\, [z_{k0}]^{x_++s-1} \\
	&
	= \mathbb{P}_{k0}\,\imath(-z_{k0}\partial_0-x_-+s)\,[z_{k0}]^{x_--s}\,[\hat{p}_0]^{x_-+s-1}\, [\hat{p}_k]^{x_+-s}\, [z_{k0}]^{x_++s-1} \\
	&
	= \mathbb{P}_{k0}\,[z_{k0}]^{x_--s}\,(-\imath z_{k0}\partial_0)\,[\hat{p}_0]^{x_-+s-1}\, [\hat{p}_k]^{x_+-s}\, [z_{k0}]^{x_++s-1} \\
	&
	= \mathbb{P}_{k0}\,[z_{k0}]^{x_--s+\Delta}\,[\hat{p}_0]^{x_-+s-1+\Delta}\, [\hat{p}_k]^{x_+-s}\, [z_{k0}]^{x_++s-1}
	= \mathbb{R}_{k0}(x_+;x_-+1,\bar{x}_-) \,,
\end{align*}
and the second one can be obtained in a similar way.
Thus we calculate the element~$(1,2)$ in~\eqref{Minv2} 
\begin{align*}
	&
	\mathbb{R}_{k0}^{-1}(u+1,\bar{u};x_-)\,\mathbb{R}_{k0}(u,\bar{u};x_-)\,
	\partial_k\,\mathbb{R}_{k0}(1-x_-;2-u,1-\bar{u}) \\
	&
	= \mathbb{R}_{k0}^{-1}(u+1,\bar{u};x_-)\,\mathbb{R}_{k0}(u,\bar{u};x_-)\,
	\partial_k\,\imath(z_{k0}\partial_k-1+u+s)\mathbb{R}_{k0}(1-x_-;1-u,1-\bar{u}) \\
	&
	= \mathbb{R}_{k0}^{-1}(u+1,\bar{u};x_-)\,\mathbb{R}_{k0}(u,\bar{u};x_-)\,\imath(z_{k0}\partial_k+u+s)\,
	\partial_k\,\mathbb{R}_{k0}(1-x_-;1-u,1-\bar{u}) \\
	&
	= \mathbb{R}_{k0}^{-1}(u+1,\bar{u};x_-)\,\mathbb{R}_{k0}(u+1,\bar{u};x_-)\,
	\partial_k\,\mathbb{R}_{k0}(1-x_-;1-u,1-\bar{u}) \\
	&
	= \partial_k\,\mathbb{R}_{k0}(1-x_-;1-u,1-\bar{u})
\end{align*}
and comparing \eqref{Minv2} with \eqref{Minv1} we conclude that
\begin{align} \nonumber
	\mathcal{M}^{(1)}_{k0}(1-u,1-x_-,1-u) =
	\left(
	\begin{array}{c|c}
		\begin{array}{l}
			-\imath(u-s)(u+s-1) \\ \times\mathbb{R}_{k0}(1-x_-;-u,1-\bar{u})
		\end{array} &
		\partial_k\,\mathbb{R}_{k0}(1-x_-;1-u,1-\bar{u})
		\\
		\hline
		0 &
		\imath\,\mathbb{R}_{k0}(1-x_-;2-u,1-\bar{u})
	\end{array}
	\right).
\end{align}
Substituting $1-u\to u \,, \, 1-x_-\to x_+$ in the last expression one obtains the desired relation~\eqref{Mk01}.

Now let us derive~\eqref{M-2}.
By definition of $\mathcal{M}^{(2)}_-$ and explicit formulas for $\mathbb{K}_-(x_+;x_-)$~\eqref{K-x+-} and $L(u_1,u_2)$~\eqref{Lfact} we have
\begin{align}
	\nonumber
	\mathcal{M}^{(2)}_-(u,u,x_-) = \, &
	[z_+]^{x_-+g-1}\,[z_-]^{x_--g}\,Z^{-1}\,[\hat{p}]^{u+x_--1}\,Z\,[z_+]^{u-g}\,[z_-]^{u+g-1} \\
	\nonumber % \label{M-2pr}
	& \times
	\begin{pmatrix}
		2u-1 & -\partial_z \\
		0 & x_-+u-1
	\end{pmatrix}
	Z^{-1}\,K_-(u)\,Z \,,
\end{align}
where we denote $z_\pm = z\pm \gamma$.
Using the commutation relation
\begin{equation} \nonumber
	\bigl[[\hat{p}]^\alpha,z\bigr] = -\imath\,\alpha\,[\hat{p}]^{\alpha-\Delta} \,,
\end{equation}
(recall the notation~\eqref{Del}) which follows from the definition of the operator~$[\hat{p}]^\alpha$~\eqref{d}, we calculate
\begin{equation} \nonumber
	Z^{-1}\,[\hat{p}]^{u+x_--1}\,Z =
	\begin{pmatrix}
		[\hat{p}]^{u+x_--1} & 0 \\
		-\imath\,(u+x_--1)\,[\hat{p}]^{u+x_--1-\Delta} & [\hat{p}]^{u+x_--1}
	\end{pmatrix}.
\end{equation}
It is also not hard to show that
\begin{align}
	& \nonumber
	[z_+]^{u-g}\,[z_-]^{u+g-1}
	\begin{pmatrix}
		2u-1 & -\partial_z \\
		0 & x_-+u-1
	\end{pmatrix} \\
	& \nonumber
	=
	\begin{pmatrix}
		2u-1 & -\partial_z+2\bigl[\gamma(g-\frac{1}{2})+z(u-\frac{1}{2})\bigr](z_+z_-)^{-1} \\
		0 & u+x_--1
	\end{pmatrix}
	[z_+]^{u-g}\,[z_-]^{u+g-1} \,, \\
	& \nonumber
	Z^{-1}\,K_-(u)\,Z =
	\begin{pmatrix}
		\gamma(g-\frac{1}{2})+z(u-\frac{1}{2}) & u-\frac{1}{2} \\
		-(u-\frac{1}{2})\,z_+z_- & \gamma(g-\frac{1}{2})-z(u-\frac{1}{2})
	\end{pmatrix}.
\end{align}
Then we multiply the last three matrices taking into account the relation~\eqref{dp} for operator~$[\hat{p}]^\alpha$ and get the following expression
\begin{align}
	& \nonumber
	\mathcal{M}^{(2)}_-(u,u,x_-) =
	[z_+]^{x_-+g-1}\,[z_-]^{x_--g} \\
	& \nonumber
	\times
	\left(
	\begin{array}{c|c}
		\imath(u-\frac{1}{2})\,[\hat{p}]^{u+x_--1+\Delta}z_+z_- &
		\begin{array}{l}
			\imath\,[\hat{p}]^{u+x_--1+\Delta}\,\bigl[z(u-\frac{1}{2})-\gamma(g-\frac{1}{2})\bigr] \\
			-2\gamma^2(u-g)(u+g-1)\,[\hat{p}]^{u+x_--1}\,(z_+z_-)^{-1}
		\end{array}
		\\
		\hline
		0 &
		2\imath\gamma^2(u-g)(u+g-1)\,[\hat{p}]^{u+x_--1-\Delta}(z_+z_-)^{-1}
	\end{array}
	\right)\\
	& \nonumber
	\times [z_+]^{u-g}\,[z_-]^{u+g-1} \,.
\end{align}
Using the formula for $\mathbb{K}_-(x_+;x_-)$ one can see that the last expression is in fact the relation~\eqref{M-2} we aimed to obtain.

The identity~\eqref{M-1} we deduce from~\eqref{M-2} by matrix inversion in the same way as we derived~\eqref{Mk01} from~\eqref{Mk02}.
On the one hand, we express the inverse of $\mathcal{M}^{(2)}_-(u,u,x_-)$ in terms of $\mathcal{M}^{(1)}_-$ using its definition~\eqref{M-2def} and formulas
\begin{align}
	\nonumber
	\mathbb{K}_-^{-1}(x_+;x_-) & = \mathbb{K}_-(1-x_-;1-x_+) \,, \\
	\nonumber
	K_-^{-1}(u) & =
	-\bigl(\gamma^2\,(u-g)(u+g-1)\bigr)^{-1} K_-(1-u) \,, \\
	\nonumber
	L^{-1}(u_1,u_2) & =
	-(u_1 u_2)^{-1} L(-u_2,-u_1) \,.
\end{align}
We arrive at
\begin{align}
	\nonumber
	\mathcal{M}^{(2)}_-{}^{-1}(u,u,x_-) & =
	\frac{Z^{-1}\,K_-(1-u)\,L(1-x_--u, 1-2u)\,\mathbb{K}_-(1-x_-;1-u)\,Z}{\gamma^2\,(u-g)(u+g-1)(2u-1)(x_-+u-1)} \\
	\label{M-inv1}
	& =
	\frac{\mathcal{M}^{(1)}_-(1-u,1-x_-,1-u)}{\gamma^2(u-g)(u+g-1)(2u-1)(u+x_--1)} \,.
\end{align}
On the other hand, one can calculate~$\mathcal{M}^{(2)}_-{}^{-1}(u,u,x_-)$ using the triangularity relation~\eqref{M-2}, the inversion formula for an upper-triangular matrix~\eqref{trinv} and the identity
\begin{multline} \nonumber
	\mathbb{K}_-^{-1}(u+1,\bar{u};x_-)\,\mathbb{K}_-(u,\bar{u};x_-)\,(z_+z_-)^{-1}\,\mathbb{K}_-^{-1}(u-1,\bar{u};x_-) \\
	= (z_+z_-)^{-1}\,\mathbb{K}_-(1-x_-;1-u,1-\bar{u}) \,,
\end{multline}
which can be easily derived by means of the explicit expression for $\mathbb{K}_-(x_+;x_-)$ and~\eqref{dp}.
The result is as follows
\begin{align}
	& \nonumber
	\mathcal{M}^{(2)}_-{}^{-1}(u,u,x_-) =
	\bigl(\gamma^2\,(u-g)(u+g-1)(2u-1)(u+x_--1)\bigr)^{-1} \\
	& \label{M-inv2}
	\times
	\left(
	\begin{array}{c|c}
		\begin{array}{l}
			-2\imath\gamma^2(u-g)(u+g-1) \\
			\times (u+x_--1)\mathbb{K}_-(1-x_-;-u,1-\bar{u})
		\end{array} &
		\begin{array}{l}
			\frac{\imath[z(u-\frac{1}{2})-\gamma(g-\frac{1}{2})]}{z^2-\gamma^2}\,\mathbb{K}_-(1-x_-;2-u,1-\bar{u}) \\
			-\frac{2\gamma^2(u-g)(u+g-1)}{z^2-\gamma^2}\,\mathbb{K}_-(1-x_-;1-u,1-\bar{u})
		\end{array} \\[12pt]
		\hline
		0 & -\imath\,\bigl(u-\frac{1}{2}\bigr)\,\mathbb{K}_-(1-x_-;2-u,1-\bar{u})
	\end{array}
	\right)\!.
\end{align}
Equating expressions~\eqref{M-inv1} and~\eqref{M-inv2} and substituting $1-x_-\to x_+$, $1-u\to u$ we obtain the needed relation~\eqref{M-1}.

\subsection{Baxter equations}

In order to prove the first Baxter equation in \eqref{baxterqeq2} we use the expression for product of transfer matrices~\eqref{bbmtt2}
\begin {align}\label{convform}
	&\hat{\mathbbm{t}}(x,s_0)\hat{t}(u)  = \tilde{\rho}_{s_0}^{-1}(u+x)\tr_{0,\mathbb{C}^2}[K_+(u) L_0(-u-x-1) \overset{0}{\mathbb{K}}_+(x,s_0) \mathbb{T}(x,s_0) L_0(u+x) T(u)] \,.
\end{align}
Recall that $L$-operator $L_0$ corresponds to spin $s_0$.
From now on, we omit the index $n$ for $Q$-operators and transfer or monodromy matrices for brevity.
By definition of $\mathbb{T}(x,s_0)$ and $T(u)$ we write 
\begin{align*}
	&
    \mathbb{T}(x,s_0) L_0(u+x) T(u) \\
    &
    = \mathbb{R}_{n0}(x) \cdots \mathbb{R}_{10}(x) \, \overset{0}{\mathbb{K}}_-(u,x_0) \, \mathbb{R}_{10}(x) \dots \mathbb{R}_{n0}(x) \\
    &
    \quad\times
    L_0(u+x) \, L_n(u) \dots L_1(u) \, K_-(u) \, L_1(u) \cdots L_n(u) \\
    &
    = \Big( \mathbb{R}_{n0}(x) L_n(u) \Big) \dots \Big( \mathbb{R}_{10}(x) L_1(u) \Big) \, \overset{0}{\mathbb{K}}_-(u,x_0) \, L_0(u+x+s_0-1,u+x-s_0) \, K_-(u) \\
    &
    \quad\times
    \Big( \mathbb{R}_{10}(x) L_1(u) \Big) \dots \Big( \mathbb{R}_{n0}(x) L_n(u) \Big) \,,
\end{align*}
where we applied the Yang-Baxter equation~\eqref{YBRLL} for $\mathbb{R}$-operator and $L$-operator $n$ times, and rewrote the $L$-operator in the parametrization~\eqref{Lfact}.
With the help of the last result we can rewrite the right hand side of~\eqref{convform} in $(x_+;x_-)$-parametrization in terms of matrices~\eqref{Mk02def}, \eqref{M-2def}, \eqref{M+2def}
\begin{align}
\nonumber
\mathbb{T}(x,s_0) L_0(u+x) T(u) = \, &
Z_0\,\mathcal{M}^{(2)}_{n0}(u,x_+,x_-) \dots \mathcal{M}^{(2)}_{10}(u,x_+,x_-) \, \overset{0}{\mathcal{M}}{}^{(2)}_-(u,x_+,x_-) \\
\label{treug}
&\times \mathcal{M}^{(2)}_{10}(u,x_+,x_-) \cdots \mathcal{M}^{(2)}_{n0}(u,x_+,x_-) \, Z_0^{-1} \,,
\end{align}
\begin{align}
	\nonumber
	K_+(u)\,L_0(-u-x-1)\,\mathbb{K}_+(x,s_0) & =
	K_+(u)L_0(-u-x+x_0-2,-u-x-x_0-1)\mathbb{K}_+(x,s_0) \\
	\label{treug+}
	& = Z_0\,\overset{0}{\mathcal{M}}{}^{(2)}_+(u,x_+,x_-)\,Z_0^{-1} \,,
\end{align}
\begin{align} \nonumber
	\tilde{\rho}_{s_0}^{-1}(u+x) = -\frac{1}{(u+x+s_0)(u+x-s_0+1)} = -\frac{1}{(u+x_+)(u+x_-)} = \beta(u,x_+,x_-) \,,
\end{align}
where the index ``$0$'' above $\mathcal{M}^{(2)}_-$ in~\eqref{treug} and $\mathcal{M}^{(2)}_+$ in~\eqref{treug+} indicates that the elements of these matrices are operators acting nontrivially in the $0$-th space $\mathrm{L}_2(\mathbb{C})$.

Thus using the definition of $Q$-operator $\hat{\mathbb{Q}}(x_+,x_-) = \hat{\mathbbm{t}}(x(x_+,x_-), s_0(x_+,x_-))$ we rewrite \eqref{convform} as
\begin{align}
	\nonumber
	\hat{\mathbb{Q}}(x_+,x_-)\,\hat{t}(u) = \beta(u,x_+,x_-) 
	\tr_{0,\mathbb{C}^2}\Bigl[
	\overset{0}{\mathcal{M}}{}^{(2)}_+(u,x_+,x_-) \mathcal{M}^{(2)}_{n0}(u,x_+,x_-) \dots \mathcal{M}^{(2)}_{10}(u,x_+,x_-) \\
	\label{Qtri}
	\times \overset{0}{\mathcal{M}}{}^{(2)}_-(u,x_+,x_-)\, \mathcal{M}^{(2)}_{10}(u,x_+,x_-) \dots \mathcal{M}^{(2)}_{n0}(u,x_+,x_-) \Bigr].
\end{align}
Note that we cancelled the leftmost and rightmost matrices $Z_0$ and $Z_0^{-1}$ thanks to the cyclic property of the trace.
Now we substitute $x_+=u$.
According to the key formulas \eqref{Mk02} -- \eqref{M+2}, matrices $\mathcal{M}^{(2)}_{k0}$, $\mathcal{M}^{(2)}_{\pm}$ become upper-triangular in this case, so it is easy to compute the trace over $\mathbb{C}^2$
\begin{align}
	& \nonumber
	\begin{aligned}
		\tr_{\mathbb{C}^2}\Bigl[ & \overset{0}{\mathcal{M}}{}^{(2)}_+(u,u,x_-) \mathcal{M}^{(2)}_{n0}(u,u,x_-) \dots \mathcal{M}^{(2)}_{10}(u,u,x_-) \\
		& \times
		\overset{0}{\mathcal{M}}{}^{(2)}_-(u,u,x_-)\, \mathcal{M}^{(2)}_{10}(u,u,x_-) \dots \mathcal{M}^{(2)}_{n0}(u,u,x_-) \Bigr] =
	\end{aligned} \\
	& \nonumber
	= (-1)^{n}\,\tilde{\gamma}^2(2u-1)(u+x_-+1)(u-\tilde{g}+1)(u+\tilde{g}) \, \hat{\mathbb{T}}(u+1,\bar{u};x_-) \\
	& \nonumber
	+ (-1)^{n}\,\gamma^2(2u+1)(u+x_--1)(u-g)(u+g-1)\Bigl((u-s)(u+s-1)\Bigr)^{2n}\hat{\mathbb{T}}(u-1,\bar{u};x_-) \,,
\end{align}
where we remind that
\begin{align*}
	\hat{\mathbb{T}}(x_+;x_-) = \, &
	\overset{0}{\mathbb{K}}_+(x_+;x_-)\,\mathbb{R}_{n0}(x_+;x_-)\dots\mathbb{R}_{10}(x_+;x_-) \\
	& \times 
	\overset{0}{\mathbb{K}}_-(x_+;x_-)\,\mathbb{R}_{10}(x_+;x_-)\dots\mathbb{R}_{n0}(x_+;x_-) \,.
\end{align*}
Substituting the last expression into~\eqref{Qtri} one immediately obtains the first Baxter equation~\eqref{baxterqeq2}.

To prove the second Baxter equation we use the same technique. We write the product $\hat{t}_n(u)\,\hat{\mathbbm{t}}_n(x,s_0)$ in the following form
\begin {align}\label{convform2}
	&\hat{t}(u)\hat{\mathbbm{t}}(x,s_0)  = \tilde{\rho}_{s}(u+x)^{-1}\tr_{0,\mathbb{C}^2}[\overset{0}{\mathbb{K}}_+(x,s_0)L_0(-u-x-1)K_+(u)\,T(u)L_0(u+x)\mathbb{T}(x,s_0) ] \,,
\end{align}
this expression can be derived in the same way as~\eqref{convform} was obtained in Section~\ref{sect:generalT}.
Applying the Yang-Baxter equation~\eqref{YBRLL} $n$ times we get
\begin{align*}
	T(u)\,L_0(u+x)\,\mathbb{T}(x,s_0) = \, & L_n(u)\,\mathbb{R}_{n0}(x) \dots L_1(u)\,\mathbb{R}_{10}(x) \\
	& \times 
	K(u)\,L_0(u+x)\,\overset{0}{\mathbb{K}}_-(x,s_0) \, L_1(u)\,\mathbb{R}_{10}(x) \dots L_n(u)\,\mathbb{R}_{n0}(x) \,.
\end{align*}
This way, we rewrite~\eqref{convform2} in $(x_+;x_-)$-parametrization in terms of matrices~\eqref{Mk01def}, \eqref{M-1def}, \eqref{M+1def}
\begin{align}
	\nonumber
	\hat{\mathbb{Q}}(x_+,x_-)\,\hat{t}(u) = \beta(u,x_+,x_-) 
	\tr_{0,\mathbb{C}^2}\Bigl[
	\overset{0}{\mathcal{M}}{}^{(1)}_+(u,x_+,x_-) \mathcal{M}^{(1)}_{n0}(u,x_+,x_-) \dots \mathcal{M}^{(1)}_{10}(u,x_+,x_-) \\
	\nonumber
	\times \overset{0}{\mathcal{M}}{}^{(1)}_-(u,x_+,x_-)\, \mathcal{M}^{(1)}_{10}(u,x_+,x_-) \dots \mathcal{M}^{(1)}_{n0}(u,x_+,x_-) \Bigr].
\end{align}
And to obtain the second Baxter equation of~\eqref{baxterqeq2} it only remains to substitute~$x_-=u$ in the last expression and use triangularity relations~\eqref{Mk01}, \eqref{M-1}, \eqref{M+1} for matrices $\mathcal{M}^{(1)}_{k0}$, $\mathcal{M}^{(1)}_\pm$ at this point to calculate the trace over~$\mathbb{C}^2$.

\subsection{Degenerate $Q$-operators}

Recall the formulas for $Q$-operator in the case of one-boundary and two-boundary spin chain
\begin{align}
	\nonumber
	\mathbb{Q}(x_+;x_-) = \, & \tr_0\bigl[\mathbb{R}_{n0}(x_+;x_-)\dots\mathbb{R}_{10}(x_+;x_-)\,\overset{0}{\mathbb{K}}_-(x_+;x_-)\,\mathbb{R}_{10}(x_+;x_-)\dots\mathbb{R}_{n0}(x_+;x_-)\bigr] \,, \\
	\nonumber
	\hat{\mathbb{Q}}(x_+;x_-) = \, & \tr_0\bigl[\overset{0}{\mathbb{K}}_+(x_+;x_-)\,\mathbb{R}_{n0}(x_+,x_-)\dots\mathbb{R}_{10}(x_+,x_-) \\
	& \nonumber
	\quad\;\;\times \overset{0}{\mathbb{K}}_-(x_+;x_-)\,\mathbb{R}_{10}(x_+;x_-)\dots\mathbb{R}_{n0}(x_+;x_-)\bigr] \,.
\end{align}
As it was mentioned, in $(x_+;x_-)$-parametrization the $\mathbb{R}$-operator can be represented as
\begin{equation} \nonumber
	\mathbb{R}_{k0}(x_+;x_-) = \mathbb{P}_{k0}\,\mathcal{R}^{(1)}_{k0}(x_-)\, \mathcal{R}^{(2)}_{k0}(x_+) \,.
\end{equation}
It will also be convenient to express $\mathbb{K}_-$ and $\mathbb{K}_+$ in terms of $\mathcal{K}$-operator~\eqref{Kop-d}
\begin{align*}
	\mathbb{K}_-(x_+;x_-) = \mathcal{K}(1-x_-,x_+) \,, \qquad
	\mathbb{K}_+(x_+;x_-) = \tilde{\mathcal{K}}(1+x_+,-x_-) \,,
\end{align*}
where operators $\mathcal{K}$ and $\tilde{\mathcal{K}}$ differ by substitution $\gamma\to \tilde{\gamma}$ and $(g,\bar{g})\to (\tilde{g},\bar{\tilde{g}})$.

It turns out that in two points $x_+ = s$ and $x_- = 1-s$ the $Q$-operator degenerates.
The reason for this is the simplification of $\mathcal{R}$-operators
\begin{align}
	\mathcal{R}^{(1)}_{k0}(1-s) = \II \,, \qquad
	\mathcal{R}^{(2)}_{k0}(s) = \II \,.
\end{align}	
In accordance with this fact, we define the degenerate one-parameter $Q$-operators in the following way
\begin{align}
	\hat{Q}_+(u):=\hat{\mathbb{Q}}(u;1-s) \,, \quad \hat{Q}_-(u):=\hat{\mathbb{Q}}(s;u) \,,
\end{align}	
and similarly in the one-boundary case.
Namely after cancelation of $\mathbb{P}$-operators we have in the one-boudary case
\begin{align}
	\nonumber
	Q_+(u) & = \tr_{0}\Bigl[\mathcal{R}^{(2)}_{0n}(u)\ldots\mathcal{R}^{(2)}_{21}(u)\,\mathcal{K}_1(s,u)\,\mathcal{R}^{(2)}_{12}(u)\ldots\mathcal{R}^{(2)}_{n0}(u)\Bigr] \,, \\
	\label{Q+}
	Q_-(u) & = \tr_{0}\Bigl[\mathcal{R}^{(1)}_{0n}(u)\ldots\mathcal{R}^{(1)}_{21}(u)\,\mathcal{K}_1(1-u,s)\,\mathcal{R}^{(1)}_{12}(u)\ldots\mathcal{R}^{(1)}_{n0}(u)\Bigr]
\end{align}
and in the two-boundary case
\begin{align}
	& \nonumber
	\hat{Q}_+(u)= \tr_{0}\Bigl[\tilde{\mathcal{K}}_0(u+1, s-1)\,\mathcal{R}^{(2)}_{0n}(u)\ldots\mathcal{R}^{(2)}_{21}(u)\,\mathcal{K}_1(s,u)\,\mathcal{R}^{(2)}_{12}(u)\ldots\mathcal{R}^{(2)}_{n0}(u)\Bigr] \,, \\
	& \nonumber
	\hat{Q}_-(u)= \tr_{0}\Bigl[\tilde{\mathcal{K}}_0(s+1, -u)\,\mathcal{R}^{(1)}_{0n}(u)\ldots\mathcal{R}^{(1)}_{21}(u)\,\mathcal{K}_1(1-u,s)\,\mathcal{R}^{(1)}_{12}(u)\ldots\mathcal{R}^{(1)}_{n0}(u)\Bigr] \,,
\end{align}
where by $\mathcal{K}_j$ we denote the operator $\mathcal{K}$ acting nontrivially in the $j$-th copy of $\mathrm{L}^2(\mathbb{C})$.
To obtain the last expressions we used the following properties of the permutation operator which are direct consequences of its definition
\begin{align}
	& \nonumber
	\mathbb{P}_{k0}\mathcal{R}^{(i)}_{k0}(x) = \mathcal{R}^{(i)}_{0k}(x)\,\mathbb{P}_{k0} \,, \;
	\mathbb{P}_{k0}\mathcal{R}^{(i)}_{0j}(x) = \mathcal{R}^{(i)}_{kj}(x)\,\mathbb{P}_{k0} \,, \;
	\mathcal{R}^{(i)}_{j0}(x)\mathbb{P}_{k0} = \mathbb{P}_{k0} \, \mathcal{R}^{(i)}_{jk}(x), \quad
	i=1,2 \,, \\
	& \label{Pprop}
	\mathbb{P}_{10}\,\mathcal{K}_0(x,y) = \mathcal{K}_1(x,y)\,\mathbb{P}_{10}, \quad \mathbb{P}_{k0}^2 = \mathds{1} \,.
\end{align}

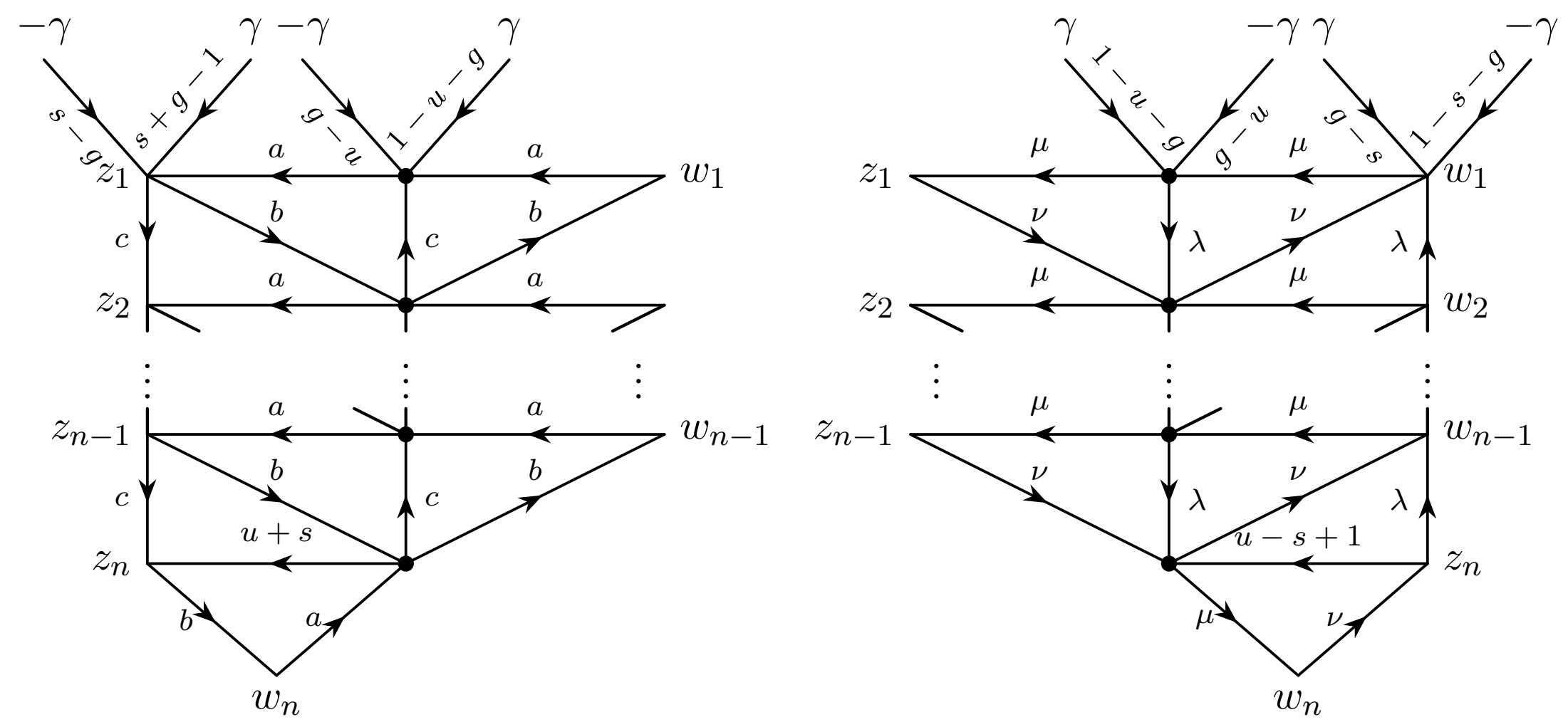
\begin{figure}
		\begin{tikzpicture}[thick, line cap = round, scale=1.2, y =-1cm]
            \def\r{0.05}
            \def\cut{0.2}
            \def\tr{0.866}
            \node [left] at (0, 0) {$z_1$};
            \node [right] at (4, 0) {$w_1$};
            \node [left] at (0, 1) {$z_2$};
            \node [left] at (0, 2) {$z_{n-1}$};
            \node [right] at (4, 2) {$w_{n-1}$};
            \node [left] at (0, 3) {$z_{n}$};
            \foreach \k in {0,2}
            {
                \draw[-<-] (0,\k) to node[midway, above = 1, font = \scriptsize]{$a$} (2,\k);
                \draw[-<-] (2,\k) to node[midway, above = 1, font = \scriptsize]{$a$} (4,\k);
                \draw[-<-] (0,\k+1) to node[midway, left = 1, font = \scriptsize]{$c$} (0,\k);
                \draw[->-] (0,\k) to node[midway, above = 1, font = \scriptsize]{$b$}  (2,\k+1);
                \draw[->-] (2,\k+1) to node[midway, right = 1, font = \scriptsize]{$c$} (2,\k);
                \draw[->-] (2,\k+1) to node[midway, above = 1, font = \scriptsize]{$b$} (4,\k);
                \draw [fill = black] (2,\k) circle (\r);
                \draw [fill = black] (2,\k+1) circle (\r);
            }

            \draw[-] (0,1) to (0, 1+\cut);
            \draw[-] (2,1) to (2, 1+\cut);
            \draw[-] (0,1) to (2*\cut, 1+\cut);
            \draw[-] (0,2) to (0, 2-\cut);
            \draw[-] (2,2) to (2, 2-\cut);
            \draw[-] (2,2) to (2-2*\cut, 2-\cut);
            \draw[-] (4,1) to (4-2*\cut, 1+\cut);
            \draw[-<-] (2,1) to node[midway, above = 1, font = \scriptsize]{$a$} (4,1);
            \draw[-<-] (0,1) to node[midway, above = 1, font = \scriptsize]{$a$} (2,1);
            \draw[-<-] (0,3) to node[midway, above = 1, font = \scriptsize]{$u+s$} (2,3);
            \draw[->-] (-0.8, -0.9) to node[midway, sloped, below, font = \scriptsize]{$s-g$} (0,0);
            \node [above] at (-0.8, -0.9) {$-\gamma$};
            \draw[->-] (0.8, -0.9) to node[midway, sloped, above, font = \scriptsize]{$s+g-1$} (0,0);
            \node [above] at (0.8, -0.9) {$\gamma$};
            \draw[->-] (1.2, -0.9) to node[midway, sloped, below, font = \scriptsize]{$g-u$}(2,0);
            \node [above] at (1.2, -0.9) {$-\gamma$};
            \draw[->-] (2.8, -0.9) to node[midway, sloped, above, font = \scriptsize]{$1-u-g$}(2,0);
            \node [above] at (2.8, -0.9) {$\gamma$};
            \draw[->-] (0, 3) to node[midway, left = 1, font = \scriptsize]{$b$}(1,3+\tr);
            \draw[-<-] (2, 3) to node[midway, left = 1, font = \scriptsize]{$a$}(1,3+\tr);
            \foreach \k in {0, 2 , 3.8}
            {
                \node at (\k, 1.5) {$\vdots$};
            }
            \node [below] at (1,3+\tr) {$w_n$}; 
        \end{tikzpicture} 
		\begin{tikzpicture}[thick, line cap = round, scale=1.2, y =-1cm, x = -1cm]
            \def\r{0.05}
            \def\cut{0.2}
            \def\tr{0.866}
            \node [right] at (0, 0) {$w_1$};
            \node [left] at (4, 0) {$z_1$};
            \node [right] at (0, 1) {$w_2$};
			\node [left] at (4, 1) {$z_2$};
            \node [right] at (0, 2) {$w_{n-1}$};
            \node [left] at (4, 2) {$z_{n-1}$};
            \node [right] at (0, 3) {$z_{n}$};
            \foreach \k in {0,2}
            {
                \draw[->-] (0,\k) to node[midway, above = 1, font = \scriptsize]{$\mu$} (2,\k);
                \draw[->-] (2,\k) to node[midway, above = 1, font = \scriptsize]{$\mu$} (4,\k);
                \draw[->-] (0,\k+1) to node[midway, left = 1, font = \scriptsize]{$\lambda$} (0,\k);
                \draw[-<-] (0,\k) to node[midway, above = 1, font = \scriptsize]{$\nu$}  (2,\k+1);
                \draw[-<-] (2,\k+1) to node[midway, right = 1, font = \scriptsize]{$\lambda$} (2,\k);
                \draw[-<-] (2,\k+1) to node[midway, above = 1, font = \scriptsize]{$\nu$} (4,\k);
                \draw [fill = black] (2,\k) circle (\r);
                \draw [fill = black] (2,\k+1) circle (\r);
            }

            \draw[-] (0,1) to (0, 1+\cut);
            \draw[-] (2,1) to (2, 1+\cut);
            \draw[-] (0,1) to (2*\cut, 1+\cut);
            \draw[-] (0,2) to (0, 2-\cut);
            \draw[-] (2,2) to (2, 2-\cut);
            \draw[-] (2,2) to (2-2*\cut, 2-\cut);
            \draw[-] (4,1) to (4-2*\cut, 1+\cut);
            \draw[->-] (2,1) to node[midway, above = 1, font = \scriptsize]{$\mu$} (4,1);
            \draw[->-] (0,1) to node[midway, above = 1, font = \scriptsize]{$\mu$} (2,1);
            \draw[->-] (0,3) to node[midway, above = 0.05, font = \scriptsize]{$u-s+1$} (2,3);
            \draw[->-] (-0.8, -0.9) to node[midway, sloped, above, font = \scriptsize]{$1-s-g$} (0,0);
            \node [above] at (-0.8, -0.9) {$-\gamma$};
            \draw[->-] (0.8, -0.9) to node[midway, sloped, below, font = \scriptsize]{$g-s$} (0,0);
            \node [above] at (0.8, -0.9) {$\gamma$};
            \draw[->-] (1.2, -0.9) to node[midway, sloped, below, font = \scriptsize]{$g-u$}(2,0);
            \node [above] at (1.2, -0.9) {$-\gamma$};
            \draw[->-] (2.8, -0.9) to node[midway, sloped, above, font = \scriptsize]{$1-u-g$}(2,0);
            \node [above] at (2.8, -0.9) {$\gamma$};
            \draw[-<-] (0, 3) to node[midway, left = 1, font = \scriptsize]{$\nu$}(1,3+\tr);
            \draw[->-] (2, 3) to node[midway, left = 1, font = \scriptsize]{$\mu$}(1,3+\tr);
            \foreach \k in {0, 2 , 3.8}
            {
                \node at (\k, 1.5) {$\vdots$};
            }
            \node [below] at (1,3+\tr) {$w_n$}; 
        \end{tikzpicture}
		\caption{
			(1) The diagrammatic representation for
			$Q_+(u)$.
			(2) The diagrammatic representation for the 
			$Q_-(u)$.
			The arrow with index $\alpha$ from $z$ to $w$ stands for $[z-w]^{-\alpha}$.
			The indices are given by the following expressions: $a=1-u-s$, $b=1-u+s$, $c=2s-1$, $\mu=u+s$, $\nu=s-u$, $\lambda=1-2s$.}
		\label{q+-1}
	\end{figure}

\begin{figure}[H]
		 \begin{tikzpicture}[thick, line cap = round, scale=1.2, y =-1cm]
            \def\r{0.05}
            \def\cut{0.2}
            \def\tr{0.866}
            \node [left] at (0, 0) {$z_1$};
            \node [right] at (4, 0) {$w_1$};
            \node [left] at (0, 1) {$z_2$};
            \node [left] at (0, 2) {$z_{n}$};
            \node [right] at (4, 2) {$w_{n}$};
            \foreach \k in {0,2}
            {
                \draw[-<-] (0,\k) to node[midway, above = 1, font = \scriptsize]{$a$} (2,\k);
                \draw[-<-] (2,\k) to node[midway, above = 1, font = \scriptsize]{$a$} (4,\k);
                \draw[-<-] (0,\k+1) to node[midway, left = 1, font = \scriptsize]{$c$} (0,\k);
                \draw[->-] (0,\k) to node[midway, above = 1, font = \scriptsize]{$b$}  (2,\k+1);
                \draw[->-] (2,\k+1) to node[midway, right = 1, font = \scriptsize]{$c$} (2,\k);
                \draw[->-] (2,\k+1) to node[midway, above = 1, font = \scriptsize]{$b$} (4,\k);
                \draw [fill = black] (2,\k) circle (\r);
                \draw [fill = black] (2,\k+1) circle (\r);
            }
			\draw [fill = black] (0,3) circle (\r);

            \draw[-] (0,1) to (0, 1+\cut);
            \draw[-] (2,1) to (2, 1+\cut);
            \draw[-] (0,1) to (2*\cut, 1+\cut);
            \draw[-] (0,2) to (0, 2-\cut);
            \draw[-] (2,2) to (2, 2-\cut);
            \draw[-] (2,2) to (2-2*\cut, 2-\cut);
            \draw[-] (4,1) to (4-2*\cut, 1+\cut);
            \draw[-<-] (2,1) to node[midway, above = 1, font = \scriptsize]{$a$} (4,1);
            \draw[-<-] (0,1) to node[midway, above = 1, font = \scriptsize]{$a$} (2,1);

            \draw[->-] (-0.8, -0.9) to node[midway, sloped, below, font = \scriptsize]{$s-g$} (0,0);
            \node [above] at (-0.8, -0.9) {$-\gamma$};
            \draw[->-] (0.8, -0.9) to node[midway, sloped, above, font = \scriptsize]{$s+g-1$} (0,0);
            \node [above] at (0.8, -0.9) {$\gamma$};
            \draw[->-] (1.2, -0.9) to node[midway, sloped, below, font = \scriptsize]{$g-u$}(2,0);
            \node [above] at (1.2, -0.9) {$-\gamma$};
            \draw[->-] (2.8, -0.9) to node[midway, sloped, above, font = \scriptsize]{$1-u-g$}(2,0);
            \node [above] at (2.8, -0.9) {$\gamma$};
			
			\draw[->-] (-0.8, 0.9+3) to node[midway, sloped, below, font = \scriptsize]{$\tilde{g}-s+1$} (0,3);
            \node [below] at (-0.8, 3+0.9) {$-\tilde{\gamma}$};
            \draw[->-] (0.8, 0.9+3) to node[midway, sloped, above, font = \scriptsize]{$2-s-\tilde{g}$} (0,3);
            \node [below] at (0.8, 0.9+3) {$\tilde{\gamma}$};
            \draw[->-] (1.2, 0.9+3) to node[midway, sloped, below, font = \scriptsize]{$1-\tilde{g}+u$}(2,3);
            \node [below] at (1.2, 0.9+3) {$-\tilde{\gamma}$};
            \draw[->-] (2.8, 0.9+3) to node[midway, sloped, above, font = \scriptsize]{$u+\tilde{g}$}(2,3);
            \node [below] at (2.8, 0.9+3) {$\tilde{\gamma}$};
			
            \foreach \k in {0, 2 , 3.8}
            {
                \node at (\k, 1.5) {$\vdots$};
            }
			\node at (2,4.5) {(1)};
        \end{tikzpicture}
		\begin{tikzpicture}[thick, line cap = round, scale=1.2, y =-1cm]
            \def\r{0.05}
            \def\cut{0.2}
            \def\tr{0.866}
            \node [left] at (0, 0) {$z_1$};
            \node [right] at (4, 0) {$w_1$};
            \node [left] at (0, 1) {$z_2$};
            \node [left] at (0, 2) {$z_{n}$};
            \node [right] at (4, 2) {$w_{n}$};
            \foreach \k in {0,2}
            {
                \draw[-<-] (2,\k) to node[midway, above = 1, font = \scriptsize]{$\mu$} (4,\k);
                \draw[-<-] (0,\k) to node[midway, above = 1, font = \scriptsize]{$\mu$} (2,\k);
                \draw[-<-] (2,\k+1) to node[midway, left = 1, font = \scriptsize]{$\lambda$} (2,\k);
                \draw[->-] (2,\k+1) to node[midway, above = 1, font = \scriptsize]{$\nu$}  (4,\k);
                \draw[->-] (4,\k+1) to node[midway, right = 1, font = \scriptsize]{$\lambda$} (4,\k);
                \draw[->-] (0,\k) to node[midway, above = 1, font = \scriptsize]{$\nu$} (2,\k+1);
                \draw [fill = black] (2,\k) circle (\r);
                \draw [fill = black] (2,\k+1) circle (\r);
            }

            \draw[-] (4,1) to (4, 1+\cut);
            \draw[-] (2,1) to (2, 1+\cut);
            \draw[-] (0,1) to (2*\cut, 1+\cut);
            \draw[-] (4,2) to (4, 2-\cut);
            \draw[-] (2,2) to (2, 2-\cut);
            \draw[-] (2,2) to (2+2*\cut, 2-\cut);
            \draw[-] (4,1) to (4-2*\cut, 1+\cut);
            \draw[-<-] (2,1) to node[midway, above = 1, font = \scriptsize]{$\mu$} (4,1);
            \draw[-<-] (0,1) to node[midway, above = 1, font = \scriptsize]{$\mu$} (2,1);

            \draw[->-] (4-0.8, -0.9) to node[midway, sloped, below, font = \scriptsize]{$g-s$} (4,0);
            \node [above] at (4-0.8, -0.9) {$-\gamma$};
            \draw[->-] (4+0.8, -0.9) to node[midway, sloped, above, font = \scriptsize]{$1-g-s$} (4,0);
            \node [above] at (4+0.8, -0.9) {$\gamma$};
            \draw[->-] (1.2, -0.9) to node[midway, sloped, above, font = \scriptsize]{$1-u-g$}(2,0);
            \node [above] at (1.2, -0.9) {$-\gamma$};
            \draw[->-] (2.8, -0.9) to node[midway, sloped, below, font = \scriptsize]{$g-u$}(2,0);
            \node [above] at (2.8, -0.9) {$\gamma$};
			
			\draw[->-] (4-0.8, 0.9+3) to node[midway, sloped, below, font = \scriptsize]{$s+1-\tilde{g}$} (4,3);
            \node [below] at (4-0.8, 3+0.9) {$-\tilde{\gamma}$};
            \draw[->-] (4+0.8, 0.9+3) to node[midway, sloped, above, font = \scriptsize]{$s+\tilde{g}$} (4,3);
            \node [below] at (4+0.8, 0.9+3) {$\tilde{\gamma}$};
            \draw[->-] (1.2, 0.9+3) to node[midway, sloped, below, font = \scriptsize]{$\tilde{g}+u$}(2,3);
            \node [below] at (1.2, 0.9+3) {$-\tilde{\gamma}$};
            \draw[->-] (2.8, 0.9+3) to node[midway, sloped, above, font = \scriptsize]{$1-u+\tilde{g}$}(2,3);
            \node [below] at (2.8, 0.9+3) {$\tilde{\gamma}$};
			\draw [fill=black] (4,3) circle (\r);
            \foreach \k in {0, 2 , 3.8}
            {
                \node at (\k, 1.5) {$\vdots$};
            }
			\node at (2,4.5) {(2)};
        \end{tikzpicture}
		\caption{
			(1) The diagrammatic representation for
			$\hat{Q}_+(u)$.
			(2) The diagrammatic representation for the 
			$\hat{Q}_-(u)$.}
		\label{q+-}	
	\end{figure}
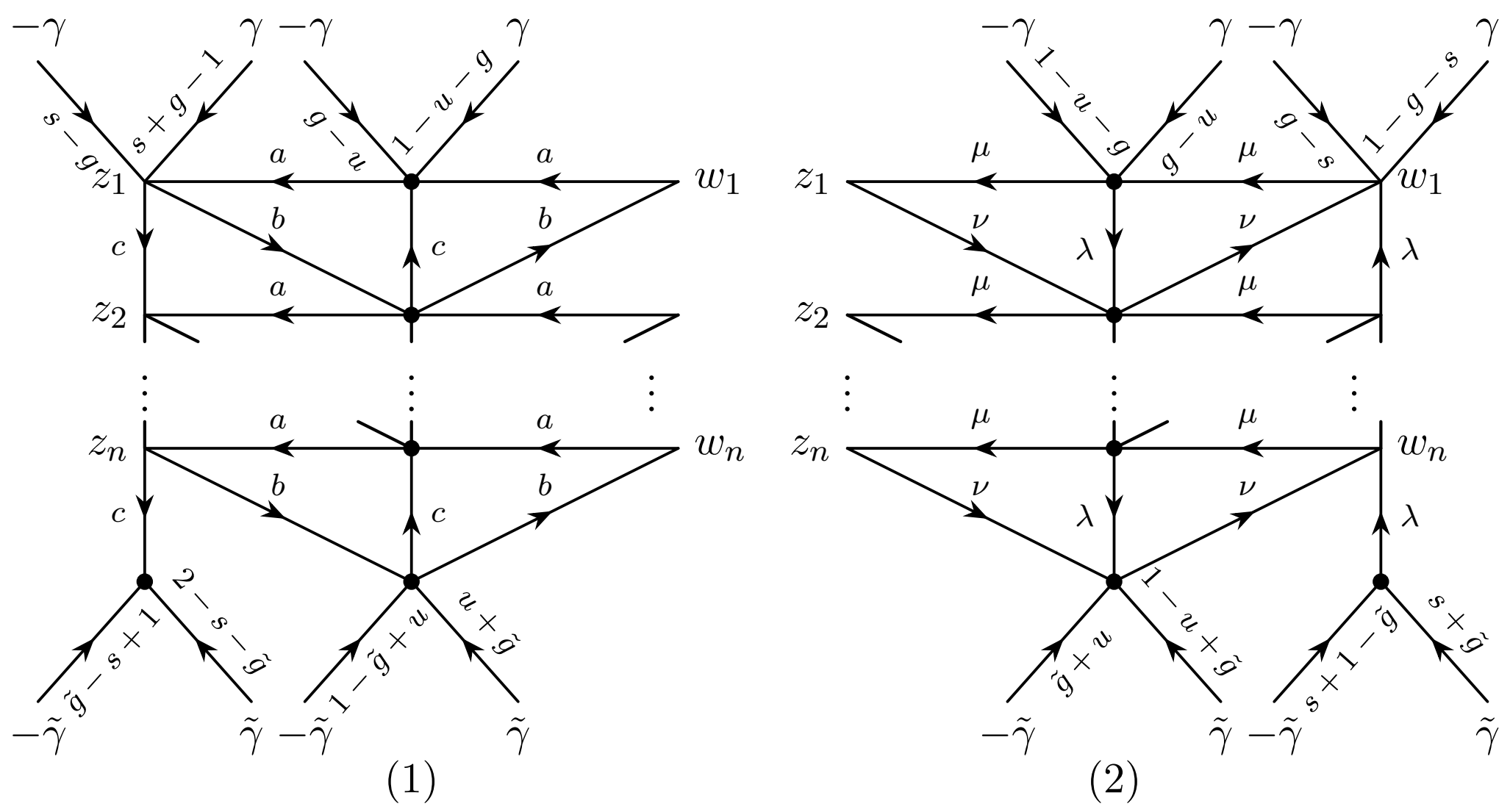

As well as the general two-parameter $Q$-operator, degenerate $Q$-operators are integral operators.
For some purposes it is convenient to represent their integral kernels with diagrams.
Using integral formulas for $\mathcal{K}$- and $\mathcal{R}$-operators~\eqref{R_kernel} and the diagram technique~\cite{DKM,DM14,DM:Gust3} described in Appendix~\ref{App-DiagTech} we obtain the diagrams shown in Figures~\ref{q+-1},~\ref{q+-}.

In the one-boundary case the trace in expression for operator $Q_+$ can be calculated explicitly and we get the following result
\begin{multline} \label{Q+expl}
	Q_+(u) \\
	= -\frac{[\imath]^{-3s-u}\,\mathbf{\Gamma}(2-2s,u+s)}{(u-s+1)(\bar{u}-\bar{s}+1)} \,
	\mathcal{R}^{(2)}_{n\,n-1}(u)\ldots\mathcal{R}^{(2)}_{21}(u)\,\mathcal{K}_1(s,u)\,\mathcal{R}^{(2)}_{12}(u)\ldots\mathcal{R}^{(2)}_{n0'}(u)\Bigr|_{z_{0'}\to z_n} .
\end{multline}
In this formula we use the auxiliary variable~$z_{0'}$. By definition~\eqref{R12def} we have
$$
	\mathcal{R}^{(2)}_{n0'}(u) = [\hat{p}_n]^{u+s-1} [z_{n0'}]^{u-s} [\hat{p}_n]^{1-2s} \,,
$$
so there are no other operators acting in the $0'$-th space, except the operator of multiplication by function of $z_{0'}$.
Therefore, the variable $z_{0'}$ can be treated as an additional parameter of the operator in the right hand side of~\eqref{Q+expl},
which is set equal to $z_n$ at the last step.
That is, at first the product of $\mathcal{R}$- and $\mathcal{K}$-operators acts on function of variables $z_1,\dots, z_n$, and then one substitutes $z_{0'}\to z_n$ in the obtained expression.

To derive~\eqref{Q+expl} we observe that the operator $\mathcal{R}^{(2)}_{0n}$ in~\eqref{Q+} acts in the $n$-th space as the operator of multiplication by function of $z_n$, and one can rewrite it in the following way
\begin{multline} \nonumber
	\mathcal{R}^{(2)}_{0n}(u) = [\hat{p}_0]^{u+s-1}[z_{0n}]^{u-s}[\hat{p}_0]^{1-2s} \\
	= \int \mathrm{d}^2z_{0'} \, \delta^2(z_{0'}-z_n) \,
	[\hat{p}_0]^{u+s-1}[z_{00'}]^{u-s}[\hat{p}_0]^{1-2s}
	= \int \mathrm{d}^2z_{0'} \, \delta^2(z_{0'}-z_n) \, \mathcal{R}^{(2)}_{00'}(u) \,,
\end{multline}
where $\delta^2(z)$ is the two-dimensional delta function
\begin{equation} \label{delta2D}
	\delta^2(z) = \delta(\Re z) \, \delta(\Im z) \,.
\end{equation}
Thus applying the cyclic property of trace we arrive at
\begin{multline} \nonumber
		Q_+(u) =
		\int \mathrm{d}^2z_{0'} \, \delta^2(z_{0'}-z_n) \,
		\mathcal{R}^{(2)}_{n\,n-1}(u)\ldots\mathcal{R}^{(2)}_{21}(u)\,\mathcal{K}_1(s,u) \\
		\times
		\mathcal{R}^{(2)}_{12}(u)\ldots\,\mathcal{R}^{(2)}_{n-1\,n}(u) \,
		\tr_{0}\Bigl[\mathcal{R}^{(2)}_{n0}(u)\mathcal{R}^{(2)}_{00'}(u)\Bigr]\,.
\end{multline}
And to obtain the final formula it remains to use the relation~\eqref{trRR2app}
\begin{equation} \nonumber
	\tr_{0}\Bigl[\mathcal{R}^{(2)}_{n0}(u)\mathcal{R}^{(2)}_{00'}(u)\Bigr]
	= -\frac{[\imath]^{-3s-u}\,\mathbf{\Gamma}(2-2s,u+s)}{(u-s+1)(\bar{u}-\bar{s}+1)}
	\mathcal{R}^{(2)}_{n0'}(u) \,.
\end{equation}

\section{Factorization of the transfer matrix} \label{sect:Qfact}
In the last section we defined the Baxter $\mathbb{Q}(u_+;u_-)$-operator as the general transfer matrix $\mathbbm{t}(u,s_0)$ with new parametrization $u_+ = u+s_0$, $u_- = u-s_0+1$.
In addition, we demonstrated that at special values of parameters $u_+$, $u_-$ the general two-parameter $Q$-operator degenerates into one-parameter operators $Q_+$, $Q_-$.
In this section we prove the following factorization formulas for one-boundary and two-boundary spin chain
\begin{align}
	& \label{Qfact1}
	\mathbb{Q}(u_+;u_-) = \frac{[u_+-s+1]\,[u_-+s]}{[1-2s]\,[u_++u_-]}\,Q_-(u_-)\,Q_+(u_+) \,, \\
	& \label{Qfact2}
	\hat{\mathbb{Q}}(u_+;u_-) = \frac{[2\tilde{\gamma}]^2\,[u_+-s+1]\,[u_-+s]\,[1-s-\tilde{g}]\,[\tilde{g}-s]}{[1-2s]\,[u_++u_-]}\,\hat{Q}_-(u_-)\,\hat{Q}_+(u_+) \,,
\end{align}
correspondingly. Here by $[x]$ we denote the product $x\bar{x}$. 
We give two proofs: using an algebraic approach of~\cite{ADV1, ADV2} and using the equivalent diagram technique formulated in Appendix~\ref{App-DiagTech}.
The advantage of the latter is that it makes the calculations visual and reformulates them in the language of simple Feynman rules.
On the other hand, an advantage of the former is the absence of numerical coefficients in majority of basic operator relations used (like the operator form of star-triangle relation~\eqref{star-tr}), so one can clearly see the origin of coefficients in~\eqref{Qfact1}, \eqref{Qfact2}.
This is in contrast to the diagrammatic method where coefficients appear after application of every graphical transformation.

\subsection{Algebraic derivation} \label{sect:algDer}

Let us begin from the algebraic method. First we obtain the formula for one-boundary case~\eqref{Qfact1}. Recall that
\begin{equation} \label{Qdef1}
	\mathbb{Q}(u_+;u_-) = \tr_0[\mathbb{R}_{n0}(u_+;u_-)\dots\mathbb{R}_{10}(u_+;u_-)\,\overset{0}{\mathbb{K}}_-(u_+;u_-)\,\mathbb{R}_{10}(u_+;u_-)\dots\mathbb{R}_{n0}(u_+;u_-)] \,,
\end{equation}
where $\mathbb{R}_{k0}(u_+;u_-) = \mathbb{P}_{k0}\,\mathcal{R}^{(1)}_{k0}(u_-)\, \mathcal{R}^{(2)}_{k0}(u_+)$, and
\begin{equation} \label{Qmp}
	\begin{aligned}
		&
		Q_-(u_-)= \tr_{0}\Bigl[\mathcal{R}^{(1)}_{0n}(u_-)\ldots\mathcal{R}^{(1)}_{21}(u_-)\,\mathcal{K}_1(1-u_-,s)\,\mathcal{R}^{(1)}_{12}(u_-)\ldots\mathcal{R}^{(1)}_{n0}(u_-)\Bigr] \,, \\
		&
		Q_+(u_+)= \tr_{0}\Bigl[\mathcal{R}^{(2)}_{0n}(u_+)\ldots\mathcal{R}^{(2)}_{21}(u_+)\,\mathcal{K}_1(s,u_+)\,\mathcal{R}^{(2)}_{12}(u_+)\ldots\mathcal{R}^{(2)}_{n0}(u_+)\Bigr] \,.
	\end{aligned}
\end{equation}
Operators $\mathbb{K}_-$, $\mathcal{K}$, $\mathcal{R}^{(1)}$ and $\mathcal{R}^{(2)}$ are defined in~\eqref{K-x+-}, \eqref{Kop-d}, \eqref{R12def}.

The derivation divides into two main steps.
At the first step, after some simple manipulations with $\mathcal{R}$- and $\mathbb{K}$-operators, which we will show a bit further, we separate the expression in~\eqref{Qdef1} into two blocks: one with operators $\mathcal{R}^{(1)}$, and the other one with $\mathcal{R}^{(2)}$
\begin{align}
	\nonumber
	\mathbb{Q}(u_+;u_-) =  \tr_0 &\Bigl[\mathcal{R}^{(1)}_{0n}(u_-)\ldots\mathcal{R}^{(1)}_{21}(u_-)\,\mathcal{K}_1(1-u_-,s)\,\mathcal{R}^{(1)}_{12}(u_-)\ldots\mathcal{R}^{(1)}_{n0}(u_-) \\
	& \label{Q1}
	\times\mathcal{R}^{(2)}_{0n}(u_+)\ldots\mathcal{R}^{(2)}_{21}(u_+)\,\mathcal{K}_1(s,u_+)\,\mathcal{R}^{(2)}_{12}(u_+)\ldots\mathcal{R}^{(2)}_{n0}(u_+)\Bigr] \,.
\end{align}
These blocks coincide with expressions in formulas for $Q_-$ and $Q_+$~\eqref{Qmp}, so we almost get the needed result~\eqref{Qfact1}.

Therefore, at the second step to transform one trace into the product of two traces we make use of the relations~\eqref{trRR1app}, \eqref{trRR2app}, \eqref{trRRRRapp}
\begin{align}
	& \label{trRR1}
	\tr_0\Bigl[\mathcal{R}^{(1)}_{10}(x)\,\mathcal{R}^{(1)}_{02}(y)\Bigr] = -\frac{[\imath]^{3s-x}\,\mathbf{\Gamma}(2s, x-s+1)}{(x+s)(\bar{x}+\bar{s})} \, \mathcal{R}^{(1)}_{12}(y) \,, \\
	& \label{trRR2}
	\tr_0\Bigl[\mathcal{R}^{(2)}_{10}(x)\,\mathcal{R}^{(2)}_{02}(y)\Bigr] = -\frac{[\imath]^{-3s-y}\,\mathbf{\Gamma}(2-2s, y+s)}{(y-s+1)(\bar{y}-\bar{s}+1)} \, \mathcal{R}^{(2)}_{12}(x) \,, \\
	& \nonumber
	\tr_0\Bigl[\mathcal{R}^{(1)}_{10}(x)\,\mathcal{R}^{(2)}_{01}(y)\,\mathcal{R}^{(1)}_{00'}(\alpha)\,\mathcal{R}^{(2)}_{0''0}(\beta)\Bigr] \\
	& \label{trRRRR}
	\qquad\qquad\qquad\qquad\;\;\;
	= -\frac{[-1]^{2s} [\imath]^{-x-y}\,\mathbf{\Gamma}(x-s+1, y+s)}{(x+y)(\bar{x}+\bar{y})}
	\,\mathcal{R}^{(1)}_{10'}(\alpha)\,\mathcal{R}^{(2)}_{0''1}(\beta) \,,
\end{align}
all of them are derived in the Appendix from one basic identity
\begin{align*}
	\tr_0\Bigl([z_{10}]^\alpha[\hat{p}_0]^{\alpha+\beta}[z_{10}]^\beta
	\Phi(z_0)\Bigr) & =
	\tr_0\Bigl([\hat{p}_0]^{\beta} [z_{10}]^{\alpha+\beta} [\hat{p}_0]^{\alpha}
	\Phi(z_0)\Bigr) \\
	& = -\frac{[\imath]^{\beta-\alpha} \mathbf{\Gamma}(\alpha+1,\beta+1)}{(\alpha+\beta+1)(\bar{\alpha}+\bar{\beta}+1)} \, \Phi(z_1) \,,
\end{align*}
where $\Phi(z_0)$ denotes the operator of multiplication by arbitrary function~$\Phi$.
Thus rewriting $\mathcal{R}^{(1)}_{0n}$ and $\mathcal{R}^{(2)}_{n0}$ in~\eqref{Q1} with the help of \eqref{trRR1} and \eqref{trRR2} (note the cyclic property of trace)
\begin{align}
	& \nonumber
	\mathcal{R}^{(1)}_{0n}(u_-) = -\frac{[\imath]^{u_--3s}\,[u_-+s]}{\mathbf{\Gamma}(2s, u_--s+1)} \, 
	\tr_{0'}\Bigl[\mathcal{R}^{(1)}_{0'n}(u_-)\,\mathcal{R}^{(1)}_{00'}(u_-)\Bigr] \,, \\
	& \label{RtrRR}
	\mathcal{R}^{(2)}_{n0}(u_+) = -\frac{[\imath]^{u_++3s}\,[u_+-s+1]}{\mathbf{\Gamma}(2-2s,u_++s)} \,
	\tr_{0''}\Bigl[\mathcal{R}^{(2)}_{0''0}(u_+)\,\mathcal{R}^{(2)}_{n0''}(u_+)\Bigr] \,.
\end{align}
we arrive at
\begin{align}
	\nonumber
	\mathbb{Q}(u_+;u_-) = & \,
	\frac{[-1]^{2s}\,[\imath]^{u_+ + u_-} [u_+-s+1] [u_-+s]}{[1-2s] \mathbf{\Gamma}(u_++s, u_--s+1)} \\
	& \label{Q2}
	\begin{aligned}
		\, \times \tr_{0,0',0''} &\Bigl[\mathcal{R}^{(1)}_{0'n}(u_-)\ldots\mathcal{R}^{(1)}_{21}(u_-)\,\mathcal{K}_1(1-u_-,s)\,\mathcal{R}^{(1)}_{12}(u_-)\ldots\mathcal{R}^{(1)}_{n-1,n}(u_-) \\
		&
		\times
		\mathcal{R}^{(1)}_{00'}(u_-)\,\mathcal{R}^{(1)}_{n0}(u_-)\,\mathcal{R}^{(2)}_{0n}(u_+)\,\mathcal{R}^{(2)}_{0''0}(u_+) \\
		& 
		\times\mathcal{R}^{(2)}_{n,n-1}(u_+)\ldots\mathcal{R}^{(2)}_{21}(u_+)\,\mathcal{K}_1(s,u_+)\,\mathcal{R}^{(2)}_{12}(u_+)\ldots\mathcal{R}^{(2)}_{n0''}(u_+)\Bigr] \,,
	\end{aligned}
\end{align}
where we have moved $\mathcal{R}^{(1)}_{00'}(u_-)$ rightwards and $\mathcal{R}^{(2)}_{0''0}(u_+)$ leftwards to $\mathcal{R}^{(1)}_{n0}(u_-)\,\mathcal{R}^{(2)}_{0n}(u_+)$ using commutativity.
This way, we have gathered all four operators which act nontrivially on functions of variable $z_0$, and it only remains to take the trace of the product of these operators over the $0$-th space by means of relation~\eqref{trRRRR}
\begin{multline*}
	\tr_0\Bigl[\mathcal{R}^{(1)}_{00'}(u_-)\,\mathcal{R}^{(1)}_{n0}(u_-)\,\mathcal{R}^{(2)}_{0n}(u_+)\,\mathcal{R}^{(2)}_{0''0}(u_+)\Bigr] =
	\tr_0\Bigl[\mathcal{R}^{(1)}_{n0}(u_-)\,\mathcal{R}^{(2)}_{0n}(u_+)\,\mathcal{R}^{(1)}_{00'}(u_-)\,\mathcal{R}^{(2)}_{0''0}(u_+)\Bigr] \\
	=
	-\frac{[-1]^{2s} [\imath]^{-u_--u_+}\,\mathbf{\Gamma}(u_--s+1, u_++s)}{[u_-+u_+]}
	\,\mathcal{R}^{(1)}_{n0'}(u_-)\,\mathcal{R}^{(2)}_{0''n}(u_+) \,,
\end{multline*}
where we have also applied the cyclicity of trace and commutativity of $\mathcal{R}^{(1)}_{00'}(u_-)$ and $\mathcal{R}^{(2)}_{0''0}(u_+)$.
Substituting the last result into~\eqref{Q2} one obtains the expression in which the traces over spaces $0'$ and $0''$ can be separated
\begin{align}
	\nonumber
	\mathbb{Q}(u_+;u_-)
	= & \,
	\frac{[u_+-s+1] \, [u_-+s]}{[1-2s] \, [u_-+u_+]} \\
	& \nonumber
	\begin{aligned}
		& \times \tr_{0'} \Bigl[\mathcal{R}^{(1)}_{0'n}(u_-)\ldots\mathcal{R}^{(1)}_{21}(u_-)\,\mathcal{K}_1(1-u_-,s)\,\mathcal{R}^{(1)}_{12}(u_-)\ldots\mathcal{R}^{(1)}_{n-1,n}(u_-)\,\mathcal{R}^{(1)}_{n0'}(u_-)\Bigr] \\
		& \times
		\tr_{0''}
		\Bigl[\mathcal{R}^{(2)}_{0''n}(u_+)\mathcal{R}^{(2)}_{n,n-1}(u_+)\ldots\mathcal{R}^{(2)}_{21}(u_+)\,\mathcal{K}_1(s,u_+)\,\mathcal{R}^{(2)}_{12}(u_+)\ldots\mathcal{R}^{(2)}_{n0''}(u_+)\Bigr]
	\end{aligned}
	\\
	= & \,
	\frac{[u_+-s+1] \, [u_-+s]}{[1-2s] \, [u_-+u_+]} \, Q_-(u_-) \, Q_+(u_+) \,,
\end{align}
and thus we obtain the desired factorization formula~\eqref{Qfact1}.

Now let us derive the intermediate expression for $Q$-operator~\eqref{Q1}.
Using formulas~\eqref{Pprop}
it is not hard to rewrite~\eqref{Qdef1} as
\begin{multline} \nonumber
	\mathbb{Q}(u_+;u_-)
	= \tr_0\Bigl[
	\mathcal{R}^{(1)}_{0n}(u_-)\,\mathcal{R}^{(2)}_{0n}(u_+)\,
	\mathcal{R}^{(1)}_{n,n-1}(u_-)\,\mathcal{R}^{(2)}_{n,n-1}(u_+) \dots \mathcal{R}^{(1)}_{21}(u_-)\,\mathcal{R}^{(2)}_{21}(u_+) \\
	\times
	\overset{1}{\mathbb{K}}_-(u_+;u_-)\,\mathcal{R}^{(1)}_{12}(u_-)\,\mathcal{R}^{(2)}_{12}(u_+)\dots
	\mathcal{R}^{(1)}_{n-1,n}(u_-)\,\mathcal{R}^{(2)}_{n-1,n}(u_+)
	\mathcal{R}^{(1)}_{n0}(u_-)\,\mathcal{R}^{(2)}_{n0}(u_+)\Bigr] \,.
\end{multline}
Then, taking into account the commutativity of operators $\mathcal{R}^{(1)}_{ij}(u_-)$ and $\mathcal{R}^{(2)}_{ki}(u_+)$ for $j \neq k$ (since they act in the space corresponding to variable $z_i$ as operators of multiplication by function of $z_i$), one can rearrange the $\mathcal{R}$-operators and gather them into products of $\mathcal{R}^{(1)}$ and~$\mathcal{R}^{(2)}$
\begin{multline}
	\nonumber
	\mathbb{Q}(u_+;u_-)
	= \tr_0\Bigl[ 
	\mathcal{R}^{(1)}_{0n}(u_-)\mathcal{R}^{(1)}_{n,n-1}(u_-)\dots\mathcal{R}^{(1)}_{21}(u_-)\,\mathcal{R}^{(2)}_{0n}(u_+)\mathcal{R}^{(2)}_{n,n-1}(u_+)\dots\mathcal{R}^{(2)}_{21}(u_+)  \\
	\times
	\overset{1}{\mathbb{K}}_-(u_+;u_-) \,
	\mathcal{R}^{(1)}_{12}(u_-)\dots
	\mathcal{R}^{(1)}_{n-1,n}(u_-)\,
	\mathcal{R}^{(1)}_{n0}(u_-)\,\mathcal{R}^{(2)}_{12}(u_+)\dots\mathcal{R}^{(2)}_{n-1,n}(u_+)\,\mathcal{R}^{(2)}_{n0}(u_+)\Bigr] \,.
\end{multline}
Next we recall~\eqref{K-x+-} ($z_{k\pm}:=z_k\pm\gamma$)
\begin{equation} \nonumber
	\overset{1}{\mathbb{K}}_-(u_+;u_-) = [z_{1+}]^{u_-+g-1}\,[z_{1-}]^{u_--g}\,[\hat{p}_1]^{u_++u_--1}\,[z_{1+}]^{u_+-g}\,[z_{1-}]^{u_++g-1}
\end{equation}
and move the leftmost power factors with $z_1$ leftwards and the rightmost ones -- rightwards
\begin{multline}
	\nonumber
	\mathbb{Q}(u_+;u_-)
	= \tr_0\Bigl[ 
	\mathcal{R}^{(1)}_{0n}(u_-)\dots\mathcal{R}^{(1)}_{21}(u_-)\,[z_{1+}]^{u_-+g-1}\,[z_{1-}]^{u_--g}\,\mathcal{R}^{(2)}_{0n}(u_+)\dots\mathcal{R}^{(2)}_{21}(u_+)  \\
	\times
	[\hat{p}_1]^{u_++u_--1} \,
	\mathcal{R}^{(1)}_{12}(u_-)\dots
	\mathcal{R}^{(1)}_{n0}(u_-)\,[z_{1+}]^{u_+-g}\,[z_{1-}]^{u_++g-1}\,\mathcal{R}^{(2)}_{12}(u_+)\dots\mathcal{R}^{(2)}_{n0}(u_+)\Bigr] \,.
\end{multline}
Applying the relation~\eqref{RpR}
\begin{multline} \nonumber
	\mathcal{R}^{(2)}_{k,k-1}(x)\dots\mathcal{R}^{(2)}_{32}(x)\,\mathcal{R}^{(2)}_{21}(x)\,[\hat{p}_1]^{x+y-1}\,\mathcal{R}^{(1)}_{12}(y)\,\mathcal{R}^{(1)}_{23}(y)\dots
	\mathcal{R}^{(1)}_{k-1,k}(y) \\
	=
	[\hat{p}_1]^{s+y-1}\,\mathcal{R}^{(1)}_{12}(y)\,\mathcal{R}^{(1)}_{23}(y)\dots
	\mathcal{R}^{(1)}_{k-1,k}(y) \,
	\mathcal{R}^{(2)}_{k,k-1}(x)\dots\mathcal{R}^{(2)}_{32}(x)\,\mathcal{R}^{(2)}_{21}(x)\,[\hat{p}_1]^{x-s}
\end{multline}
proven in Appendix we finally put together all operators $\mathcal{R}^{(1)}$ and $\mathcal{R}^{(2)}$
\begin{multline} \label{Q3}
	\mathbb{Q}(u_+;u_-) \\
	= \tr_0\Bigl[ 
	\mathcal{R}^{(1)}_{0n}(u_-)\dots\mathcal{R}^{(1)}_{21}(u_-)\,[z_{1+}]^{u_-+g-1}\,[z_{1-}]^{u_--g} \,
	[\hat{p}_1]^{s+u_--1} \mathcal{R}^{(1)}_{12}(u_-)\dots \mathcal{R}^{(1)}_{n0}(u_-) \, \\
	\times
	\mathcal{R}^{(2)}_{0n}(u_+)\dots\mathcal{R}^{(2)}_{21}(u_+)\,[\hat{p}_1]^{u_+-s} \,
	[z_{1+}]^{u_+-g}\,[z_{1-}]^{u_++g-1}\,\mathcal{R}^{(2)}_{12}(u_+)\dots\mathcal{R}^{(2)}_{n0}(u_+)\Bigr] \,.
\end{multline}
To obtain the required formula~\eqref{Q1} from~\eqref{Q3} it remains to extract the $\mathcal{K}$-operators
\begin{align}
	& \nonumber
	[z_{1+}]^{u_-+g-1}\,[z_{1-}]^{u_--g} \,
	[\hat{p}_1]^{s+u_--1} = \mathcal{K}(1-u_-,s)\,[z_{1+}]^{g-s}[z_{1-}]^{1-g-s} \,, \\
	& \nonumber
	[\hat{p}_1]^{u_+-s} \,
	[z_{1+}]^{u_+-g}\,[z_{1-}]^{u_++g-1} =
	[z_{1+}]^{s-g}\,[z_{1-}]^{s+g-1}\mathcal{K}(s,u_+)
\end{align}
and cancel out mutually inverse factors $[z_{1+}]^{g-s}[z_{1-}]^{1-g-s}$ and $[z_{1+}]^{s-g}\,[z_{1-}]^{s+g-1}$ thanks to their commutativity with $\mathcal{R}^{(1)}_{12}(u_-)$ and $\mathcal{R}^{(2)}_{21}(u_+)$ (since the latter act as operators of multiplication by function in the space of functions of $z_1$).

Finally, we prove the factorization formula~\eqref{Qfact2} for two-boundary $Q$-operator
\begin{multline} \label{Qdef2}
	\hat{\mathbb{Q}}(u_+;u_-) \\
	= \tr_0[\overset{0}{\mathbb{K}}_+(u_+;u_-)\mathbb{R}_{n0}(u_+;u_-)\dots\mathbb{R}_{10}(u_+;u_-)\,\overset{0}{\mathbb{K}}_-(u_+;u_-)\,\mathbb{R}_{10}(u_+;u_-)\dots\mathbb{R}_{n0}(u_+;u_-)] \,.
\end{multline}
The operator~\eqref{Qdef2} differs from the one-boundary $Q$-operator only by the additional factor $\mathbb{K}_+(u_+;u_-)$ under the sign of trace,
so the main idea of the proof is similar to the one-boundary case: we
separate operators $\mathcal{R}^{(1)}$ from $\mathcal{R}^{(2)}$ as in~\eqref{Q1} and then use~\eqref{RtrRR} to gather all the operators acting nontrivially in the $0$-th space
\begin{align}
	\nonumber
	\hat{\mathbb{Q}}(u_+;u_-) = & \,
	\frac{[-1]^{2s}\,[\imath]^{u_+ + u_-} [u_+-s+1] [u_-+s]}{[1-2s] \mathbf{\Gamma}(u_++s, u_--s+1)} \\
	& \label{Q4}
	\begin{aligned}
		\, \times \tr_{0',0''} &\Bigl[\mathcal{R}^{(1)}_{0'n}(u_-)\ldots\mathcal{R}^{(1)}_{21}(u_-)\,\mathcal{K}_1(1-u_-,s)\,\mathcal{R}^{(1)}_{12}(u_-)\ldots\mathcal{R}^{(1)}_{n-1,n}(u_-) \\
		&
		\times
		\tr_0\Bigl( \overset{0}{\mathbb{K}}_+(u_+;u_-)\,\mathcal{R}^{(1)}_{00'}(u_-)\,\mathcal{R}^{(1)}_{n0}(u_-)\,\mathcal{R}^{(2)}_{0n}(u_+)\,\mathcal{R}^{(2)}_{0''0}(u_+) \Bigr) \\
		& 
		\times\mathcal{R}^{(2)}_{n,n-1}(u_+)\ldots\mathcal{R}^{(2)}_{21}(u_+)\,\mathcal{K}_1(s,u_+)\,\mathcal{R}^{(2)}_{12}(u_+)\ldots\mathcal{R}^{(2)}_{n0''}(u_+)\Bigr].
	\end{aligned}
\end{align}
The trace over the $0$-th space divides into the product of two traces
\begin{align}
	& \nonumber
	\tr_0\Bigl( \overset{0}{\mathbb{K}}_+(u_+;u_-)\,\mathcal{R}^{(1)}_{00'}(u_-)\,\mathcal{R}^{(1)}_{n0}(u_-)\,\mathcal{R}^{(2)}_{0n}(u_+)\,\mathcal{R}^{(2)}_{0''0}(u_+) \Bigr) \\
	& \nonumber
	= \frac{[2\tilde{\gamma}]^2\,[u_-+s]\,[u_+-s+1]\,[s-\tilde{g}]\,[s+\tilde{g}-1]}{[u_++u_-]\,[2s-1]} \\
	& \label{trKRRRR}
	\quad \times \tr_a\Bigl[\mathcal{R}^{(1)}_{na}(u_-)\,\tilde{\mathcal{K}}_a(s+1,-u_-)\,\mathcal{R}^{(1)}_{a0'}(u_-)\Bigr] \,
	\tr_b\Bigl[\mathcal{R}^{(2)}_{0''b}(u_+)\,\tilde{\mathcal{K}}_b(u_++1,s-1)\,\mathcal{R}^{(2)}_{bn}(u_+)\Bigr] \,,
\end{align}
where we introduced two additional spaces indexed by $a$ and $b$ in order to write the traces and we remind that operators $\tilde{\mathcal{K}}(x,y)$ and $\mathcal{K}(x,y)$ differ by substitution $\gamma\to\tilde{\gamma}$, $(g,\bar{g})\to(\tilde{g},\bar{\tilde{g}})$.
We prove~\eqref{trKRRRR} further in this section. After substituting it into~\eqref{Q4} we use the cyclic property of trace and~\eqref{RtrRR}
\begin{align}
	& \nonumber
	\begin{aligned}
		\tr_{0',a}\Bigl[\mathcal{R}^{(1)}_{0'n}(u_-)\ldots\mathcal{R}^{(1)}_{21}(u_-)\,\mathcal{K}_1(1-u_-,s)\,\mathcal{R}^{(1)}_{12}(u_-)\ldots\mathcal{R}^{(1)}_{n-1,n}(u_-) \\
		\times
		\mathcal{R}^{(1)}_{na}(u_-)\,\tilde{\mathcal{K}}_a(s+1,-u_-)\,\mathcal{R}^{(1)}_{a0'}(u_-)\Bigr]
	\end{aligned}
	\\
	& \nonumber
	\begin{aligned}
		=
		\tr_{0',a}\Bigl[
		\tilde{\mathcal{K}}_a(s+1,-u_-)\,\mathcal{R}^{(1)}_{a0'}(u_-)\,
		\mathcal{R}^{(1)}_{0'n}(u_-)\ldots\mathcal{R}^{(1)}_{21}(u_-)\,\mathcal{K}_1(1-u_-,s) \\
		\times
		\mathcal{R}^{(1)}_{12}(u_-)\ldots\mathcal{R}^{(1)}_{n-1,n}(u_-) \, \mathcal{R}^{(1)}_{na}(u_-)\Bigr]
	\end{aligned}
	\\
	& \nonumber
	\begin{aligned}
		= -\frac{[\imath]^{3s-u_-}\,\mathbf{\Gamma}(2s, u_--s+1)}{[u_-+s]}
		\tr_{a}\Bigl[
		\tilde{\mathcal{K}}_a(s+1,-u_-)\,\mathcal{R}^{(1)}_{an}(u_-)\,
		\mathcal{R}^{(1)}_{n,n-1}(u_-)\ldots\mathcal{R}^{(1)}_{21}(u_-) \\
		\times
		\mathcal{K}_1(1-u_-,s) \,
		\mathcal{R}^{(1)}_{12}(u_-)\ldots\mathcal{R}^{(1)}_{n-1,n}(u_-) \, \mathcal{R}^{(1)}_{na}(u_-)\Bigr]
	\end{aligned}
	\\
	& \label{tr0'a}
	= -\frac{[\imath]^{3s-u_-}\,\mathbf{\Gamma}(2s, u_--s+1)}{[u_-+s]} \, \hat{Q}_-(u_-)
\end{align}
and similarly for the trace over $0''$-th and $b$-th spaces
\begin{align}
	& \nonumber
	\begin{aligned}
		\tr_{0'',b}\Bigl[ &
		\mathcal{R}^{(2)}_{0''b}(u_+)\,\tilde{\mathcal{K}}_b(u_++1,s-1)\,\mathcal{R}^{(2)}_{bn}(u_+) \\
		&
		\times
		\mathcal{R}^{(2)}_{n,n-1}(u_+)\ldots\mathcal{R}^{(2)}_{21}(u_+)\,\mathcal{K}_1(s,u_+)\,\mathcal{R}^{(2)}_{12}(u_+)\ldots\mathcal{R}^{(2)}_{n0''}(u_+) \Bigr]
	\end{aligned}
	\\
	& \nonumber
	\begin{aligned}
		= -\frac{[\imath]^{-3s-u_+}\,\mathbf{\Gamma}(2-2s, u_++s)}{[u_+-s+1]}
		\tr_{b}\Bigl[ &
		\tilde{\mathcal{K}}_b(u_++1,s-1)\,\mathcal{R}^{(2)}_{bn}(u_+) \,
		\mathcal{R}^{(2)}_{n,n-1}(u_+)\ldots\mathcal{R}^{(2)}_{21}(u_+) \\
		&
		\times
		\mathcal{K}_1(s,u_+)\,\mathcal{R}^{(2)}_{12}(u_+)\ldots\mathcal{R}^{(2)}_{n-1,n}(u_+)\,\mathcal{R}^{(2)}_{nb}(u_+)\Bigr]
	\end{aligned}
	\\
	& \label{tr0''b}
	= -\frac{[\imath]^{-3s-u_+}\,\mathbf{\Gamma}(2-2s, u_++s)}{[u_+-s+1]} \, \hat{Q}_+(u_+) \,.
\end{align}
Thus after transformations~\eqref{tr0'a} and~\eqref{tr0''b} the expression~\eqref{Q4} turns into the desired factorization formula~\eqref{Qfact2}.

Now to derive the key identity~\eqref{trKRRRR} used in the proof above we apply the relation~\eqref{trppApp} proven in the Appendix
\begin{equation} \label{trpp}
	\tr_0\Bigl[[\hat{p}_0]^\alpha\,\Phi(z_0)\,[\hat{p}_0]^{-\alpha-2}\,\Psi(z_0)\Bigr] =
	-\frac{1}{\pi^2\,(\alpha+1)(\bar{\alpha}+1)} \,
	\int \mathrm{d}^2z \, \Phi(z) \, \int \mathrm{d}^2z \, \Psi(z) \,,
\end{equation}
where $\Phi$ and $\Psi$ are arbitrary functions.
Using the formulas for $\mathcal{R}$- and $\mathbb{K}$-operators \eqref{R12def}, \eqref{K+x+-}, \eqref{Kop-d}
one can rewrite the traces in~\eqref{trKRRRR} in the form of the left hand side of~\eqref{trpp}
\begin{align}
	& \nonumber
	\begin{aligned}
		&
		\tr_0\Bigl( \overset{0}{\mathbb{K}}_+(u_+;u_-)\,\mathcal{R}^{(1)}_{00'}(u_-)\,\mathcal{R}^{(1)}_{n0}(u_-)\,\mathcal{R}^{(2)}_{0n}(u_+)\,\mathcal{R}^{(2)}_{0''0}(u_+) \Bigr) \\
		& = [-1]^{u_+-s}
		[\hat{p}_{0'}]^{2s-1}[\hat{p}_{0''}]^{u_++s-1} \, t_1(z_n, z_{0'}, z_{0''}) \, [\hat{p}_{0'}]^{u_--s}[\hat{p}_{0''}]^{1-2s} \,,
	\end{aligned} 
	\\
	& \nonumber
	\tr_a\Bigl[\mathcal{R}^{(1)}_{na}(u_-)\,\tilde{\mathcal{K}}_a(s+1,-u_-)\,\mathcal{R}^{(1)}_{a0'}(u_-)\Bigr] =
	[\hat{p}_{0'}]^{2s-1} \, t_2(z_n, z_{0'}) \, [\hat{p}_{0'}]^{u_--s} \,, \\
	& \nonumber
	\tr_b\Bigl[\mathcal{R}^{(2)}_{0''b}(u_+)\,\tilde{\mathcal{K}}_b(u_++1,s-1)\,\mathcal{R}^{(2)}_{bn}(u_+)\Bigr] =
	[-1]^{u_++s-1}
	[\hat{p}_{0''}]^{u_++s-1} \, t_3(z_n, z_{0''}) \, [\hat{p}_{0''}]^{1-2s} \,,
\end{align}
where
\begin{align}
	& \nonumber
	t_1(z_n, z_{0'}, z_{0''}) = \tr_0\Bigl[[\hat{p}_0]^{-u_+-u_--1}\,\Phi(z_0, z_n, z_{0'})\,[\hat{p}_0]^{u_++u_--1}\,\Psi(z_0, z_n, z_{0''})\Bigr] , \\
	& \nonumber
	t_2(z_n, z_{0'}) =
	\tr_a\Bigl[[\hat{p}_a]^{u_-+s-1}\,\Phi_1(z_a, z_n)\,[\hat{p}_a]^{-u_--s-1}\,\Phi(z_a, z_n, z_{0'})\Bigr] , \\
	& \nonumber
	t_3(z_n, z_{0''}) =
	\tr_b\Bigl[[\hat{p}_b]^{-u_++s-2}\,\Psi_1(z_b, z_n)\,[\hat{p}_b]^{u_+-s}\,\Psi(z_b, z_n, z_{0''})\Bigr] , \\
	& \nonumber
	\Phi(z_0, z_n, z_{0'}) =
	[z_0+\tilde{\gamma}]^{-u_--\tilde{g}} [z_0-\tilde{\gamma}]^{-u_-+\tilde{g}-1} [z_{00'}]^{u_--1+s} [z_{n0}]^{u_--s} \,, \\
	& \nonumber
	\Psi(z_0, z_n, z_{0''}) =
	[z_0+\tilde{\gamma}]^{-u_++\tilde{g}-1} [z_0-\tilde{\gamma}]^{-u_+-\tilde{g}} [z_{0''0}]^{u_+-s} [z_{n0}]^{u_++s-1} \,, \\
	& \nonumber
	\Phi_1(z_a, z_n) = [z_{na}]^{2s-1}[z_a+\tilde{\gamma}]^{-s+\tilde{g}-1}[z_a-\tilde{\gamma}]^{-s-\tilde{g}} \,, \\
	& \nonumber
	\Psi_1(z_b, z_n) = [z_{bn}]^{1-2s}[z_b+\tilde{\gamma}]^{s-\tilde{g}-1}[z_b-\tilde{\gamma}]^{s+\tilde{g}-2} \,.
\end{align}
This way,~\eqref{trKRRRR} is equivalent to
\begin{multline} \label{t123}
	t_1(z_n, z_{0'}, z_{0''}) \\
	=
	\frac{[2\tilde{\gamma}]^2\,[-1]^{2s}\,[u_-+s]\,[u_+-s+1]\,[s-\tilde{g}]\,[s+\tilde{g}-1]}{[u_++u_-]\,[2s-1]} \,
	t_2(z_n, z_{0'}) \, t_3(z_n, z_{0''}) \,.
\end{multline}
To prove~\eqref{t123} one just needs to rewrite traces $t_1$, $t_2$, $t_3$ using~\eqref{trpp}
\begin{align}
	& \label{t1}
	t_1(z_n, z_{0'}, z_{0''}) =
	-\frac{1}{\pi^2\,[u_++u_-]} \,
	\int \mathrm{d}^2z_0 \, \Phi(z_0, z_n, z_{0'}) \, \int \mathrm{d}^2z_0 \, \Psi(z_0, z_n, z_{0''}) \,, \\
	& \nonumber
	t_2(z_n, z_{0'}) \, t_3(z_n, z_{0''}) =
	-\frac{[-1]^{2s}\,[2\tilde{\gamma}]^{-2}\,[1-2s]}{\pi^2\,[u_-+s]\,[u_+-s+1]\,[s-\tilde{g}]\,[s+\tilde{g}-1]} \\
	& \label{t2t3}
	\qquad\qquad\qquad\qquad\quad\;\;\; \times
	\int \mathrm{d}^2z_a \, \Phi(z_a, z_n, z_{0'}) \, \int \mathrm{d}^2z_b \, \Psi(z_b, z_n, z_{0''})
\end{align}
and compare the results of~\eqref{t1} and~\eqref{t2t3}.
Note that to obtain~\eqref{t2t3} we also applied the star-triangle relation~\eqref{Star} to integrate~$\Phi_1$ and~$\Psi_1$ and used the properties of gamma function~\eqref{gamma-diff},~\eqref{gamma-refl}.

\subsection{Diagrammatic derivation}

An alternative approach to proving the factorization formula is to verify the corresponding relation for the kernels of integral operators using the diagrammatic technique. For the sake of brevity, we prove the two-boundary part only.

The main idea is to start with the product $\hat{Q}_-(u_-)\hat{Q}_+(u_+)$ and use the diagrammatic technique to obtain the diagram for $\hat{\mathbbm{t}}(u,s_0)$.
Gluing together the diagrams for $\hat{Q}_-(u_-)$ and $\hat{Q}_+(u_+)$ (see Figure~\ref{q+-}), we obtain
\begin{figure}[H]  
	\centering
	% Левый рисунок
	\begin{minipage}[c]{0.48\textwidth}
		\centering
		\begin{tikzpicture}[thick, line cap = round, scale=1.2, y =-1cm]
			\def\r{0.04}
			\def\cut{0.2}
			\draw[-<-] (0,0) to node[midway, above = 1,sloped, font = \scriptsize]{$2u+1$} (2,0);
			\draw[->-] (-0.8, -0.9) to node[midway, sloped, below, font = \scriptsize]{$\pi$} (0,0);
			\node [above] at (-0.8, -0.9) {$-\gamma$};
			\draw[->-] (0.8, -0.9) to node[midway, sloped, below, font = \scriptsize]{$\rho$} (0,0);
			\node [above] at (0.8, -0.9) {$\gamma$};
			\draw[->-] (1.2, -0.9) to node[midway, sloped, below, font = \scriptsize]{$\sigma$}(2,0);
			\node [above] at (1.2, -0.9) {$-\gamma$};
			\draw[->-] (2.8, -0.9) to node[midway, sloped, above, font = \scriptsize]{$\tau$}(2,0);
			\node [above] at (2.8, -0.9) {$\gamma$};
			
			\draw[->-] (-0.8, 5 + 0.9) to node[midway, sloped, below, font = \scriptsize]{$\phi$} (0,5);
			\node [below] at (-0.8, 5+ 0.9) {$-\tilde{\gamma}$};
			\draw[->-] (0.8, 5 + 0.9) to node[midway, sloped, below, font = \scriptsize]{$\chi$} (0,5);
			\node [below] at (0.8, 5+0.9) {$\tilde{\gamma}$};
			\draw[->-] (1.2, 5+0.9) to node[midway, sloped, below, font = \scriptsize]{$\psi$}(2,5);
			\node [below] at (1.2, 5+0.9) {$-\tilde{\gamma}$};
			\draw[->-] (2.8, 5+0.9) to node[midway, sloped, above, font = \scriptsize]{$\omega$}(2,5);
			\node [below] at (2.8, 5+0.9) {$\tilde{\gamma}$};
			\foreach \k in {0,3}
			{
				\draw[->-] (0,\k) to node[midway, above = 1,sloped, font = \scriptsize]{$\alpha$} (-1,\k+1);
				\draw[->-] (0,\k) to node[midway, above = 1,sloped, font = \scriptsize]{$\beta$} (1,\k+1);
				\draw[->-] (-1,\k+1) to node[midway, above = 1,sloped, font = \scriptsize]{$\gamma$} (0,\k+2);
				\draw[->-] (1,\k+1) to node[midway, above = 1,sloped, font = \scriptsize]{$\delta$} (0,\k+2);
				\draw[-<-] (2,\k) to node[midway, above = 1,sloped, font = \scriptsize]{$\epsilon$} (1,\k+1);
				\draw[-<-] (2,\k) to node[midway, above = 1,sloped, font = \scriptsize]{$\zeta$} (3,\k+1);
				\draw[-<-] (1,\k+1) to node[midway, above = 1,sloped, font = \scriptsize]{$\eta$} (2,\k+2);
				\draw[-<-] (3,\k+1) to node[midway, above = 1,sloped, font = \scriptsize]{$\theta$} (2,\k+2);
				\draw [fill = black] (0,\k) circle (\r);
				\draw [fill = black] (2,\k) circle (\r);
				\draw [fill = black] (1,\k+1) circle (\r);
				\draw [fill = black] (0,\k+2) circle (\r);
				\draw [fill = black] (2,\k+2) circle (\r);
			}
			\foreach \k in {0,2}
			{
				\node at (\k,2.5){\Large{$\vdots$}};
				\draw[-] (\k,2) to (\k-\cut,2+\cut);
				\draw[-] (\k,2) to (\k+\cut,2+\cut);
				\draw[-] (\k,3) to (\k-\cut,3-\cut);
				\draw[-] (\k,3) to (\k+\cut,3-\cut);
			}
			\draw[->-] (0,5) to node[midway, above = 1,sloped, font = \scriptsize]{$-2u-1$} (2,5);
			\node[left] at (-1,1) {$z_1$};
			\node[left] at (-1,4) {$z_{n}$};
			\node[right] at (3,1) {$w_1$};
			\node[right] at (3,4) {$w_{n}$};
			\node at (1,7) {(1)};
		\end{tikzpicture}
	\end{minipage}%
	\hfill
	% Правый рисунок
	\begin{minipage}[c]{0.48\textwidth}
		\centering
		\begin{tikzpicture}[thick, scale=1.5, y =-1cm]
			\def\r{0.04}
			\def\cut{0.2}
			\foreach \k in {0,2}
			{
				\draw[->-] (0,\k) to node[midway, above = 1,sloped, font = \scriptsize]{$\mu$} (-1,\k);
				\draw[-<-] (0,\k) to node[midway, above = 1,sloped, font = \scriptsize]{$\mu$} (1,\k);
				\draw[->-] (0,\k) to node[midway, right = 1, font = \scriptsize]{$\lambda$} (0,\k+1);
				\draw[->-] (-1,\k) to node[midway, above = 1,sloped, font = \scriptsize]{$\nu$} (0,\k+1);
				\draw[-<-] (1,\k) to node[midway, above = 1,sloped, font = \scriptsize]{$\nu$} (0,\k+1);
				\draw[-<-] (1,\k) to node[midway, above = 1,sloped, font = \scriptsize]{$a$} (2,\k);
				\draw[-<-] (2,\k) to node[midway, above = 1,sloped, font = \scriptsize]{$a$} (3,\k);
				\draw[-<-] (3,\k) to node[midway, above = 1,sloped, font = \scriptsize]{$b$} (2,\k+1);
				\draw[-<-] (2,\k) to node[midway, left = 1, font = \scriptsize]{$c$} (2,\k+1);
				\draw[-<-] (2,\k+1) to node[midway, above = 1,sloped, font = \scriptsize]{$b$} (1,\k);
				\draw [fill = black] (0,\k) circle (\r);
				\draw [fill = black] (2,\k) circle (\r);
				\draw [fill = black] (1,\k) circle (\r);
				\draw [fill = black] (0,\k+1) circle (\r);
				\draw [fill = black] (2,\k+1) circle (\r);
			}
			\draw[-<-] (0,1) to node[midway, above = 1,sloped, font = \scriptsize]{$\mu$} (1,1);
			\draw[-<-] (1,1) to node[midway, above = 1,sloped, font = \scriptsize]{$a$} (2,1);
			\draw [fill = black] (1,1) circle (\r);
			\draw[->-] (-0.8, -0.9) to node[midway, sloped, above, font = \scriptsize]{$1-u_--g$} (0,0);
			\node [above] at (-0.8, -0.9) {$-\gamma$};
			\draw[->-] (0.8, -0.9) to node[midway, sloped, above, font = \scriptsize]{$g-u_-$} (0,0);
			\node [above] at (0.8, -0.9) {$\gamma$};
			\draw[->-] (1.2, -0.9) to node[midway, sloped, above, font = \scriptsize]{$g-u_+$}(2,0);
			\node [above] at (1.2, -0.9) {$-\gamma$};
			\draw[->-] (2.8, -0.9) to node[midway, sloped, above, font = \scriptsize]{$1-u_+-g$}(2,0);
			\node [above] at (2.8, -0.9) {$\gamma$};
			\draw[-] (0,1) to (\cut,1+\cut);
			\draw[-] (2,1) to (2-\cut,1+\cut);
			\draw[-] (0,2) to (\cut,2-\cut);
			\draw[-] (2,2) to (2-\cut,2-\cut);
			\draw[-] (0,1) to (-1,1);
			\draw[-] (2,1) to (3,1);
			\draw[-] (-1,1) to (-1+\cut,1+\cut);
			\draw[-] (3,1) to (3-\cut,1+\cut);
			\node at (0,1.5) {\Large{$\vdots$}};
			\node at (2,1.5) {\Large{$\vdots$}};
			\draw[->-] (-1, 3) to node[midway, sloped, above, font = \scriptsize]{$\tilde{g}+u_-$} (0,3);
			\node [left] at (-1, 3) {$-\tilde{\gamma}$};
			\draw[->-] (-1, 4) to node[midway, sloped, below, font = \scriptsize]{$1-\tilde{g}+u_-$} (0,3);
			\node [left] at (-1, 4) {$\tilde{\gamma}$};
			\draw[->-] (3, 3) to node[midway, sloped, above = 0.02, font = \scriptsize]{$\tilde{g}+u_+$}(2,3);
			\node [right] at (3, 3) {$\tilde{\gamma}$};
			\draw[->-] (3, 4) to node[midway, sloped, below, font = \scriptsize]{$1-\tilde{g}+u_+$}(2,3);
			\node [right] at (3, 4) {$-\tilde{\gamma}$};
			\node[left] at (-1,0) {$z_1$};
			\node[left] at (-1,2) {$z_{n}$};
			\node[right] at (3,0) {$w_1$};
			\node[right] at (3,2) {$w_{n}$};
			\node[left] at (-1,1) {$z_{2}$};
			\node[right] at (3,1) {$w_2$};
			\draw[-<-] (1, 2) to node[midway, sloped, below, font = \scriptsize]{$1-2s$} (0.4,3.2);
			\draw[->-] (1, 2) to node[midway, sloped, below, font = \scriptsize]{$2s-1$} (1.6,3.2);
			\draw[->-] (0.4-1,3.2+1) to node[midway, sloped, below, font = \scriptsize]{$s+1-\tilde{g}$} (0.4,3.2);
			\node [below] at (0.4-1,3.2+1) {$-\tilde{\gamma}$};
			\draw[->-] (0.4+0.5,3.2+1) to node[midway, sloped, below, font = \scriptsize]{$s+\tilde{g}$} (0.4,3.2);
			\node [below left ] at (0.4+0.5,3.2+1) {$\tilde{\gamma}$};
			\draw[->-] (1.6-0.5,3.2+1) to node[midway, sloped, above = 0.02, font = \scriptsize]{$\tilde{g}-s+1$}(1.6,3.2);
			\node [below right ] at (1.6-0.5,3.2+1) {$-\tilde{\gamma}$};
			\draw[->-] (1.6+1,3.2+1) to node[midway, sloped, below, font = \scriptsize]{$2-s-\tilde{g}$}(1.6,3.2);
			\node [below] at (1.6+1,3.2+1) {$\tilde{\gamma}$};
			\draw [fill = black] (0.4,3.2) circle (\r);
			\draw [fill = black] (1.6,3.2) circle (\r);
			\node at (1,5.3) {(2)};
		\end{tikzpicture}
	\end{minipage}
	
	\caption{
		(1) Diagrammatic representation of $\mathbbm{t}_n(u,s_0)$.
		(2) Diagram for the product $Q_-(u_-)Q_+(u_+)$.
		The indices are given by the following expressions: $\alpha=1+u+s-s_0$, $\beta=1-u-s-s_0$, $\gamma=-1-u-s-s_0$, $\delta=1+u-s+s_0$, $\epsilon=-1-u+s+s_0$, $\zeta=1+u-s+s_0$, 
		$\eta=1+u+s-s_0$, $\theta=1-u-s-s_0$, $\pi=-u-s_0-g$, $\rho=-u+s_0-1+g$, $\sigma=g-u-s_0$, $\tau=1-u-s_0-g$, $\phi=1+u-s_0+\tilde{g}$, $\chi=u+2-s_0-\tilde{g}$, $\psi=1+u+s_0-\tilde{g}$,
		$\omega=\tilde{g}+u+s_0$.
	}   
	\label{Qgeneral}
\end{figure}
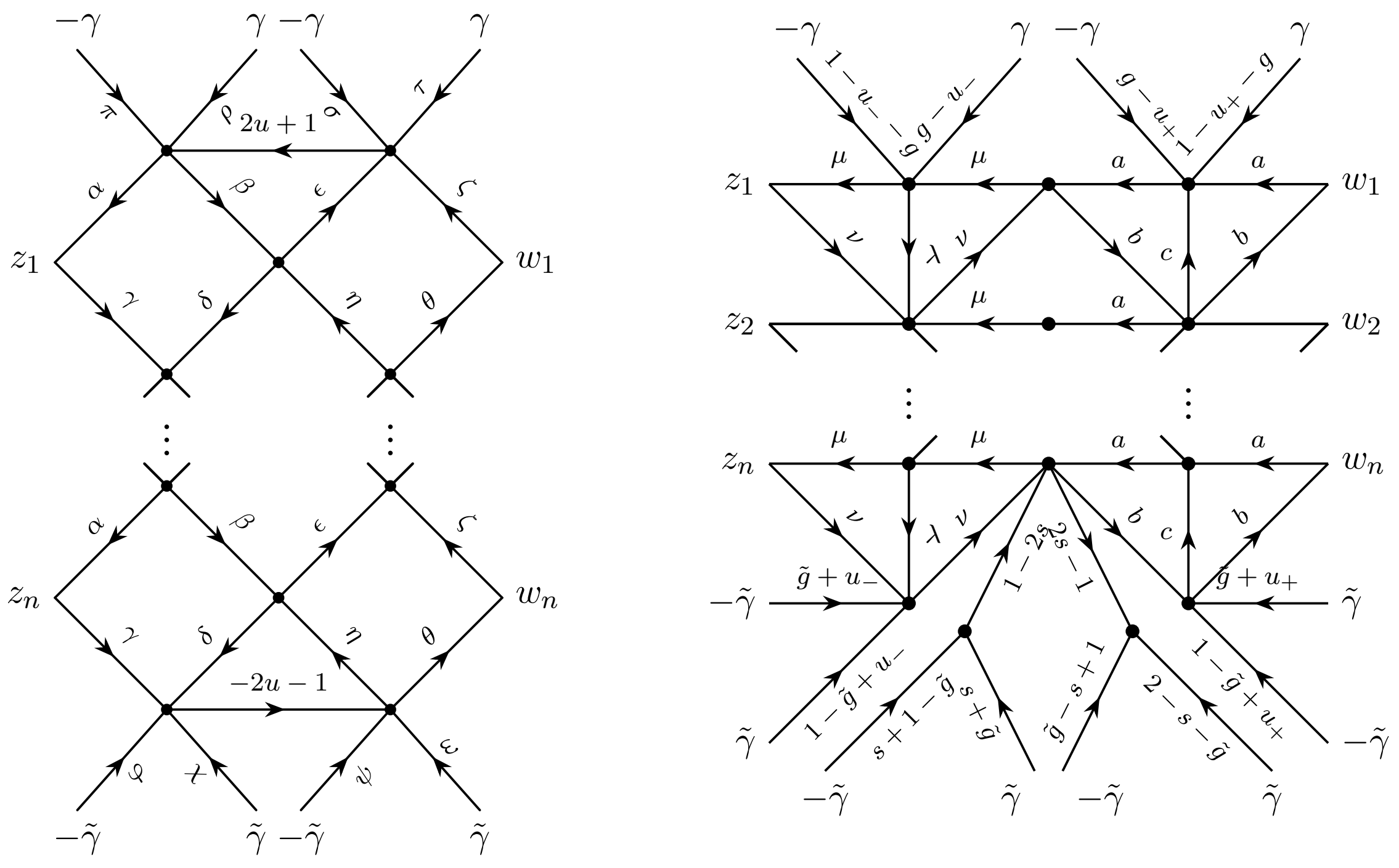   

First, we transform the two triangles at the bottom of this diagram.

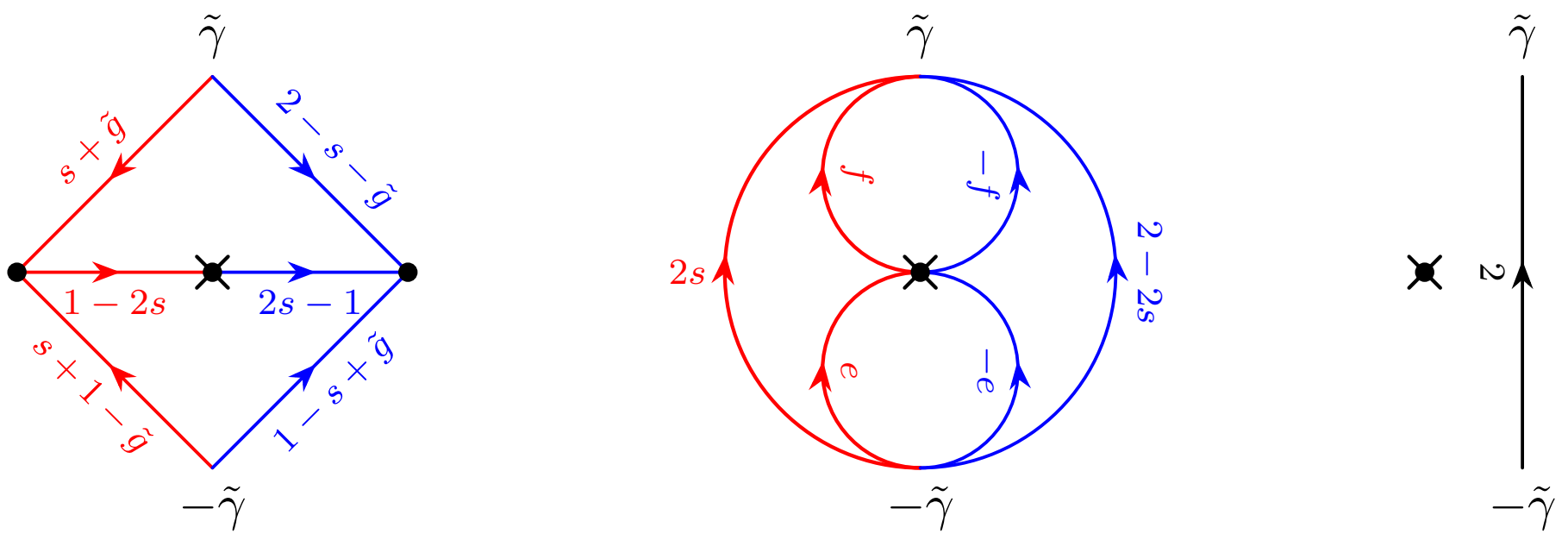
\begin{figure}[H]
	\centering
	\begin{tikzpicture}[thick, line cap = round, scale=1.5, y = -1cm]
		\def\r{0.04}
		\def\cut{0.08}
		\draw[red, ->-] (0,0) to node[midway, sloped, above, font = \scriptsize]{$s+\tilde{g}$}(-1,1);
		\draw[red, ->-] (-1,1) to node[midway, sloped, below, font = \scriptsize]{$1-2s$}(0,1);
		\draw[blue, ->-] (0,1) to node[midway, sloped, below, font = \scriptsize]{$2s-1$}(1,1);
		\draw[blue, ->-] (0,0) to node[midway, sloped, above, font = \scriptsize]{$2-s-\tilde{g}$}(1,1);
		\node [above] at (0,0) {$\tilde{\gamma}$};
		\draw[red, ->-] (0,2) to node[midway, sloped, below, font = \scriptsize]{$s+1-\tilde{g}$}(-1,1);
		\draw[blue, ->-] (0,2) to node[midway, sloped, below, font = \scriptsize]{$1-s+\tilde{g}$}(1,1);
		\node [below] at (0,2) {$-\tilde{\gamma}$};
		\draw [fill = black] (-1,1) circle (\r);
		\draw [fill = black] (1,1) circle (\r);
		\draw [fill = black] (0,1) circle (\r);
		\draw[-] (0,1) to (\cut,1+\cut);
		\draw[-] (0,1) to (\cut,1-\cut);
		\draw[-] (0,1) to (-\cut,1+\cut);
		\draw[-] (0,1) to (-\cut,1-\cut);
	\end{tikzpicture}
	\hspace{1.5cm}
	\begin{tikzpicture}[thick, line cap = round, scale=1.5, y = -1cm]
		\def\r{0.04}
		\def\cut{0.08}
		\draw[red, -<-, thick] (0,0) arc (-90:-270:1) node[midway, left, font = \scriptsize]{$2s$};
		\draw[blue, -<-] (0,0) arc (-90:90:1) node[midway, above, sloped, font = \scriptsize]{$2-2s$}; 
		\draw[red, -<-, thick] (0,0) arc (-90:-270:0.5) node[midway, above, sloped, font = \scriptsize]{$f$};
		\draw[blue, -<-] (0,0) arc (-90:90:0.5) node[midway, below, sloped, font = \scriptsize]{$-f$};
		\draw[red, -<-, thick] (0,1) arc (-90:-270:0.5) node[midway, above, sloped, font = \scriptsize]{$e$};
		\draw[blue, -<-] (0,1) arc (-90:90:0.5) node [midway, below, sloped,  font = \scriptsize]{$-e$};
		\draw [fill = black] (0,1) circle (\r);
		\node [above] at (0,0) {$\tilde{\gamma}$};
		\node [below] at (0,2) {$-\tilde{\gamma}$};
		\draw[-] (0,1) to (\cut,1+\cut);
		\draw[-] (0,1) to (\cut,1-\cut);
		\draw[-] (0,1) to (-\cut,1+\cut);
		\draw[-] (0,1) to (-\cut,1-\cut);
	\end{tikzpicture}
	\hspace{1.5cm}
	\begin{tikzpicture}[thick, line cap = round, scale=1.5, y = -1cm]
		\def\r{0.04}
		\def\cut{0.08}
		\node [above] at (0.5,0) {$\tilde{\gamma}$};
		\draw[-<-] (0.5,0) to node[midway, sloped, below, font = \scriptsize]{$2$}(0.5,2);
		\node [below] at (0.5,2) {$-\tilde{\gamma}$};
		\draw [fill = black] (0,1) circle (\r);
		\draw[-] (0,1) to (\cut,1+\cut);
		\draw[-] (0,1) to (\cut,1-\cut);
		\draw[-] (0,1) to (-\cut,1+\cut);
		\draw[-] (0,1) to (-\cut,1-\cut);
	\end{tikzpicture}
	\caption{
		Transformation of the bottom block. We transform the red and blue stars into triangles using \eqref{Star}, after which the lines cancel each other. Here, $f = \tilde{g}-s$ and $e = 1-s-\tilde{g}$.
	}
\end{figure}        

Next, we transform the main building block of the diagram in Figure~\ref{Qgeneral}.

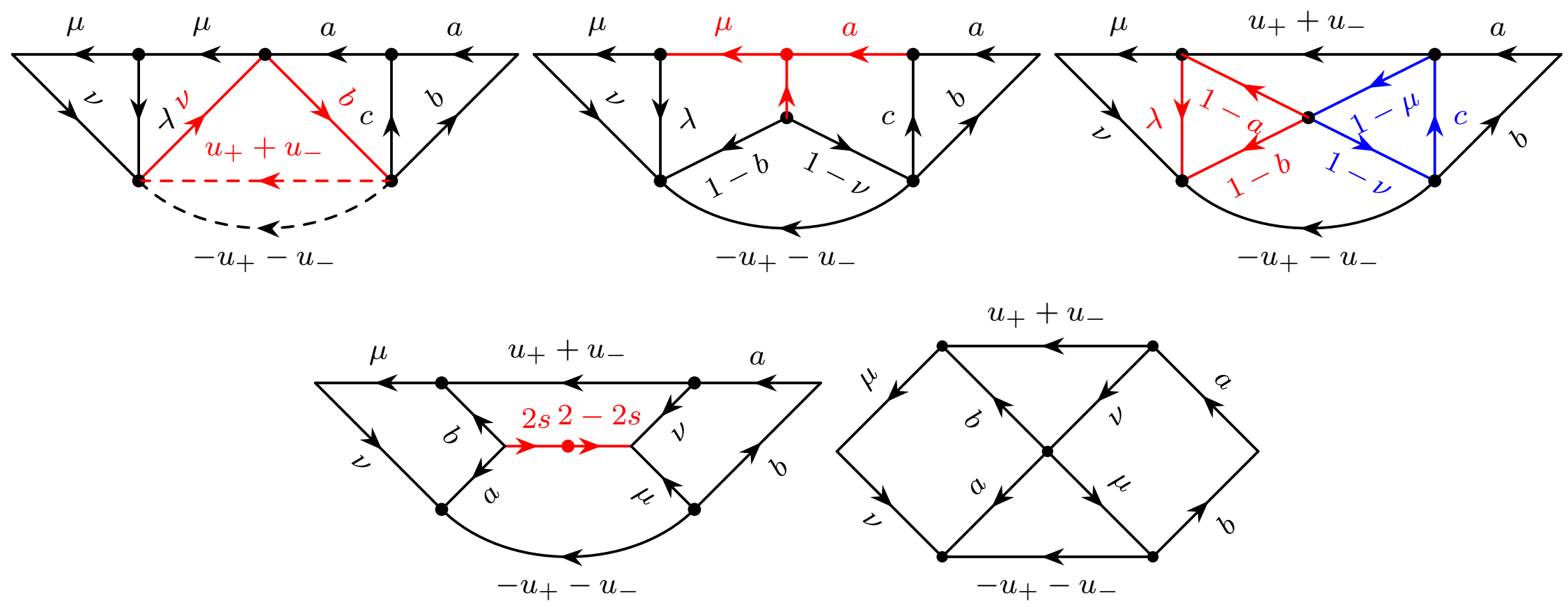
\begin{figure}[H]
	\centering
	\begin{tikzpicture}[thick, line cap = round, scale=1.2, y =-1cm]
		\def\r{0.04}
		\def\cut{0.2}
		\foreach \k in {0}
		{
			\draw[->-] (0,\k) to node[midway, above = 1,sloped, font = \scriptsize]{$\mu$} (-1,\k);
			\draw[-<-] (0,\k) to node[midway, above = 1,sloped, font = \scriptsize]{$\mu$} (1,\k);
			\draw[->-] (0,\k) to node[midway, right = 1, font = \scriptsize]{$\lambda$} (0,\k+1);
			\draw[->-] (-1,\k) to node[midway, above = 1,sloped, font = \scriptsize]{$\nu$} (0,\k+1);
			\draw[red, -<-] (1,\k) to node[midway, above = 1,sloped, font = \scriptsize]{$\nu$} (0,\k+1);
			\draw[-<-] (1,\k) to node[midway, above = 1,sloped, font = \scriptsize]{$a$} (2,\k);
			\draw[-<-] (2,\k) to node[midway, above = 1,sloped, font = \scriptsize]{$a$} (3,\k);
			\draw[-<-] (3,\k) to node[midway, above = 1,sloped, font = \scriptsize]{$b$} (2,\k+1);
			\draw[-<-] (2,\k) to node[midway, left = 1, font = \scriptsize]{$c$} (2,\k+1);
			\draw[red, -<-] (2,\k+1) to node[midway, above = 1,sloped, font = \scriptsize]{$b$} (1,\k);
			\draw [fill = black] (0,\k) circle (\r);
			\draw [fill = black] (2,\k) circle (\r);
			\draw [fill = black] (1,\k) circle (\r);
			\draw [fill = black] (0,\k+1) circle (\r);
			\draw [fill = black] (2,\k+1) circle (\r);
			\draw[dashed,red, -<-] (0,1) to  node[midway, above = 1,sloped, font = \scriptsize]{$u_++u_-$} (2,1);
			\draw[dashed, -<-] (0,1) .. controls (0.5,1.5) and (1.5,1.5) .. node[midway, below = 1,sloped, font = \scriptsize]{$-u_+-u_-$} (2,1);
		}
	\end{tikzpicture}
	\begin{tikzpicture}[thick, line cap = round, scale=1.2, y =-1cm]
		\def\r{0.04}
		\def\cut{0.2}
		\foreach \k in {0}
		{
			\draw[->-] (0,\k) to node[midway, above = 1,sloped, font = \scriptsize]{$\mu$} (-1,\k);
			\draw[red, -<-] (0,\k) to node[midway, above = 1,sloped, font = \scriptsize]{$\mu$} (1,\k);
			\draw[->-] (0,\k) to node[midway, right = 1, font = \scriptsize]{$\lambda$} (0,\k+1);
			\draw[->-] (-1,\k) to node[midway, above = 1,sloped, font = \scriptsize]{$\nu$} (0,\k+1);
			\draw[-<-] (0,\k+1) to node[midway, below = 1,sloped, font = \scriptsize]{$1-b$} (1,0.5);
			\draw[red, -<-] (1,\k) to node[midway, above = 1,sloped, font = \scriptsize]{$a$} (2,\k);
			\draw[-<-] (2,\k) to node[midway, above = 1,sloped, font = \scriptsize]{$a$} (3,\k);
			\draw[-<-] (3,\k) to node[midway, above = 1,sloped, font = \scriptsize]{$b$} (2,\k+1);
			\draw[-<-] (2,\k) to node[midway, left = 1, font = \scriptsize]{$c$} (2,\k+1);
			\draw[-<-] (2,\k+1) to node[midway, below = 1,sloped, font = \scriptsize]{$1-\nu$} (1,0.5);
			\draw [fill = black] (0,\k) circle (\r);
			\draw [fill = black] (2,\k) circle (\r);
			\draw [fill = red, draw = red] (1,\k) circle (\r);
			\draw [fill = black] (0,\k+1) circle (\r);
			\draw [fill = black] (2,\k+1) circle (\r);
			\draw [fill = black] (1,\k+0.5) circle (\r);
			\draw[red, ->-] (1,0.5) to  node[midway, above = 1,sloped, font = \tiny]{$ $} (1,0);
			\draw[-<-] (0,1) .. controls (0.5,1.5) and (1.5,1.5) .. node[midway, below = 1,sloped, font = \scriptsize]{$-u_+-u_-$} (2,1);
		}
	\end{tikzpicture}
	\begin{tikzpicture}[thick, line cap = round, scale=1.2, y =-1cm]
		\def\r{0.04}
		\def\cut{0.2}
		\foreach \k in {0}
		{
			\draw[->-] (0,\k) to node[midway, above = 1,sloped, font = \scriptsize]{$\mu$} (-1,\k);
			\draw[-<-] (0,\k) to node[midway, above = 1,sloped, font = \scriptsize]{$u_++u_-$} (2,\k);
			\draw[red, ->-] (0,\k) to node[midway, left = 1, font = \scriptsize]{$\lambda$} (0,\k+1);
			\draw[->-] (-1,\k) to node[midway, below = 1,sloped, font = \scriptsize]{$\nu$} (0,\k+1);
			\draw[red, -<-] (0,\k+1) to node[midway, below = 1,sloped, font = \scriptsize]{$1-b$} (1,0.5);
			\draw[blue, -<-] (1,0.5) to node[midway, below = 1,sloped, font = \scriptsize]{$1-\mu$} (2,\k);
			\draw[-<-] (2,\k) to node[midway, above = 1,sloped, font = \scriptsize]{$a$} (3,\k);
			\draw[-<-] (3,\k) to node[midway, below = 1,sloped, font = \scriptsize]{$b$} (2,\k+1);
			\draw[blue, -<-] (2,\k) to node[midway, right = 1, font = \scriptsize]{$c$} (2,\k+1);
			\draw[blue, -<-] (2,\k+1) to node[midway, below = 1,sloped, font = \scriptsize]{$1-\nu$} (1,0.5);
			\draw [fill = black] (0,\k) circle (\r);
			\draw [fill = black] (2,\k) circle (\r);
			\draw [fill = black] (0,\k+1) circle (\r);
			\draw [fill = black] (2,\k+1) circle (\r);
			\draw [fill = black] (1,0.5) circle (\r);
			\draw[red, ->-] (1,0.5) to  node[midway, below = 1,sloped, font = \scriptsize]{$1-a$} (0,0);
			\draw[-<-] (0,1) .. controls (0.5,1.5) and (1.5,1.5) .. node[midway, below = 1,sloped, font = \scriptsize]{$-u_+-u_-$} (2,1);
		}
	\end{tikzpicture}
	\begin{tikzpicture}[thick, line cap = round, scale=1.2, y =-1cm]
		\def\r{0.04}
		\def\cut{0.2}
		\foreach \k in {0}
		{
			\draw[->-] (0,\k) to node[midway, above = 1,sloped, font = \scriptsize]{$\mu$} (-1,\k);
			\draw[-<-] (0,\k) to node[midway, above = 1,sloped, font = \scriptsize]{$u_++u_-$} (2,\k);
			\draw[red, ->-] (0.5,0.5) to node[midway, above = 1, font = \scriptsize]{$2s$} (1,0.5);
			\draw[->-] (-1,\k) to node[midway, below = 1,sloped, font = \scriptsize]{$\nu$} (0,\k+1);
			\draw[-<-] (0,\k+1) to node[midway, below = 1,sloped, font = \scriptsize]{$a$} (0.5,0.5);
			\draw[-<-] (1.5,0.5) to node[midway, below = 1,sloped, font = \scriptsize]{$\nu$} (2,\k);
			\draw[-<-] (2,\k) to node[midway, above = 1,sloped, font = \scriptsize]{$a$} (3,\k);
			\draw[-<-] (3,\k) to node[midway, below = 1,sloped, font = \scriptsize]{$b$} (2,\k+1);
			\draw[red, ->-] (1,0.5) to node[midway, above = 1, font = \scriptsize]{$2-2s$} (1.5,0.5);
			\draw[->-] (2,\k+1) to node[midway, below = 1,sloped, font = \scriptsize]{$\mu$} (1.5,0.5);
			\draw [fill = black] (0,\k) circle (\r);
			\draw [fill = black] (2,\k) circle (\r);
			\draw [fill = black] (0,\k+1) circle (\r);
			\draw [fill = black] (2,\k+1) circle (\r);
			\draw [fill = red, draw = red] (1,0.5) circle (\r);
			\draw[->-] (0.5,0.5) to  node[midway, below = 1,sloped, font = \scriptsize]{$b$} (0,0);
			\draw[-<-] (0,1) .. controls (0.5,1.5) and (1.5,1.5) .. node[midway, below = 1,sloped, font = \scriptsize]{$-u_+-u_-$} (2,1);
		}
	\end{tikzpicture}
	\begin{tikzpicture}[thick, line cap = round, scale=1, y =-1cm]
		\def\r{0.04}
		\def\cut{0.2}
		\foreach \k in {0}
		{
			\draw[->-] (0,\k) to node[midway, above = 1,sloped, font = \scriptsize]{$\mu$} (-1,1);
			\draw[-<-] (0,\k) to node[midway, above = 1,sloped, font = \scriptsize]{$u_++u_-$} (2,\k);
			\draw[->-] (-1,1) to node[midway, below = 1,sloped, font = \scriptsize]{$\nu$} (0,2 );
			\draw[-<-] (0,2) to node[midway, above = 1,sloped, font = \scriptsize]{$a$} (1,1);
			\draw[-<-] (1,1) to node[midway, below = 1,sloped, font = \scriptsize]{$\nu$} (2,\k);
			\draw[-<-] (2,\k) to node[midway, above = 1,sloped, font = \scriptsize]{$a$} (3,1);
			\draw[-<-] (3,1) to node[midway, below = 1,sloped, font = \scriptsize]{$b$} (2,2);
			\draw[-<-] (2,2) to node[midway, above = 1,sloped, font = \scriptsize]{$\mu$} (1,1);
			\draw [fill = black] (0,\k) circle (\r);
			\draw [fill = black] (2,\k) circle (\r);
			\draw [fill = black] (0,\k+2) circle (\r);
			\draw [fill = black] (2,\k+2) circle (\r);
			\draw [fill = black] (1,1) circle (\r);
			\draw[->-] (1,1) to  node[midway, below = 1,sloped, font = \scriptsize]{$b$} (0,0);
			\draw[-<-] (0,2) to node[midway, below = 1,sloped, font = \scriptsize]{$-u_+-u_-$} (2,2);
		}
	\end{tikzpicture}
	\caption{
		Transformation of the main building block. First, we transform the central unique triangle into a star; in the second step, we transform the upper star into a triangle. Next, we transform the blue and red unique triangles into stars, and in the final step, we use relation \eqref{delta2}.
	}    
\end{figure}  

Combining all these steps, we obtain the diagram for the transfer matrix (see Figure~\ref{Qgeneral}). Recalling the relation $\hat{\mathbb{Q}}(u_+, u_-) = \hat{\mathbbm{t}}(u,s_0)$, we arrive at \eqref{Qfact2}.

\section{Separation of variables} \label{sect:SoV}

Our goal is to find the spectrum of the transfer matrix $t_n(u)$:
\begin{equation}\label{ep}
    \begin{aligned}
        \tilde{t}_n(u)\Phi_{\bm{x}_{n-1}}^{\lambda}(\bm{z}_{n}) 
        &= \lambda \bigl(u^2-x_1^2\bigr) \ldots \bigl(u^2-x_{n-1}^2\bigr) \Phi_{\bm{x}_{n-1}}^{\lambda}(\bm{z}_{n}) \,, \\
        \bar{\tilde{t}}_n(\bar{u})\Phi_{\bm{x}_{n-1}}^{\lambda}(\bm{z}_{n}) 
        &= \bar{\lambda} \bigl(\bar{u}^2-\bar{x}_1^2\bigr) \ldots \bigl(\bar{u}^2-\bar{x}_{n-1}^2\bigr) \Phi_{\bm{x}_{n-1}}^{\lambda}(\bm{z}_{n}) \,,
    \end{aligned}
\end{equation}
where
\begin{equation*}
    \tilde{t}_n(u) = \frac{t_n(u) - 2\gamma\bigl(g-\frac{1}{2}\bigr)\bigl(s-\frac{1}{2}\bigr)^{2n}}{2\gamma\bigl(u^{2}-\frac{1}{4}\bigr)} = u^{2n-2}H_1+u^{2n-4}H_2+\ldots+H_n \,.
\end{equation*}
Following \cite{IS}, we look for the eigenfunctions of the transfer matrix $t_n(u)$ in the form 
\begin{equation}\label{int}
    \Phi_{\bm{x}_{n-1}}^{\lambda}(\bm{z}_n) = \int \mathcal{D}\bm{y}_n \, \mu(\bm{y}_n) \Psi_{\bm{y}_n}(\bm{z}_n) \, C(\lambda, \bm{x}_{n-1}|\,\bm{y}_{n}) \,,
\end{equation}  
where $\{\Psi_{\bm{y}_n}(\bm{z}_n)\}$ is a set of eigenfunctions of the operator $B_n(u)$~\eqref{Top} constructed in \cite{ADV2}, and $C(\lambda, \bm{x}_{n-1}|\,\bm{y}_n)$ is the kernel of the integral transformation.
The spectral variables of $B$-operator's eigenfunctions have the form 
\begin{equation} \label{y}
	(y_k, \bar{y}_k) = \bigl(\tfrac{n_k}{2}+\imath\nu_k, -\tfrac{n_k}{2}+\imath\nu_k\bigr) \,, \qquad
	n_k \in \mathbb{Z}+\sigma \,, \; \nu_k \in \mathbb{R} \,.
\end{equation}
By $\int \mathcal{D}\bm{y}_n$ we denote the following
\begin{equation} \label{intx}
    \int \mathcal{D}\bm{y}_n = \prod\limits_{k=1}^{n}\int\mathcal{D}y_k \,, \qquad
    \int\mathcal{D}y_k = \sum\limits_{n_k\in\mathbb{Z}+\sigma}\int\limits_{\mathbb{R}}\mathrm{d}\nu_k \,,
\end{equation}
and the integration measure is given by
\begin{equation}\label{measure}
    \mu(\bm{y}_n) =
    \frac{\prod\limits_{k=1}^n|y_k|^2 \prod\limits_{1\leq i<j\leq n}|y_i^2-y_j^2|^2}{4^n\,\pi^{2n(n+1)}\,n!\,|\gamma|^{2n}} \,.
\end{equation}
We compute the action of the transfer matrix $t_n(u)$ on the eigenfunction $\Psi_{\bm{y}_{n}}(\bm{z}_n)$ and obtain a multiparametric difference equation for the kernel $C(\lambda, \bm{x}_{n-1}|\,\bm{y}_n)$. This equation admits separation of variables.

\subsection{Eigenfunctions of the operator $B_n(u)$}

Recall that the monodromy matrix $T_n(u)$ has the form
\begin{equation*}
    T_n(u) = L_n(u) \cdots L_1(u) K_-(u) L_1(u) \cdots L_n(u) = \begin{pmatrix}
        A_n(u) & B_n(u) \\
        C_n(u) & D_n(u)
    \end{pmatrix} .
\end{equation*}
From the reflection equation \eqref{refl_eq} for $T_n(u)$, it follows that the operators $B_n(u)$ and $\bar{B}_n(\bar{u})$ form a commutative family:
\begin{align*}
    [B_n(u), B_n(v)] = 0 \,, \quad [\bar{B}_n(\bar{u}), \bar{B}_n(\bar{v})] = 0 \,, \quad [B_n(u), \bar{B}_n(\bar{v})] = 0 \,.
\end{align*}
Therefore, these operators share a common system of eigenfunctions, which was constructed in \cite{ADV2}. In this section, we present these eigenfunctions and list the properties required below. The operators $B_n(u)$ and $\bar{B}_n(\bar{u})$ are polynomials in $u$ and $\bar{u}$ of degree $2n+1$, proportional to $(u-\frac{1}{2})$ and $(\bar{u}-\frac{1}{2})$, respectively. The eigenfunctions $\Psi_{\bm{x}_n}$ are labeled by the zeros of these polynomials:
\begin{equation} \nonumber
    \begin{aligned}
        B_{n}(u)\,\Psi_{\bm{x}_n}(\bm{z}_n) &= 
        \left(u - \tfrac{1}{2} \right)\left(u^2-x^2_1\right)\cdots
        \left(u^2-x^2_{n}\right)\,
        \Psi_{\bm{x}_n}(\bm{z}_n) \,, \\
        \bar{B}_{n}(\bar{u})\,\Psi_{\bm{x}_n}(\bm{z}_n) &= 
        \left(\bar{u} - \tfrac{1}{2} \right)\left(\bar{u}^2-\bar{x}^2_1\right)\cdots
        \left(\bar{u}^2-\bar{x}^2_{n}\right)\,
        \Psi_{\bm{x}_n}(\bm{z}_n) \,.
    \end{aligned}
\end{equation}
These eigenfunctions can be constructed iteratively using the raising integral operators $\Lambda_k(x)$ for $k=2,3,\ldots$, expressed in terms of $\mathcal{R}$-operators as
\begin{align*}
    \Lambda_k(x) &= \lambda_k(x)\,\mathcal{R}_{k\,k-1}(x)\,
    \mathcal{R}_{k-1\, k-2}(x)
    \ldots \mathcal{R}_{2 1}(x)\,
    \mathcal{K}_1(s,x)\,
    \mathcal{R}_{1 2}(x) \, \mathcal{R}_{2 3}(x) \ldots 
    \mathcal{R}_{k-1\, k}(x)\,, \\ 
    \Lambda_1(x) &= \lambda_1(x)\,\mathcal{K}_1(s,x) \,,
\end{align*} 
where $\mathcal{R}(u):=\mathcal{R}^{(2)}(u)$ and $\lambda_k(x)$ is the normalization coefficient of the form
\begin{equation} \label{nu}
    \lambda_k(x) = \pi^{4k-3}\,[-2\imath\gamma]^{s-x}\,c^{2(k-1)}(-x-s)\,\mathbf{\Gamma}(g-x) \,.
\end{equation}
To obtain the eigenfunctions, one applies the raising operators to the function identically equal to $1$
\begin{equation} \label{BLambda}
    \Psi_{\bm{x}_n}(\bm{z}_n) = \Psi_{x_1, \ldots, x_n}(z_1, \dots, z_n) = 
    \Lambda_n(x_n) \, \Lambda_{n-1}(x_{n-1}) \cdots \Lambda_{1}(x_1) \cdot 1 \,.
\end{equation}  
With this choice of normalization factor, the $\Lambda$-operators obey the exchange relation
\begin{equation*}
    \Lambda_n(x)\,\Lambda_{n-1}(y) =  
    \Lambda_n(y)\,\Lambda_{n-1}(x) \,, \qquad
    \Lambda_n(x) = \Lambda_n(-x) \,.    
\end{equation*} 
By definition \eqref{BLambda}, this implies the symmetry of $\Psi_{\bm{x}_n}$ under permutations and reflections of spectral variables
\begin{align} \nonumber
    \Psi_{\tau \bm{x}_n}(\bm{z}_n) = 
    \Psi_{\bm{x}_n}(\bm{z}_n)\,, \qquad
    \Psi_{\sigma_k \bm{x}_n}(\bm{z}_n) = 
    \Psi_{\bm{x}_n}(\bm{z}_n)\,,
\end{align}
where $\tau \bm{x}_n = (x_{\tau(1)},\bar{x}_{\tau(1)}, \ldots , x_{\tau(n)},\bar{x}_{\tau(n)})$ and $\sigma_k \bm{x}_n = (x_{1},\bar{x}_1,\ldots , -x_{k},-\bar{x}_{k}, \ldots , x_{n},\bar{x}_{n})$.

The complete set consists of eigenfunctions $\Psi_{\bm{x}_n}$ labeled by parameters of the form $(x_k, \bar{x}_k) = \bigl(\tfrac{n_k}{2}+\imath\nu_k, -\tfrac{n_k}{2}+\imath\nu_k\bigr)$, where $n_k \in \mathbb{Z}+\sigma \,, \; \nu_k \in \mathbb{R}$.
The orthogonality relation has the form
\begin{equation} \label{orth}
	\langle\Psi_{\bm{y}_n}|\Psi_{\bm{x}_n}\rangle = 
\int \mathrm{d}^2\bm{z}_n \, 
\overline{\Psi_{\bm{y}_n}(\bm{z}_n)} \, \Psi_{\bm{x}_n}(\bm{z}_n) = 
	\mu^{-1}(\bm{x}_n)\, \delta^{(2)}(\bm{x}_n,\bm{y}_n) \,,
\end{equation}
where $\mu(\bm{x}_n)$ is the measure \eqref{measure} and the symmetric delta function $\delta^{(2)}(\bm{x}_n,\bm{y}_n)$ is 
defined in the following way
\begin{align}\nonumber
	\delta^{(2)}(\bm{x}_n,\bm{y}_n) =
	\frac{1}{2^n\,n!}
	\sum\limits_{
		\begin{smallmatrix}
			\tau\in \mathfrak{S}_n \\ \sigma_1,\ldots,\sigma_n=\pm 1
	\end{smallmatrix}}
	\delta^{(2)}(x_1-\sigma_1\,y_{\tau(1)}) \ldots
	\delta^{(2)}(x_n-\sigma_n\,y_{\tau(n)}) \,,
\end{align}
here $\mathfrak{S}_n$ denotes the symmetric group of degree $n$. The competeness relation reads
\begin{equation} \nonumber
	\int \mathcal{D}\bm{x}_n \, \mu(\bm{x}_n) \,
	\Psi_{\bm{x}_n}(\bm{z}_n) \, \overline{\Psi_{\bm{x}_n}(\bm{w}_n)} =
	\delta^2(z_1-w_1) \ldots \delta^2(z_n-w_n) \,,
\end{equation}
where we remind that $\delta^2(z)$ is the two-dimensional delta function in the complex plane~\eqref{delta2D}.

\subsection{Transfer Matrix Action}

To obtain the difference equation for the integral kernel $C(\lambda, \bm{x}_{n-1}|\,\bm{y}_n)$, we need to establish how the transfer matrix $t_n(u)$ acts on the eigenfunctions $\Psi_{\bm{x}_n}(\bm{z}_n)$ of the $B$-element.

First of all, the one-boundary monodromy matrix possesses the important property~\cite{Skl:Boundary}
\begin{equation} \nonumber
    \begin{pmatrix}
        A_n(-u) & B_n(-u) \\
        C_n(-u) & D_n(-u)
    \end{pmatrix} = \frac{1}{2u-1} \begin{pmatrix}
        -A_n(u)+2uD_n(u) & -(2u+1)B_n(u) \\
        -(2u+1)C_n(u) & 2uA_n(u)-D_n(u)
    \end{pmatrix} .
\end{equation}
In particular, this implies
\begin{equation*} 
    A_n(u) = \frac{1}{u}\Bigl\{ \bigl(u-\tfrac{1}{2}\bigr) D_n(-u) + \tfrac{1}{2} D_n(u) \Bigr\} ,
\end{equation*} 
and thus
\begin{equation} \label{tD}
    t_n(u) = \frac{1}{u}\Bigl\{ \bigl(u-\tfrac{1}{2}\bigr) D_n(-u) + \bigl(u + \tfrac{1}{2}\bigr) D_n(u) \Bigr\} .
\end{equation}  
Therefore, it is sufficient to derive the action of the $D$-operator on $\Psi_{\bm{x}_n}(\bm{z}_n)$. To do so, we use the following relation employed in the construction of the eigenfunctions $\Psi_{\bm{x}_n}(\bm{z}_n)$
\begin{multline*}
    \Lambda_n (x)\,
    L_{n}\left(u+x-1,u_{2}\right)\,T_{n-1}(u)\,
    L_{n}\left(u_{1},u-x\right) \\
    = L_{n}\left(u_1,u-x\right)\,T_{n-1}(u)\, 
    L_{n}\left(u+x-1,u_{2}\right)\,
    \Lambda_n(x)
\end{multline*}
(recall that $u_1=u+s-1$, $u_2=u-s$).
This is a consequence of the key relation for the $\mathcal{R}$-operator \eqref{defR} and the reflection equation \eqref{Kdef}. Using the unitarity of the $L$-operator~\eqref{unitarity}, we rewrite this relation as
\begin{align*}
	&
    \Lambda_n (x)\,
    L_{n}\left(u+x-1,u_{2}\right)\,T_{n-1}(u)\,
    L_{n}\left(u_{1},u-x\right) \\
    &
    = L_{n}\left(u_1,u-x\right)\, L_n\left(u_1,u_2\right)^{-1} T_{n}(u)\, 
    L_n\left(u_1,u_2\right)^{-1}L_{n}\left(u+x-1,u_{2}\right)\,
    \Lambda_n(x) \\
    &
    = \begin{pmatrix}
        1 & 0 \\
        \frac{x-s}{u_2}\,z_n & \frac{u-x}{u_2}
    \end{pmatrix} \, T_n(u) \, 
    \begin{pmatrix}
        \frac{u+x-1}{u_1} & 0 \\
        \frac{x-s}{u_2}\,z_n & 1    
    \end{pmatrix} \, \Lambda_n (x) \,.
\end{align*}
Evaluating the $(2,2)$ matrix element of the above relation yields
\begin{multline*}
    (u-x)(x+s)\,\Lambda_n (x)\, z_n\, B_{n-1}(u) + (u-x)(u-s)\,\Lambda_n(x) \,D_{n-1}(u) \\ 
    = \frac{x-s}{u-s}\,z_n\,B_n(u)\,\Lambda_n(x) + \frac{u-x}{u-s}\,D_n(u)\,\Lambda_n(x) \,.
\end{multline*} 
The next step is to act on a function $\psi(z_1, \ldots, z_{n-1})$ independent of $z_n$. In \cite{ADV2} it was shown that
\begin{equation*}
    B_{n}(u)\,\Lambda_{n}\left(x\right)\,\psi(z_1\ldots z_{n-1}) =
    \left(u^2-x^2\right)\,\Lambda_{n}\left(x\right)\,B_{n-1}(u)\,
    \psi(z_1\ldots z_{n-1}) \,.
\end{equation*}
Substituting this into the previous relation and canceling the factor $(u-x)$, we obtain
\begin{multline}\label{1id}
    (u-s)\Lambda_n(x)\,D_{n-1}(u)\,\psi(z_1\ldots z_{n-1}) = \Bigl(\frac{x-s}{u-s}\,(u+x)\,z_n\,\Lambda_n(x)\,B_{n-1}(u) + \\ 
    + u_2^{-1}\,D_n(u)\,\Lambda_n(x) - (x+s)\,\Lambda_n(x)\,z_n\,B_{n-1}(u)\Bigr)\,\psi(z_1\ldots z_{n-1}) \,.
\end{multline}  
To derive the action of the $D$-operator on $\Psi_{\bm{x}_n}(\bm{z}_n)$, we use the triangularity property previously employed to derive the Baxter equation. Recall the formula~\eqref{ZLRZ2}
\begin{multline}\label{Rtr}
    \mathcal{M}_{ij}\left(u,x\right) := Z_i^{-1}\,
    \mathcal{R}_{ij}(x)\,L_i(u_1,u_2)\,Z_j = \\ 
    \begin{pmatrix}
        \imath\mathcal{R}_{ij}(x+1,\bar{x}) +
        (u-x)\mathcal{R}_{ij}(x,\bar{x}) & -\mathcal{R}_{ij}(x,\bar{x})\,\partial_{i} \\
        -(u-x)\,z_{ij}\,\mathcal{R}_{ij}(x,\bar{x}) & -\imath(x+s-1)(x-s)
        \mathcal{R}_{ij}(x-1,\bar{x})+(u-x)\mathcal{R}_{ij}(x,\bar{x})
    \end{pmatrix} ,
\end{multline}
where, again, for simplicity we write $\mathcal{R}_{ij}(x)$ instead of $\mathcal{R}^{(2)}_{ij}(x)$.

Let us consider the expression
\begin{multline*}
    \Omega\left(u,x\right) = \mathcal{M}_{n,n-1}\left(u,x\right) \ldots \mathcal{M}_{21}\left(u,x\right)\,Z_1^{-1}\,\mathcal{K}_1\left(s,x\right) \times \\
    \times L_1\left(u+x-1,u_2\right) K_-(u)\,Z_1\,\mathcal{M}_{12}\left(u,x\right) \ldots \mathcal{M}_{n-1,n}\left(u,x\right) .
\end{multline*}
Multiplying out the left-hand side of \eqref{Rtr}, we get
\begin{multline*}
    \Omega\left(u,x\right) = Z_n^{-1}\,\mathcal{R}_{n,n-1}(u) \ldots \mathcal{R}_{21}(u)\,\mathcal{K}_1\left(s,x\right) L_n(u) \ldots L_2(u) \times \\
    \times L_1\left(u+x-1,u_2\right) \mathcal{R}_{12}\left(u,x\right) \ldots \mathcal{R}_{n-1,n}\left(u,x\right) K(u) L_1(u) \ldots L_{n-1}(u) Z_n \,.
\end{multline*} 
One can see that the key relation for the $\mathcal{R}$-operator \eqref{defR} is equivalent to
\begin{equation*}
    \mathcal{R}_{ij}(x) \, L_j(u+x-1, u_2) \, 
    L_i(u_1, u_2) =  
    L_j(u_1, u_2) \, L_i(u+x-1, u_2) \, 
    \mathcal{R}_{ij}(x) \,.
\end{equation*}
Applying this relation $n$ times, we see that $\Omega\left(u,x\right)$ takes the form
\begin{align*}
    \Omega\left(u,x\right) = \, & Z_n^{-1}\,\mathcal{R}_{n,n-1}(u) \ldots \mathcal{R}_{21}(u)\,\mathcal{K}_1\left(s,x\right)\,\mathcal{R}_{12}\left(u,x\right) \ldots \mathcal{R}_{n-1,n}\left(u,x\right) \\
    &
    \times L_n\left(u+x-1,u_2\right)\,L_{n-1}(u) \ldots L_1(u)\,K_-(u)\,L_1(u) \ldots L_{n-1}(u)\,Z_n \\
    = \, & \lambda_n^{-1}(x)\,Z_n^{-1}\,\Lambda_n(x)L_n\left(u+x-1,u_2\right)\,T_{n-1}(u)\,Z_n \,.
\end{align*}
In particular, the $(2,2)$ element of $\Omega\left(u,x\right)$ reads
\begin{multline*}
    \Omega_{22}\left(u,x\right) = \lambda_n^{-1}(x)\Bigl(- z_n\,\Lambda_n(x)\,(u+x+z_n\partial_n)\,B_{n-1}(u)+\Lambda_n(x)\,(z_n^2\partial_n+(x+s)z_n)\,B_{n-1}(u) \\
    - z_n\,\Lambda_n(x)\,\partial_n\,D_{n-1}(u)+\Lambda_n(x)\,(u-s-z_n\partial_n)\,D_{n-1}(u)\Bigr) .
\end{multline*} 
Acting on a function $\psi(z_1, \ldots, z_{n-1})$, we obtain
\begin{multline*}
    \Omega_{22}\left(u,x\right)\,\psi(z_1 \ldots z_{n-1}) = \lambda_n^{-1}(x)\Bigl(-(u+x)\, z_n\,\Lambda_n(x)\,B_{n-1}(u) + \\ 
    + (x+s)\,\Lambda_n(x)\,z_n\,B_{n-1}(u) + (u-s)\,\Lambda_n(x)\,D_{n-1}(u)\Bigr)\,\psi(z_1 \ldots z_{n-1}) .
\end{multline*} 
Now, substituting \eqref{1id} into the last expression and setting $x = u$, we get
\begin{equation}\label{D1}
    \Omega_{22}\left(u,u\right)\,\psi(z_1 \ldots z_{n-1}) = \lambda_n^{-1}(u)\,u_2^{-1}\,D_n(u)\,\Lambda_n(u)\,\psi(z_1 \ldots z_{n-1}) .
\end{equation}

On the other hand, setting $x = u$ and multiplying out the right-hand side of \eqref{Rtr}, we obtain
\begin{align*}
	&
	\Omega(u,u) \\
	&
	=
    \begin{pmatrix}
        -\imath^{n-1}\mathcal{R}_{n,n-1}(u+1)\ldots\mathcal{R}_{21}(u+1) & \ast \\
        0 & -\imath^{n-1}(u_1u_2)^{n-1}\mathcal{R}_{n,n-1}(u+1)\ldots\mathcal{R}_{21}(u-1)
    \end{pmatrix} \\
	&
	\times \overset{1}{\mathcal{M}}{}^{(2)}_-(u,u,1-s) \\
	&
    \times \begin{pmatrix}
        -\imath^{n-1}\mathcal{R}_{12}(u+1)\ldots\mathcal{R}_{n-1,n}(u+1) & \ast \\
        0 & -\imath^{n-1}(u_1u_2)^{n-1}\mathcal{R}_{12}(u+1)\ldots\mathcal{R}_{n-1,n}(u-1)
    \end{pmatrix} ,
\end{align*} 
where $\mathcal{R}(u\pm 1) := \mathcal{R}(u\pm 1,\bar{u})$ and the product of $\mathcal{K}_1$, $L_1$ operators and $K_-$ matrix gathered into the matrix $\mathcal{M}^{(2)}_-$~\eqref{M-2def} (recall that $\mathcal{K}_1(s,u)=\overset{1}{\mathbb{K}}_-(u; 1-s)$)
\begin{equation} \nonumber
	\overset{1}{\mathcal{M}}{}^{(2)}_-(u,u,1-s) =
	Z_1^{-1}\,\mathcal{K}_1\left(s,u\right)\,L_1\left(2u-1,u_2\right)\,K_-(u)\,Z_1 \,.
\end{equation}
Using the triangularity relation for~$\mathcal{M}^{(2)}_-$~\eqref{M-2} we deduce from the last relation that
\begin{align}
	\nonumber
    \Omega_{22}(u,u) = \, & (-1)^{n-1}\,2\imath\,\gamma^2(u-s)(u-g)(u+g-1)(u_1u_2)^{2n-2} \\
    \nonumber
    &
    \times \mathcal{R}_{n,n-1}(u+1)\ldots\mathcal{R}_{21}(u-1)\,\mathcal{K}(s,u-1)\,\mathcal{R}_{12}(u+1)\ldots\mathcal{R}_{n-1,n}(u-1) \\
    \label{D2}
    = \, & (-1)^{n-1}\,2\imath\,\gamma^2(u-s)(u-g)(u+g-1)(u_1u_2)^{2n-2}\lambda_n^{-1}(u-1,\bar{u})\,\Lambda_n(u-1,\bar{u}) .
\end{align}
Comparing \eqref{D1} with \eqref{D2} and using the explicit expression \eqref{nu} for $\lambda_n(u)$ we arrive at
\begin{equation}\label{D_act}
    D_n(u)\,\Lambda_n(u)\,\psi(z_1 \ldots z_{n-1}) = \alpha_n(u)\,\Lambda_n(u-1)\,\psi(z_1 \ldots z_{n-1}) \,,
\end{equation}  
where
\begin{equation} \label{alpha_n}
    \alpha_n(u) = \gamma\,(u-s)^{2n}\,(u+g-1) .
\end{equation}

Now, \eqref{BLambda} and \eqref{D_act} imply 
\begin{equation*}
    D_n(x_n)\,\Psi_{\bm{x}_n} = \alpha_n(x_n)\,\Psi_{\bm{x}_n^{-n}} \,,
\end{equation*} 
where
\begin{equation} \label{x_shift}
	\bm{x}_n^{\pm k} = (x_{1},\bar{x}_{1}, \ldots, x_k \pm 1, \bar{x}_k, \ldots, x_{n},\bar{x}_{n})
\end{equation}
Using the symmetry of $\Psi_{\bm{x}_n}$ under permutations and reflections, one easily obtains
\begin{align*}
    D_n(x_k)\,\Psi_{\bm{x}_n} = \alpha_n(x_k)\,\Psi_{\bm{x}_n^{-k}} \,, \qquad D_n(-x_k)\,\Psi_{\bm{x}_n} = \alpha_n(-x_k)\,\Psi_{\bm{x}_n^{+k}} \,.
\end{align*}    
In addition, the relation \eqref{T_n(1/2)} yields
\begin{align*}
    D_n\left(\tfrac{1}{2}\right)\,\Psi_{\bm{x}_n} = \alpha \, \Psi_{\bm{x}_n} \,,
\end{align*}    
where $\alpha = \gamma\bigl(g-\frac{1}{2}\bigr)\bigl(s-\frac{1}{2}\bigr)^{2n}$. Since $D_n(u)$ is a polynomial in $u$ of degree $2n$ and its action is known at $2n+1$ points, we can determine its action at any point using the Lagrange interpolation formula
\begin{multline*}
D_n(u)\Psi_{\bm{x}_n} 
= \alpha\left(\prod_{i=1}^{n}\frac{u^2-x_i^2}{\frac{1}{4}-x_i^2}\right)\Psi_{\bm{x}_n}
+ \sum_{i=1}^{n}\left(\prod_{k \neq i}\frac{u^2-x_k^2}{x_i^2-x_k^2}\right)\Biggl\{ 
\frac{u-\frac{1}{2}}{x_i-\frac{1}{2}}\,\frac{u+x_i}{2x_i}\,\alpha_n(x_i)\,\Psi_{\bm{x}_n^{-i}} \\
+ \frac{u-\frac{1}{2}}{x_i+\frac{1}{2}}\,\frac{u-x_i}{2x_i}\,\alpha_n(-x_i)\,\Psi_{\bm{x}_n^{+i}} 
\Biggr\} \,.
\end{multline*}
Finally, applying the last result and \eqref{tD} we compute the action of transfer matrix~$t_n(u)$ on~$\Psi_{\bm{x}_n}$
\begin{align}
\nonumber
t_n(u)\Psi_{\bm{x}_n} 
&= 2\alpha\left(\prod_{i=1}^{n}\frac{u^2-x_i^2}{\frac{1}{4}-x_i^2}\right)\Psi_{\bm{x}_n} + \\
\nonumber
&\quad + \left(u^2-\tfrac{1}{4}\right)\sum_{i=1}^{n}\left(\prod_{k \neq i}\frac{u^2-x_k^2}{x_i^2-x_k^2}\right)\Biggl\{ 
\frac{\alpha_n(x_i)}{x_i(x_i-\frac{1}{2})}\,\Psi_{\bm{x}_n^{-i}} +  \frac{\alpha_n(-x_i)}{x_i(x_i+\frac{1}{2})}\,\Psi_{\bm{x}_n^{+i}} 
\Biggr\} \,.
\end{align}

\subsection{Kernel Equation}

Now that we know the action of the transfer matrix $t_n(u)$ on $\Psi_{\bm{x}_n}$, we substitute the ansatz for the transfer matrix eigenfunctions~\eqref{int} into the spectral problem~\eqref{ep}.
Recalling that $\tilde{t}_n(u) = \frac{t_n(u) - 2\,\alpha}{2\gamma(u^{2}-\frac{1}{4})}$, we get the following
\begin{multline*}
    \tilde{\tau}^{\lambda}_{\bm{x}_{n-1}}(u)\Phi_{\bm{x}_{n-1}}^{\lambda} = \int \mathcal{D}\bm{y}_n \,\mu(\bm{y}_n)\,C(\lambda, \bm{x}_{n-1}|\,\bm{y}_{n})\Biggl\{\frac{2\alpha}{u^2-\frac{1}{4}}\left(\prod_{i=1}^{n}\frac{u^2-y_i^2}{\frac{1}{4}-y_i^2}-1\right) \Psi_{\bm{y}_n} + \\ 
    + \sum_{i=1}^{n}\left(\prod_{k \neq i}\frac{u^2-y_k^2}{y_i^2-y_k^2}\right) 
    \frac{\alpha_n(y_i)}{y_i(y_i-\frac{1}{2})}\,\Psi_{\bm{y}_n^{-i}} + \frac{\alpha_n(-y_i)}{y_i(y_i+\frac{1}{2})}\,\Psi_{\bm{y}_n^{+i}} \Biggr\} ,
\end{multline*} 
where $\tilde{\tau}^{\lambda}_{\bm{x}_{n-1}}(u) = 2\gamma\lambda \bigl(u^2-x_1^2\bigr) \ldots \bigl(u^2-x_{n-1}^2\bigr)$ and the shifts of spectral variables $\bm{y}_n^{\pm i}$ were defined in~\eqref{x_shift}.
Next, we shift the integration variables and apply the identities
\begin{equation*}
    \frac{\mu(\bm{y}_n^{-i})}{\mu(\bm{y}_n)} = \frac{y_i-1}{y_i}\prod\limits_{k \neq i} \frac{(y_i - 1)^2-y_k^2}{y_i^2-y_k^2} \,, \qquad \frac{\mu(\bm{y}_n^{+i})}{\mu(\bm{y}_n)} = \frac{y_i+1}{y_i}\prod\limits_{k \neq i} \frac{(y_i + 1)^2-y_k^2}{y_i^2-y_k^2} \,.
\end{equation*}    
Thus, we have
\begin{multline} \label{tauPhi}
    \tilde{\tau}^{\lambda}_{\bm{x}_{n-1}}(u)\Phi_{\bm{x}_{n-1}}^{\lambda} = \int \mathcal{D}\bm{y}_n \,\mu(\bm{y}_n)\,\Psi_{\bm{y}_n}\Biggl\{\frac{2\,\alpha}{u^2-\frac{1}{4}}\left(\prod_{i=1}^{n}\frac{u^2-y_i^2}{\frac{1}{4}-y_i^2}-1\right) C(\lambda, \bm{x}_{n-1}|\,\bm{y}_{n}) + \\ 
    + \sum_{i=1}^{n}\left(\prod_{k \neq i}\frac{u^2-y_k^2}{y_i^2-y_k^2}\right)
    \frac{\alpha_n(y_i+1)}{y_i(y_i+\frac{1}{2})}\,C(\lambda, \bm{x}_{n-1}|\,\bm{y}_{n}^{+i}) + \frac{\alpha_n(-y_i+1)}{y_i(y_i-\frac{1}{2})}\,C(\lambda, \bm{x}_{n-1}|\,\bm{y}_{n}^{-i}) \Biggr\} .
\end{multline}

Using the identity
\begin{equation*}
    \frac{1}{u^2-\frac{1}{4}}\left(\prod_{i=1}^{n}\frac{u^2-y_i^2}{\frac{1}{4}-y_i^2}-1\right) = \sum_{i=1}^{n}\left(\prod_{k \neq i}\frac{u^2-y_k^2}{y_i^2-y_k^2}\right)\frac{1}{\frac{1}{4}-y_i^2} \,,
\end{equation*}
the orthogonality relation \eqref{orth}
and setting $u = y_i$, we see that the relation~\eqref{tauPhi} holds iff
\begin{align*}
    \tilde{\tau}^{\lambda}_{\bm{x}_{n-1}}(y_i)\,C(\lambda, \bm{x}_{n-1}|\,\bm{y}_{n}) &= \frac{2\,\alpha}{\frac{1}{4}- y_i^2}\,C(\lambda, \bm{x}_{n-1}|\,\bm{y}_{n}) + \\
    &\quad +\frac{\alpha_n(y_i+1)}{y_i\,(y_i+\frac{1}{2})}\,C(\lambda, \bm{x}_{n-1}|\,\bm{y}_{n}^{+i}) + \frac{\alpha_n(-y_i+1)}{y_i\,(y_i-\frac{1}{2})}\,C(\lambda, \bm{x}_{n-1}|\,\bm{y}_{n}^{-i})
\end{align*}
for $i=1,\dots,n$.
Denoting by $\tau_n(u\,|\lambda,\bm{x}_{n-1})$ the eigenvalue of the transfer matrix $t_n(u)$
\begin{equation} \label{tau}
	\tau_n(u\,|\lambda,\bm{x}_{n-1}) = 2\gamma\lambda\,\bigl(u^2-\frac{1}{4}\bigr)\,(u^2-x_1^2) \ldots (u^2-x_{n-1}^2) + 2\alpha
\end{equation}
we arrive at
\begin{equation}\label{sov}
    \begin{split}
        y_i\,\tau_n(y_i\,|\lambda,\bm{x}_{n-1})\,C(\lambda, \bm{x}_{n-1}|\,\bm{y}_{n}) 
        &= \left(y_i-\tfrac{1}{2}\right)\,\alpha_n(y_i+1)\,C(\lambda, \bm{x}_{n-1}|\,\bm{y}_{n}^{+i}) \\
        &\quad + \left(y_i+\tfrac{1}{2}\right)\,\alpha_n(-y_i+1)\,C(\lambda, \bm{x}_{n-1}|\,\bm{y}_{n}^{-i}) \,.
    \end{split}
\end{equation}
This multiparametric system of equations admits separation of variables. Namely, assuming the factorization property
\begin{equation*}
    C(\lambda, \bm{x}_{n-1}|\,\bm{y}_{n}) = \prod\limits_{i=1}^{n} c(\lambda, \bm{x}_{n-1}|\,y_i, \bar{y}_i) \,,
\end{equation*} 
the functions $c(\lambda, \bm{x}_{n-1}|y_i, \bar{y}_i)$ satisfy the one-parameter Baxter equation
\begin{equation}\label{sov2}
    \begin{split}
        y\,\tau_n(y\,|\lambda,\bm{x}_{n-1})\,c(\lambda, \bm{x}_{n-1}|y, \bar{y}) 
        &= \left(y-\tfrac{1}{2}\right)\,\alpha_n(y+1)\,c(\lambda, \bm{x}_{n-1}|\,y+1, \bar{y}) \\
        &\quad + \left(y+\tfrac{1}{2}\right)\,\alpha_n(-y+1)\,c(\lambda, \bm{x}_{n-1}|\,y-1, \bar{y}) \,.
    \end{split} 
\end{equation}
Using the same argument, one can derive a similar difference equation with respect to the anti-holomorphic variable
\begin{equation}\label{sov2bar}
    \begin{split}
        \bar{y}\,\bar{\tau}_n(\bar{y}\,|\bar{\lambda},\bm{x}_{n-1})\,c(\lambda, \bm{x}_{n-1}|y, \bar{y}) 
        &= \left(\bar{y}-\tfrac{1}{2}\right)\,\bar{\alpha}_n(\bar{y}+1)\,c(\lambda, \bm{x}_{n-1}|\,y, \bar{y}+1) \\
        &\quad + \left(\bar{y}+\tfrac{1}{2}\right)\,\bar{\alpha}_n(-\bar{y}+1)\,c(\lambda, \bm{x}_{n-1}|\,y, \bar{y}-1) \,,
    \end{split} 
\end{equation}
where
\begin{align*}
	&
	\bar{\tau}_n(\bar{u}\,|\bar{\lambda},\bm{x}_{n-1}) = 2\bar{\gamma}\bar{\lambda}\,\bigl(\bar{u}^2-\tfrac{1}{4}\bigr)\,(\bar{u}^2-\bar{x}_1^2) \ldots (\bar{u}^2-\bar{x}_{n-1}^2) + 2\bar{\alpha} \,, \\
	&
	\bar{\alpha}_n(\bar{u}) = \bar{\gamma}(\bar{u}-\bar{s})^{2n}\,(\bar{u}+\bar{g}-1) \,, \quad
	\bar{\alpha} = \bar{\gamma}(\bar{g}-\tfrac{1}{2})(\bar{s}-\tfrac{1}{2})^{2n} \,.
\end{align*}

\subsection{Relation to Baxter $\mathbb{Q}$-operator}

Recall the definiton of the $Q_+$-operator
\begin{align*}
	Q_+(u,\bar{u}):=\mathbb{Q}(u,\bar{u},1-s,1-\bar{s}) \,.
\end{align*}
According to \eqref{qtcom}, \eqref{qqcom} \eqref{baxterqeq1} these operators form a commutative family, commute with the transfer matrix $t_n(u)$ and satisfy the following Baxter equation
\begin{align*}
	Q_+(u,\bar{u})\,t_n(u)=\delta_1(u)Q_+(u+1,\bar{u})+\delta_2(u)Q_+(u-1,\bar{u}) \,,
\end{align*}
where 
\begin{align*}
	\delta_1(u)&=\frac{(-1)^n\imath(u-\frac{1}{2})(u-s)}{2u(u-s+1)} \,, \\
	\delta_2(u)&=\frac{(-1)^n\,\imath\,4\gamma^2(u-\frac{1}{2})(u-s)(u-g)(u+g-1)((u-s)(u+s-1))^{2n}}{2u(u-s+1)} \,.
\end{align*}
and the similar equation with respect to the antiholomorphic variable.
It is not hard to check that if one defines
\begin{align} \label{Qren}
Q(u,\bar{u}) = \nu(u,\bar{u}) \, Q_+(u,\bar{u}) \,, \quad
\nu(u,\bar{u}) := \left[ \frac{\imath\,(-1)^n}{2\gamma} \right]^{u-s} \frac{\mathbf{\Gamma}(u + 2 - s)}{\mathbf{\Gamma}(u + g)
	\,\mathbf{\Gamma}^{2n+1}(u + 1 - s)} \,,
\end{align}
then the Baxter equations take the form
\begin{equation}\label{reBaxt}
	\begin{aligned}
		u\,Q(u,\bar{u})\,t_n(u) 
        &= \left(u-\tfrac{1}{2}\right)\,\alpha_n(u+1)\,Q(u+1,\bar{u})+\left(u+\tfrac{1}{2}\right)\,\alpha_n(-u+1)\,Q(u-1,\bar{u}) \,, \\
		\bar{u}\,Q(u,\bar{u})\,\bar{t}_n(u)
        &= \left(\bar{u}-\tfrac{1}{2}\right)\,\bar{\alpha}_n(\bar{u}+1)\,Q(u,\bar{u}+1)+\left(\bar{u}+\tfrac{1}{2}\right)\,\bar{\alpha}_n(-\bar{u}+1)\,Q(u,\bar{u}-1) \,.	
	\end{aligned}
\end{equation}
From the commutativity properties it follows that the Baxter $Q$-operator and the transfer matrix $t_n(u)$ share the common set of the eigenfunctions.
\begin{align*}
	Q(u)\,\Phi^{\lambda}_{\bm{x_{n-1}}}(\bm{z_n})&=q^{\lambda}_{\bm{x_{n-1}}}(u)\,\Phi^{\lambda}_{\bm{x_{n-1}}}(\bm{z_n}) \,,\\
	t_n(u)\,\Phi^{\lambda}_{\bm{x_{n-1}}}(\bm{z_n})&=\tau_n(u\,|\lambda,\bm{x}_{n-1})\,\Phi^{\lambda}_{\bm{x_{n-1}}}(\bm{z_n})\,.
\end{align*}
Thus acting by left hand sides and right hand sides of equations~\eqref{reBaxt} on function~$\Phi^{\lambda}_{\bm{x_{n-1}}}$ we conclude that the eigenvalues of the $Q$-operator satisfy Baxter equations with the transfer matrix~$t_n(u)$ replaced by its eigenvalue $\tau_n(u\,|\lambda,\bm{x}_{n-1})$, and these Baxter equations
coincide with one-dimensional difference equations~\eqref{sov2}, \eqref{sov2bar} obtained for the kernel of SoV transform.

Therefore, assuming the property of factorization, solution of the equation \eqref{sov} can be written as
\begin{equation*}
	C(\lambda, \bm{x}_{n-1}|\,\bm{y}_{n}) = \prod\limits_{i=1}^{n} q^{\lambda}_{\bm{x_{n-1}}}(y_i,\bar{y}_i) \,.
\end{equation*}
Since the transfer matrix eigenfunctions are orthogonal to each other for different sets of quantum numbers with respect to the scalar product \eqref{sp}, using the orthogonality of $B$-element eigenfunctions \eqref{orth} we find that the eigenfunctions of $Q$-operator satisfy the orthogonality relation
\begin{equation*}
	\int \mathcal{D}\bm{y}_n \prod\limits_{i=1}^{n} q^{\lambda\,\ast}_{\bm{x_{n-1}}}(y_i,\bar{y}_i)\,q^{\lambda'}_{\bm{x'_{n-1}}}(y_i,\bar{y}_i) \sim \delta^{(2)}(\lambda-\lambda')\,\delta^{(2)}(\bm{x}_{n-1},\bm{x'}_{n-1}) \,.
\end{equation*}

\section{Conclusion} \label{sect:conclusion}

Let us gather the main results.
The general two-parameter $\mathbb{Q}$-operator in both cases of one-boundary and two boundary spin chain is constructed from $\mathbb{R}$-operator~\eqref{RYB} and $\mathbb{K}_{\pm}$-operators~\eqref{Kvs}, \eqref{Kvs+} that satisfy the Yang-Baxter and reflection equations:
\begin{align*}
	&\mathbb{R}_{12}(u-v)\,\mathbb{R}_{13}(u)\,\mathbb{R}_{23}(v) = \mathbb{R}_{23}(v)\,\mathbb{R}_{13}(u)\,\mathbb{R}_{12}(u-v) \,, \\[0.5em]
	&\mathbb{R}_{12}(u-v)\,\overset{1}{\mathbb{K}}_-(u,s_1)\,\mathbb{R}_{12}(u+v)\,\overset{2}{\mathbb{K}}_-(v,s_2) \\
	&\qquad = \overset{2}{\mathbb{K}}_-(v,s_2)\,\mathbb{R}_{12}(u+v)\,\overset{1}{\mathbb{K}}_-(u,s_1)\,\mathbb{R}_{12}(u-v) \,, \\[0.5em]
	&\mathbb{R}_{12}(v-u)\,\overset{1}{\mathbb{K}}_+(u,s_1)\,\mathbb{R}_{12}(-u-v-2)\,\overset{2}{\mathbb{K}}_+(v,s_2) \\
	&\qquad = \overset{2}{\mathbb{K}}_+(v,s_2)\,\mathbb{R}_{12}(-u-v-2)\,\overset{1}{\mathbb{K}}_+(u,s_1)\,\mathbb{R}_{12}(v-u) \,.
\end{align*}
They are integral operators given by the explicit expressions
\begin{align*}
	\mathbb{R}_{12}(u) &= \mathbb{P}_{12}\,\check{R}_{12}(u) = \mathbb{P}_{12}\,[z_{12}]^{u+1-s_1-s_2}\,[\hat{p}_1]^{u+s_2-s_1}\,[\hat{p}_2]^{u+s_1-s_2}\,[z_{12}]^{u+s_1+s_2-1} , \\[0.6em]
	\mathbb{K}_-(v,s) &= [z+\gamma]^{v-s+g}\,[z-\gamma]^{v-s+1-g}\,[\hat{p}]^{2v}\,[z+\gamma]^{v+s-g}\,[z-\gamma]^{v+s+g-1} , \\[0.6em]
	\mathbb{K}_+(v,s) &= \tilde{\mathbb{K}}_-(-1-v,s) \,.
\end{align*}
The operator $\mathbb{R}_{12}(u)$ acts in the tensor product of two $\mathrm{L}^2(\mathbb{C})$ spaces of principal series representations with spins $(s_1,\bar{s}_1)$ and $(s_2,\bar{s}_2)$, where the spin parameters are of the form~\eqref{sparam}.
By $\mathbb{P}_{12}$ we denote the permutation operator~\eqref{perm}, $z_{12} :=z_1-z_2$ and the definitions of double power $[z]^\alpha$ and integral operator $[\hat{p}]^\alpha$ are given in~\eqref{power}, \eqref{d}.
Operators $\mathbb{K}_\pm$ act in the $\mathrm{L}^2(\mathbb{C})$ space, and the indices ``$1$'' and ``$2$'' above the letter $\mathbb{K}$ show in which copy of this space the $\mathbb{K}$-operator acts nontrivially.
In addition to the spectral parameter $v$ and spin parameter $s$ the operator $\mathbb{K}_-(v,s)$ is determined by the parameters $\gamma, \bar{\gamma}$, $(g,\bar{g})$ of boundary matrices $K_-$, $\bar{K}_-$~\eqref{K-}, where $\gamma, \bar{\gamma}$ are arbitrary complex numbers conjugated to each other and $(g,\bar{g})$ have the same form~\eqref{gparam} as spins.
The first and second reflection equations differ by substitution $u\to-u-1,\,v\to-v-1$, therefore the solution
$\mathbb{K}_+$ of the second one is expressed in terms of the solution of the first one. Namely, we express $\mathbb{K}_+$ in terms of the operator $\tilde{\mathbb{K}}_-$ which differs from $\mathbb{K}_-$ by replacing $\gamma$ and $(g,\bar{g})$ with similar parameters $\tilde{\gamma}$, $(\tilde{g},\bar{\tilde{g}})$ of boundary matrices $K_+, \bar{K}_+$~\eqref{K+}.

The operators $\mathbb{K}_{\pm}$ also satisfy reflection equations with $L$-operator~\eqref{L} and $K_\pm$-matrices
\begin{align*} 
	&L(u-v)\,K_-(u)\,L(u+v)\,\mathbb{K}_-(v,s) = 
	\mathbb{K}_-(v,s)\,L(u+v)\,K_-(u)\,L(u-v) \,,\\
	&L(v-u+1)\,K_+(u)\,L(-u-v-1)\,\mathbb{K}_+(v,s) = 
	\mathbb{K}_+(v,s)\,L(-u-v-1)\,K_+(u)\,L(v-u+1) \,.
\end{align*}
All mentioned identities play the key role in the proof of pairwise commutativity of $\mathbb{Q}$-operators and commutativity of $\mathbb{Q}$-operator and transfer-matrix.

The transfer matrices $t_n(u)$ and $\hat{t}_n(u)$ for the one-boundary and two-boundary chains, respectively, are defined as the traces of corresponding monodromy matrices~\eqref{Top}, \eqref{Top1}
\begin{align*}
	&
	T_n(u) = L_n(u) \cdots L_1(u) K_-(u) L_1(u) \cdots L_n(u) = \begin{pmatrix}
		A_n(u) & B_n(u) \\
		C_n(u) & D_n(u)
	\end{pmatrix}, \;\;
	\hat{T}_n(u) = K_+(u)\,T_n(u) \,, \\[0.6em]
	&
	t_n(u) = \tr_{\mathbb{C}^2} T_n(u) \,, \quad \hat{t}_n(u) = \tr_{\mathbb{C}^2} \hat{T}_n(u) \,,
\end{align*}
they act in the Hilbert space $H=V_1\otimes\cdots\otimes V_n$, where $V_k=\mathrm{L}^2(\mathbb{C})$.
The spin chains are homogeneous, that is at each site of these chains we take a copy of one and the same representation with spins $(s,\bar{s})$.
We also define the infinite-dimensional counterparts of $t_n(u)$, $\hat{t}_n(u)$ -- the general transfer matrices
\begin{align*}
	&\mathbbm{t}_n(u,s_0) = \operatorname{tr}_0[\mathbb{R}_{n0}(u)\cdots \mathbb{R}_{10}(u)\overset{0}{\mathbb{K}}_-(u,s_0)\mathbb{R}_{10}(u)\cdots \mathbb{R}_{n0}(u)] \,, \\
	&\hat{\mathbbm{t}}_n(u,s_0) = \operatorname{tr}_0[\overset{0}{\mathbb{K}}_+(u,s_0)\mathbb{R}_{n0}(u)\cdots \mathbb{R}_{10}(u)\overset{0}{\mathbb{K}}_-(u,s_0)\mathbb{R}_{10}(u)\cdots \mathbb{R}_{n0}(u)] \,,
\end{align*}
where the trace of an integral operator is defined in~\eqref{trInt}. Each operator $\mathbb{R}_{k0}$ acts nontrivially in $V_k$ and the infinite-dimensional auxiliary space $V_0$ of principal series representation with spins $(s_0,\bar{s}_0)$, over which the trace is taken.

The $\mathbb{Q}$-operators for the one-boundary and two-boundary chain, defined as 
\begin{align*}
	\mathbb{Q}_n(x_+;x_-) := \mathbbm{t}_n(x(x_+,x_-), s_0(x_+,x_-)) \,, \quad
	\hat{\mathbb{Q}}_n(x_+;x_-) := \hat{\mathbbm{t}}_n(x(x_+,x_-), s_0(x_+,x_-)) 
\end{align*}
with $x_+ = x+s_0$ and $x_- = x-s_0+1$, possess all characteristic properties of the two-parameter Baxter $\mathbb{Q}$-operator
\begin{align*}
	\mathbb{Q}_n(x_+;x_-)\,t_n(u) &= t_n(u)\,\mathbb{Q}_n(x_+;x_-) \,, \\
	\hat{\mathbb{Q}}_n(x_+;x_-)\,\hat{t}_n(u) &= \hat{t}_n(u)\,\hat{\mathbb{Q}}_n(x_+;x_-) \,, \\[1em]
	\mathbb{Q}_n(x_+;x_-)\,\mathbb{Q}_n(y_+;y_-) &= \mathbb{Q}_n(y_+;y_-)\,\mathbb{Q}_n(x_+;x_-) \,, \\
	\hat{\mathbb{Q}}_n(x_+;x_-)\,\hat{\mathbb{Q}}_n(y_+;y_-) &= \hat{\mathbb{Q}}_n(y_+;y_-)\,\hat{\mathbb{Q}}_n(x_+;x_-) \,, \\[1em]
	\mathbb{Q}_n(u,\bar{u};x_-)\,t_n(u) &= \Delta_1(u,x_-)\,\mathbb{Q}_n(u + 1,\bar{u};x_-) + \Delta_2(u,x_-)\,\mathbb{Q}_n(u - 1,\bar{u};x_-) \,, \\
	t_n(u)\,\mathbb{Q}_n(x_+;u,\bar{u}) &= \Delta_1(u,x_+)\,\mathbb{Q}_n(x_+;u + 1,\bar{u}) + \Delta_2(u,x_+)\,\mathbb{Q}_n(x_+;u - 1,\bar{u}) \,, \\[1em]
	\hat{\mathbb{Q}}_n(u,\bar{u};x_-)\,\hat{t}_n(u) &= \hat{\Delta}_1(u,x_-)\,\hat{\mathbb{Q}}_n(u + 1,\bar{u};x_-) + \hat{\Delta}_2(u,x_-)\,\hat{\mathbb{Q}}_n(u - 1,\bar{u};x_-) \,, \\
	\hat{t}_n(u)\,\hat{\mathbb{Q}}_n(x_+;u,\bar{u}) &= \hat{\Delta}_1(u,x_+)\,\hat{\mathbb{Q}}_n(x_+;u + 1,\bar{u}) + \hat{\Delta}_2(u,x_+)\,\hat{\mathbb{Q}}_n(x_+;u - 1,\bar{u}) \,,
\end{align*}
where the coefficients $\Delta_1, \Delta_2, \hat{\Delta}_1, \hat{\Delta}_2$ are defined in~\eqref{DeltaCoeff}.
The relations between $\bar{t}_n(u)$ and $\mathbb{Q}_n(x_+;x_-)$ are similar to their holomorphic analogues with $t_n(u)$, and the same holds in the two-boundary case.

At specific points $u_+ = s$ and $u_- = 1-s$ two-parameter $\mathbb{Q}$-operators simplify to the one-parameter Baxter operators
\begin{align*}
	Q_+(u) &:= \mathbb{Q}(u;1-s) \,, \quad Q_-(u) := \mathbb{Q}(s;u) \,, \\
	\hat{Q}_+(u) &:= \hat{\mathbb{Q}}(u;1-s) \,, \quad \hat{Q}_-(u) := \hat{\mathbb{Q}}(s;u) \,.
\end{align*}
Each of them is expressed in terms of one of the operators $\mathcal{R}^{(1)}$ and $\mathcal{R}^{(2)}$~\eqref{R12def} from the factorization formula~\eqref{Rfactor}, which have simpler structure than $\mathbb{R}$-operator,
\begin{align}
	& \nonumber
	\hat{Q}_+(u)= \tr_{0}\Bigl[\tilde{\mathcal{K}}_0(u+1, s-1)\,\mathcal{R}^{(2)}_{0n}(u)\ldots\mathcal{R}^{(2)}_{21}(u)\,\mathcal{K}_1(s,u)\,\mathcal{R}^{(2)}_{12}(u)\ldots\mathcal{R}^{(2)}_{n0}(u)\Bigr]\,, \\
	& \nonumber
	\hat{Q}_-(u)= \tr_{0}\Bigl[\tilde{\mathcal{K}}_0(s+1, -u)\,\mathcal{R}^{(1)}_{0n}(u)\ldots\mathcal{R}^{(1)}_{21}(u)\,\mathcal{K}_1(1-u,s)\,\mathcal{R}^{(1)}_{12}(u)\ldots\mathcal{R}^{(1)}_{n0}(u)\Bigr] \,.
\end{align}
The $\mathcal{K}$ and $\tilde{\mathcal{K}}$ operators are given by~\eqref{Kop-d}, \eqref{Kop-d-tilde}, they differ by substitution $\gamma\to\tilde{\gamma}$, $(g,\bar{g})\to (\tilde{g},\bar{\tilde{g}})$.
For the one-boundary chain the expressions are similar, one just needs to remove $\tilde{\mathcal{K}}$-operators.
Moreover, in that case the trace in the expression for operator~$Q_+$ can be calculated explicitly, and the latter takes the form~\eqref{Q+expl}.
Baxter $\mathbb{Q}$-operators admit the following factorization
\begin{equation*}
	\begin{aligned}
		\mathbb{Q}(u_+;u_-) &= \frac{[u_+-s+1]\,[u_-+s]}{[1-2s]\,[u_++u_-]}\,Q_-(u_-)\,Q_+(u_+) \,, \\
		\hat{\mathbb{Q}}(u_+;u_-) &= \frac{[u_+-s+1]\,[u_-+s]\,[1-s-\tilde{g}]\,[\tilde{g}-s]\,[2\tilde{\gamma}]^2}{[1-2s]\,[u_++u_-]}\,\hat{Q}_-(u_-)\,\hat{Q}_+(u_+) \,.
	\end{aligned}
\end{equation*}     
where by $[x]$ we mean the product $x\bar{x}$.

The eigenfunctions of the one-boundary model are defined by the spectral problem of the form
\begin{equation*} 
	\begin{aligned}
		\tilde{t}(u)\Phi_{\bm{x}_{n-1}}^{\lambda}(\bm{z}_{n}) 
		&= \lambda \bigl(u^2-x_1^2\bigr) \ldots \bigl(u^2-x_{n-1}^2\bigr) \Phi_{\bm{x}_{n-1}}^{\lambda}(\bm{z}_{n}) \,, \\
		\bar{\tilde{t}}(\bar{u})\Phi_{\bm{x}_{n-1}}^{\lambda}(\bm{z}_{n}) 
		&= \bar{\lambda} \bigl(\bar{u}^2-\bar{x}_1^2\bigr) \ldots \bigl(\bar{u}^2-\bar{x}_{n-1}^2\bigr) \Phi_{\bm{x}_{n-1}}^{\lambda}(\bm{z}_{n}) \,,
	\end{aligned}
\end{equation*}
where
\begin{equation*} 
	\tilde{t}(u) = \frac{t_n(u) - 2\gamma\bigl(g-\frac{1}{2}\bigr)\bigl(s-\frac{1}{2}\bigr)^{2n}}{2\gamma \bigl(u^{2} - \frac{1}{4}\bigr)} = u^{2n-2}H_1+u^{2n-4}H_2+\ldots+H_n
\end{equation*}
and $\{H_i\}$ are the commuting Hamiltonians of the one-boundary spin chain.
We use the following compact notation for tuples of variables
\begin{align*}
	\bm{z}_n = (z_1,\bar{z}_1, \ldots, z_n, \bar{z}_n) \,, \qquad  
	\bm{x}_n = (x_1,\bar{x}_1, \ldots, x_n, \bar{x}_n) \,.
\end{align*}

Following the SoV approach, we look for the eigenfunctions in the form
\begin{equation*}
	\Phi_{\bm{x}_{n-1}}^{\lambda}(\bm{z}_n) = \int \mathcal{D}\bm{y}_n \, \mu(\bm{y}_n) \Psi_{\bm{y}_n}(\bm{z}_n) \, C(\lambda, \bm{x}_{n-1}|\,\bm{y}_{n}) \,,
\end{equation*} 
where $\Psi_{\bm{y}_n}(\bm{z}_n)$ are eigenfunctions~\eqref{BLambda} of the operator $B_n(u)$.
The range of values for the integration variables and the integration measure are given by~\eqref{y} and \eqref{measure}, respectively. The integral is defined in~\eqref{intx}.

For the kernel $C(\lambda, \bm{x}_{n-1}|\,\bm{y}_{n})$ we obtain the Baxter equation
\begin{equation*}
	\begin{split}
		y_i\,\tau_n(y_i\,|\lambda,\bm{x}_{n-1})\,C(\lambda, \bm{x}_{n-1}|\,\bm{y}_{n}) 
		&= \left(y_i-\tfrac{1}{2}\right)\,\alpha_n(y_i+1)\,C(\lambda, \bm{x}_{n-1}|\,\bm{y}_{n}^{+i}) \\
		&\quad + \left(y_i+\tfrac{1}{2}\right)\,\alpha_n(-y_i+1)\,C(\lambda, \bm{x}_{n-1}|\,\bm{y}_{n}^{-i}) \,,
	\end{split}
\end{equation*}
where we use the notation $\bm{y}_{n}^{+i}=(y_1,\bar{y}_1, \ldots, y_{i-1}, \bar{y}_{i-1}, y_i+1, \bar{y}_i, y_{i+1}, \bar{y}_{i+1}, \ldots, y_n, \bar{y}_n)$.
The function $\tau_n(u\,|\lambda,\bm{x}_{n-1})$~\eqref{tau} is the eigenvalue of $t_n(u)$, and the coefficient $\alpha_n$ is given by~\eqref{alpha_n}.
Assuming the factorization of the kernel,
\begin{equation*}
	C(\lambda, \bm{x}_{n-1}|\,\bm{y}_{n}) = \prod\limits_{i=1}^{n} c(\lambda, \bm{x}_{n-1}|\,y_i, \bar{y}_i) \,,
\end{equation*} 
we obtain the one-dimensional Baxter equations:
\begin{align*}
	y\,\tau_n(y\,|\lambda,\bm{x}_{n-1})\,c(\lambda, \bm{x}_{n-1}|y, \bar{y}) 
	&= \left(y-\tfrac{1}{2}\right)\,\alpha_n(y+1)\,c(\lambda, \bm{x}_{n-1}|\,y+1, \bar{y}) \\
	&\quad + \left(y+\tfrac{1}{2}\right)\,\alpha_n(-y+1)\,c(\lambda, \bm{x}_{n-1}|\,y-1, \bar{y}) \,, \\[1em]
	\bar{y}\,\bar{\tau}_n(\bar{y}\,|\bar{\lambda},\bm{x}_{n-1})\,c(\lambda, \bm{x}_{n-1}|y, \bar{y}) 
	&= \left(\bar{y}-\tfrac{1}{2}\right)\,\bar{\alpha}_n(\bar{y}+1)\,c(\lambda, \bm{x}_{n-1}|\,y, \bar{y}+1) \\
	&\quad + \left(\bar{y}+\tfrac{1}{2}\right)\,\bar{\alpha}_n(-\bar{y}+1)\,c(\lambda, \bm{x}_{n-1}|\,y, \bar{y}-1) \,.
\end{align*}
Comparing this result with the Baxter equation for an operator $Q(u)=\nu(u)\,Q_+(u)$~\eqref{Qren} one can see that the kernel can be written as 
\begin{equation*}
	C(\lambda, \bm{x}_{n-1}|\,\bm{y}_{n}) = \prod\limits_{i=1}^{n} q^{\lambda}_{\bm{x_{n-1}}}(y_i,\bar{y}_i) \,.
\end{equation*} 
where $q^{\lambda}_{\bm{x_{n-1}}}(u,\bar{u})$ are the eigenvalues of the operator $Q(u)$.

The Baxter $Q$-operator for the open XXX spin chain with spin $\frac{1}{2}$ and diagonal $K$-matrices was constructed in \cite{Frassek_2015}, these results were generalized to arbitrary compact spin in~\cite{Frassek2022}. As shown in \cite{DM09,CDS}, the general $\mathbb{R}$-operator acting in the continuous series representations can be reduced to finite-dimensional $SL(2,\mathbb{C})$ representations. A natural question is whether our formula for the $\mathbb{Q}$-operator can be extended to finite-dimensional representation spaces.
And if such reduction exists, an interesting problem would be to relate it to the results of the mentioned works.

The SoV approach for the B-type Toda chain was originally developed in~\cite{IS}, where the action of the two-boundary transfer matrix on the eigenfunctions of the one-boundary $B$-operator was derived via the Mellin--Barnes integral representation. However, their analysis was restricted to triangular boundary $K$-matrices. In contrast, because the $K$-matrices in our model are non-triangular, the action of the two-boundary transfer matrix on the eigenfunction $\Psi_{\bm{x}_{n}}(\bm{z}_n)$ induces double shifts $\Psi_{\bm{x}_{n}^{+i,+k}}(\bm{z}_n)$, rendering the corresponding kernel equation considerably more involved. The natural approach is to perform gauge transformation to the two-boundary transfer matrix $\hat{t}_n(u)$ making the $K_+$-matrix lower-triangular and then expand the transfer matrix eigenfunctions over the set of eigenfunctions of  the transformed $B$-element.

As shown in~\cite{KL3}, the solution to the kernel equation for the periodic Toda chain can be expressed in terms of ratios of infinite-dimensional determinants. It would be of great interest to construct an explicit solution to the more intricate kernel equation obtained in this work.

In \cite{DKM, DKM2} the analytical properties of the Baxter $Q$-operator eigenvalues were established for the periodic $SL(2,\mathbb{C})$ spin chain and the open $SL(2,\mathbb{R})$ spin chain, respectively. Because the transfer matrix eigenfunctions in the Separation of Variables representation factorize into products of $Q$-operator eigenvalues, these results directly determine the analytical properties of the eigenfunctions themselves. Extending this analysis to the $Q$-operator eigenvalues for the model considered in this work presents a compelling direction for future research.

Another challenging direction for future study is generalizing the obtained results to higher-rank $SL(N,\mathbb{C})$ spin chains corresponding to principal series representations of the group.
In these models, the $L$-operator is expressed in terms of representation generators, analogously to the $N=2$ case~\cite{DM06}.
The general $SL(N,\mathbb{C})$-invariant $\mathbb{R}$-operator acting on the tensor product of two principal series representations 
and satisfying the Yang-Baxter equation was constructed in~\cite{DM06,DM09} using the factorization approach similar to~\eqref{Rfactor}, \eqref{defR}.
The SoV method was applied to the periodic $SL(3,\mathbb{C})$ chain in~\cite{DV19}.
Although the generic $\mathbb{C}$-valued matrix solutions of reflection equations with Yang's $R$-matrix in $\mathbb{C}^N\otimes\mathbb{C}^N$ are known~\cite{MNopen},
the infinite-dimensional $\mathbb{K}_\pm$-operators corresponding to these generic $K_\pm$-matrices have not yet been constructed.
To date, some algebraic expressions for $\mathbb{K}$-operators have been obtained in the case of diagonal $K$-matrices~\cite{Frassek2026}.
The method used to construct the $\mathbb{K}_-$-operator in~\cite{ABDV}
is based on permutations of the $L$-operator's parameters as well as the aforementioned factorization approach, and it appears extendable to higher ranks by analogy with solutions of the Yang-Baxter equation.
Such a generalization would represent a significant step in the study of higher-rank spin chains with non-diagonal integrable boundaries.

The $SL(N,\mathbb{C})$-invariant $\mathbb{R}$-operator is realized as an integral operator, and the $\mathbb{K}_\pm$-operators can presumably be represented in a similar way, as in the $SL(2,\mathbb{C})$ model.
The diagrammatic technique used to manipulate integral expressions is not available in the case of $N > 2$.
In Section~\ref{sect:algDer} of this work, we extend the algebraic approach developed in~\cite{ADV2,ADV1} by introducing formulas involving traces of integral operators. 
A generalization of this algebraic approach to higher-rank cases appears to be a particularly promising tool for calculations involving integral operators such as the $\mathbb{R}$- and $\mathbb{K}_\pm$-operators. By analogy with spin chains without boundary interaction~\cite{DM06,VDK16}, these operators are expected to serve as basic building blocks in the construction of Baxter $Q$-operators and eigenfunctions.

\section*{Acknowledgements}

We are grateful to S.~Derkachov and N.~Belousov for productive discussions.
The work of P.~Antonenko was supported by the 
Theoretical Physics and Mathematics Advancement Foundation BASIS and by the Ministry of Science and Higher Education of the Russian Federation, agreement 075-15-2025-344 dated 29/04/2025 for Saint Petersburg Leonhard Euler International Mathematical Institute at PDMI RAS.

\appendix

\section{The chain rule and the star-triangle relation}
\label{App-DiagTech}
In this section, we formulate the integral identities used throughout this paper and describe the corresponding diagrammatic technique.

\begin{figure}[H]
    \centering
    \begin{tikzpicture}[thick, line cap = round, scale=1]
        \draw[->-] (0,0) to node[midway,above]{$a$} (2,0);
        \node[left] at (0,0) {$z$}; \node[right] at (2,0) {$w$};
    \end{tikzpicture}
    \caption{Diagrammatic representation of $[z-w]^{-a}$.}  \label{propag}
\end{figure}
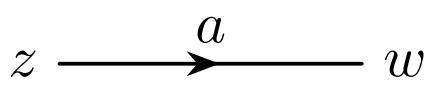

The directed line (Figure~\ref{propag}) corresponds to the complex double power
\begin{equation*}
    \frac{1}{[z-w]^a} \equiv \frac{1}{(z-w)^a (\bar{z}-\bar{w})^{\bar a}} =
    \frac{(\bar{z}-\bar{w})^{a-\bar a}}{|z-w|^{2a}} \,,
\end{equation*}
where we always assume $a-\bar a \in \mathbb{Z}$, ensuring that this function is well defined.
Reversing the arrow introduces a sign factor:
\begin{equation*}
    \frac{1}{[z-w]^a} =
    \frac{(-1)^{a-\bar a}}{[w-z]^{a}} = \frac{[-1]^{a}}{[w-z]^{a}} \,.
\end{equation*}

We now state two useful integral identities: the chain rule and the star-triangle relation. These relations can be written in several equivalent forms and can also be represented diagrammatically.
Detailed proofs can be found in \cite[Appendix~A]{DKM}.

\begin{figure}[H]
    \centering
    \begin{tikzpicture}[thick, line cap = round, scale=1.2]
        % Diagram for chain relation
        \def\rad{0.08} % radius of the circle of bold vertex
        \draw[->-] (0,0) to node[midway,above]{$a$} (2,0);
        \draw[->-] (2,0) to node[midway,above]{$b$} (4,0);
        \draw[fill = black] (2,0) circle (\rad);
        \node at (6,0) {$=\pi (-1)^{c-\bar{c}} \,\mathbf{\Gamma}^{-1}(a,b,c)$};
        \draw[->-] (8,0) to node[midway,above]{$a+b-1$} (11,0);
    \end{tikzpicture}
    \caption{The chain relation, $a+b+c=2$.}
    \label{Rules1}
\end{figure}
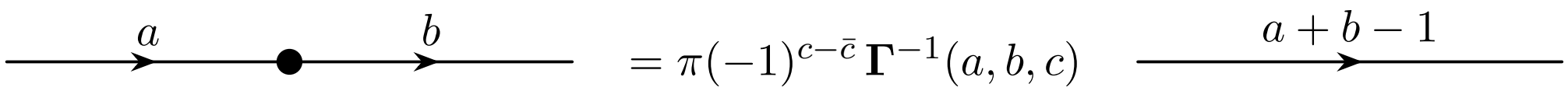

The first identity is the \textit{chain relation}, depicted in Figure~\ref{Rules1}.
The bold vertex denotes integration over the complex plane.

In integral form, this identity reads
\begin{align}\label{Chain}
    \int \mathrm{d}^2 w \frac{1}{[z-w]^a [w-z_0]^{b}} =
    \frac{\pi [-1]^c}{\mathbf{\Gamma}(a,b,c)}
    \frac{1}{[z-z_0]^{a+b-1}} \,.
\end{align}
Here, the gamma function of the complex field~\cite[Section~1.4]{GGR},~\cite[Section~1.3]{N} is defined as
\begin{equation*}
    \bm{\Gamma}(a) = \frac{\Gamma(a)}{\Gamma(1-\bar{a})} \,.
\end{equation*}
Using the standard properties of the ordinary Euler gamma function, the following relations immediately follow:
\begin{align}\label{gamma-diff}
    & \bm{\Gamma}(a + 1) = - a \bar{a} \bm{\Gamma}(a) \,, \\
    & \bm{\Gamma}(a) = [-1]^{a} \bm{\Gamma}(\bar{a}) \,, \\ \label{gamma-refl}
    & \bm{\Gamma}(a) \bm{\Gamma}(1-a) = [-1]^a \,.
\end{align}

Relation \eqref{Chain} can also be expressed as the action of the operator $[\hat{p}]^{a-1}$ on the function $[z-z_0]^{-b}$:
\begin{equation}\nonumber
    [\hat{p}]^{a-1}\,[z-z_0]^{-b} =
    \frac{[\imath]^{a}\,[-1]^{-a-b}}{\mathbf{\Gamma}(b,2-a-b)}\,
    [z-z_0]^{1-a-b} \,.
\end{equation}

\begin{figure}[h]
    \centering
     \begin{tikzpicture}[thick, line cap = round, scale=1.5]
        % Diagram for star-triangle relation
        %The length of triangle's side is equal to 2
        \def\rad{0.08} % radius of the circle of bold vertex
        \def\si{0.866} % sin(pi/3)
        \def\yc{0.577} % y-coordinate of the "star" center: 1*tg(pi/6)
        \def\yt{1.732} % y-coordinate of the top vertex center: 2*sin(pi/3)
        \def\shi{3.35} %shift of the coefficient by x axis from the coordinate origin
        \def\sh{5} %shift of the triangle by x axis from the coordinate origin
        \draw[->-] (1,\yc) to node[midway,above=1]{$c$} (0,0);
        \draw[->-] (1,\yc) to node[midway,above=1]{$b$} (2,0);
        \draw[->-] (1,\yc) to node[midway,right]{$a$} (1,\yt);
        \draw[fill = black] (1,\yc) circle (\rad);
        \node at (\shi,\yc) {$=\pi \, \mathbf{\Gamma}^{-1}(a,b,c)$};
        \draw[->-] (\sh,0) to node[midway,above]{$1-b\qquad$} (\sh+1,\yt);
        \draw[->-] (\sh+1,\yt) to node[midway,above]{$\qquad\; 1-c$} (\sh+2,0);
        \draw[->-] (\sh+2,0) to node[midway,below]{$1-a$} (\sh,0);
    \end{tikzpicture}
    \caption{Star-triangle relation, $a+b+c=2$.}
    \label{Rules}
\end{figure}
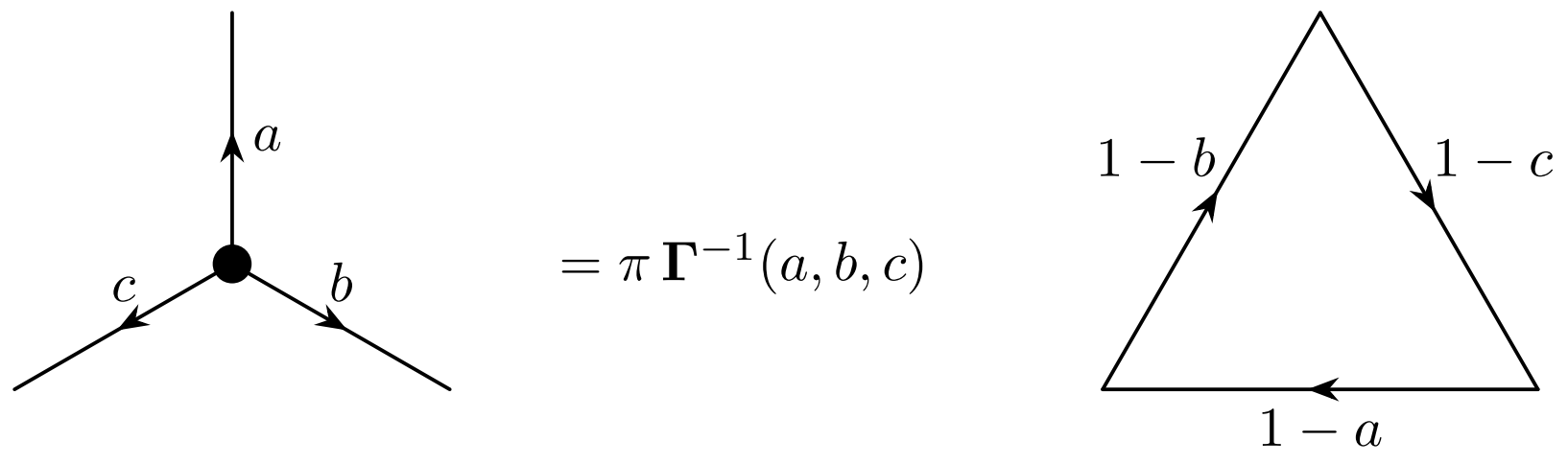

The second identity is the \textit{star-triangle relation}, shown in Figure~\ref{Rules}.

This identity can be expressed as the integral relation
\begin{align}\label{Star}
\int \mathrm{d}^2 w \frac{1}{[w-z_1]^a[w-z_2]^b [w-z_3]^c}
 = \frac{\pi}{\mathbf{\Gamma}(a,b,c)}
\frac{1}{[z_{12}]^{1-c}[z_{31}]^{1-b}[z_{23}]^{1-a}} \,,
\end{align}
where $a+b+c=2$ and $\bar a+\bar b+\bar c=2$.
It can also be written in operator form \cite{DM09,DM14,Isa}:
\begin{align}\label{star-tr}
[\hat{p}]^{a} [z-z_0]^{a+b} [\hat{p}]^{b} =
[z-z_0]^{b} [\hat{p}]^{a+b} [z-z_0]^{a} \,.
\end{align}

Finally, we present two formulas for the two-dimensional delta function in the complex plane. The first one reads
\begin{equation}\label{delta1}
    \delta^2(z) = \lim_{\varepsilon\to 0}\frac{\varepsilon}{\pi} \frac{1}{[z]^{1-\varepsilon}} \,,
\end{equation}
and the second one
\begin{equation} \label{delta2}
    \int \mathrm{d}^2 w \frac{1}{[z_1-w]^{2-\alpha}[w-z_2]^\alpha} =
    \frac{\pi^2}{\mathbf{\Gamma}(\alpha,2-\alpha)} \, \delta^2(z_1-z_2)
    = -\frac{\pi^2\,[-1]^\alpha}{(\alpha-1)(\bar{\alpha}-1)} \, \delta^2(z_1-z_2)
\end{equation}
is derived by combining the chain relation~\eqref{Chain} with~\eqref{delta1}.

\section{Trace and transposition of integral operators}
\label{app:TrTransp}

Let us prove the properties of trace and transposition of integral operators which we use in the main text.
First three key relations are the cyclic property and the invariance of the trace under the transposition of its argument, along with the anti-homomorphism property of transposition.
Namely, if $\mathcal{A}$ and $\mathcal{B}$ are some integral operators in the tensor product of $\mathrm{L}^2(\mathbb{C})$ spaces $V_0\otimes V_1\otimes V_2$ of functions of variables $z_0, z_1, z_2$ such that $\mathcal{A}$ acts trivially (as an identity operator) in $V_2$ and $\mathcal{B}$ acts trivially in $V_1$, then
\begin{align}
	& \label{tr-cycle}
	\tr_0\bigl[\mathcal{A}\,\mathcal{B}\bigr] = \tr_0\bigl[\mathcal{B}\,\mathcal{A}\bigr] \,, \\
	& \label{transp}
	\bigl[\mathcal{A}\,\mathcal{B}\bigr]^{t_0} = \mathcal{B}^{t_0} \mathcal{A}^{t_0}
\end{align}
and for an integral operator $\mathcal{C}$ in $V_0\otimes V_1$ we have
\begin{equation} \label{tr-transp}
	\tr_0[\mathcal{C}] = \tr_0[\mathcal{C}^{t_0}] \,.
\end{equation}
These formulas (as well as their derivations) can be easily generalized to the case of operators acting on functions of arbitrary number of variables, that is, one can take instead of $V_1$ and $V_2$ a tensor product of any number of $\mathrm{L}^2(\mathbb{C})$ copies.
Operators~$\mathcal{A}$,~$\mathcal{B}$ and~$\mathcal{C}$ may also act nontrivially only in the space~$V_0$.
An important point in~\eqref{tr-cycle} and~\eqref{transp} is that the space $V_0$, over which we take the trace or with respect to which we consider the transposition, must be the only space in which both operators $\mathcal{A}$ and $\mathcal{B}$ act nontrivially.

All relations~\eqref{tr-cycle},~\eqref{transp} and~\eqref{tr-transp} follow directly from the definition of operations of trace and transposition.
Since $\mathcal{A}$ acts trivially in $V_2$, we may treat it as an integral operator in $V_0\otimes V_1$ with kernel $A(z_0,z_1|w_0,w_1)$.
Similarly, $\mathcal{B}$ can be treated as an operator in $V_0\otimes V_2$ with kernel $\mathcal{B}(z_0,z_2|w_0,w_2)$.
The kernels of $\mathcal{A}\,\mathcal{B}$ and $\mathcal{B}\,\mathcal{A}$ are then equal to
\begin{align}
	& \label{ABker}
	\bigl[\mathcal{A}\,\mathcal{B}\bigr](z_0,z_1,z_2|w_0,w_1,w_2) = \int\,\mathrm{d}^2w'_0 \, \mathcal{A}(z_0,z_1|w_0',w_1)\,\mathcal{B}(w_0',z_2|w_0,w_2) \,, \\
	& \label{BAker}
	\bigl[\mathcal{B}\,\mathcal{A}\bigr](z_0,z_1,z_2|w_0,w_1,w_2) = \int\,\mathrm{d}^2w'_0 \, \mathcal{B}(z_0,z_2|w_0',w_2) \, \mathcal{A}(w_0',z_1|w_0,w_1) \,.
\end{align}
To obtain the kernels of integral operators $\tr_0\bigl[\mathcal{A}\,\mathcal{B}\bigr]$ and $\tr_0\bigl[\mathcal{B}\,\mathcal{A}\bigr]$ one needs to take $z_0=w_0$ and integrate over $w_0$ in~\eqref{ABker} and~\eqref{BAker}, respectively.
The result in both cases will evidently be the same, which yields~\eqref{tr-cycle}.

From~\eqref{ABker} we deduce that the kernel of~$\bigl[\mathcal{A}\,\mathcal{B}\bigr]^{t_0}$ reads
\begin{equation} \nonumber
	\bigl[\mathcal{A}\,\mathcal{B}\bigr]^{t_0}(z_0,z_1,z_2|w_0,w_1,w_2) = \int\,\mathrm{d}^2w'_0 \, \mathcal{A}(w_0,z_1|w_0',w_1)\,\mathcal{B}(w_0',z_2|z_0,w_2) \,,
\end{equation}
and it is not hard to show that the kernel of~$\mathcal{B}^{t_0} \mathcal{A}^{t_0}$ has exactly the same form
\begin{align}
	\nonumber
	\bigl[\mathcal{B}^{t_0}\mathcal{A}^{t_0}\bigr](z_0,z_1,z_2|w_0,w_1,w_2) 
	& = \int\,\mathrm{d}^2w'_0 \, \mathcal{B}^{t_0}(z_0,z_2|w_0',w_2) \, \mathcal{A}^{t_0}(w_0',z_1|w_0,w_1) \\
	\nonumber
	& = \int\,\mathrm{d}^2w'_0 \, \mathcal{B}(w_0',z_2|z_0,w_2) \, \mathcal{A}(w_0,z_1|w_0',w_1)
\end{align}
whence follows the identity~\eqref{transp}.

To show~\eqref{tr-transp} we observe that the kernels of~$\mathcal{C}$ and~$\mathcal{C}^{t_0}$ differ only by permutation of variables corresponding to the space~$V_0$
\begin{equation} \nonumber
	\mathcal{C}^{t_0}(z_0,z_1|w_0,w_1) = \mathcal{C}(w_0,z_1|z_0,w_1) \,.
\end{equation}
Therefore, when we equate $z_0=w_0$ in $\mathcal{C}(z_0,z_1|w_0,w_1)$ and $\mathcal{C}^{t_0}(z_0,z_1|w_0,w_1)$ and integrate over $w_0$ to obtain, correspondingly, the kernels of $\tr_0[\mathcal{C}]$ and $\tr_0[\mathcal{C}^{t_0}]$, we get the same result. Thus, the latter operators coincide and we get~\eqref{tr-transp}.

We also use other two helpful relations.
Consider two integral operators $\mathcal{A}$ and $\mathcal{B}$ in the tensor product of $\mathrm{L}^2(\mathbb{C})$ copies~$V_0\otimes V_1$.
The first relation is the infinite-dimensional generalization of~\eqref{trAtBt}
\begin{equation} \label{tr-tt}
	\tr_0\bigl[\mathcal{A}^{t_0}\mathcal{B}^{t_0}\bigr] = \tr_0\bigl[\mathcal{A}\,\mathcal{B}\bigr] \,.
\end{equation}
And the second one is quite obvious: if the operator~$\mathcal{B}$ acts trivially in $V_0$, then
\begin{equation} \label{transp-triv}
	\bigl[\mathcal{A}\,\mathcal{B}\bigr]^{t_0} = \mathcal{A}^{t_0} \mathcal{B} \,, \quad
	\bigl[\mathcal{B}\,\mathcal{A}\bigr]^{t_0} = \mathcal{B}\,\mathcal{A}^{t_0} \,.
\end{equation}
Again, one can replace~$V_1$ with the tensor product of arbitrary number of~$\mathrm{L}^2(\mathbb{C})$ spaces, but for clarity we consider the operators acting on functions of two variables.
And again, we check everything just using the definitions of trace and transposition.
Kernels of operators $\mathcal{A}^{t_0}$, $\mathcal{B}^{t_0}$ and $\mathcal{A}$, $\mathcal{B}$ are related in the following way
\begin{equation} \nonumber
	\mathcal{A}^{t_0}(z_0,z_1|w_0,w_1) = \mathcal{A}(w_0,z_1|z_0,w_1) \,,
	\quad
	\mathcal{B}^{t_0}(z_0,z_1|w_0,w_1) = \mathcal{B}(w_0,z_1|z_0,w_1) \,.
\end{equation}
Therefore, the kernel of~$\mathcal{A}^{t_0}\mathcal{B}^{t_0}$ is equal to
\begin{align}
	\nonumber
	\bigl[\mathcal{A}^{t_0}\mathcal{B}^{t_0}\bigr](z_0,z_1|w_0,w_1) & =
	\int \mathrm{d}^2 w_0'\,\mathrm{d}^2 w_1' \,
	\mathcal{A}^{t_0}(z_0,z_1|w_0',w_1') \, \mathcal{B}^{t_0}(w_0',w_1'|w_0,w_1) \\
	\nonumber
	& =
	\int \mathrm{d}^2 w_0'\,\mathrm{d}^2 w_1' \,
	\mathcal{A}(w_0',z_1|z_0,w_1') \, \mathcal{B}(w_0,w_1'|w_0',w_1) \,.
\end{align}
Consequently, the kernel of its trace over the space~$V_0$ reads
\begin{align}
	\nonumber
	\Bigl[\tr_0\bigl[\mathcal{A}^{t_0}\mathcal{B}^{t_0}\bigr]\Bigr](z_1|w_1)
	& =
	\int \mathrm{d}^2 w_0 \, \mathrm{d}^2 w_0'\,\mathrm{d}^2 w_1' \,
	\mathcal{A}(w_0',z_1|w_0,w_1') \, \mathcal{B}(w_0,w_1'|w_0',w_1) \\
	\nonumber
	& = \int \mathrm{d}^2 w_0' \, \bigl[\mathcal{A}\,\mathcal{B}\bigr](w_0',z_1|w_0',w_1)
	= \Bigl[\tr_0\bigl[\mathcal{A}\,\mathcal{B}\bigr]\Bigr](z_1|w_1) \,,
\end{align}
and thus we obtain~\eqref{tr-tt}.

If the operator~$\mathcal{B}$ acts trivially in~$V_0$, then its kernel simplifies
\begin{equation*}
	\mathcal{B}(z_0,z_1|w_0,w_1)=\delta^2(z_0-w_0)\,\mathcal{B}(z_1|w_1) \,,
\end{equation*}
and we find that expressions for kernels of~$\bigl[\mathcal{A}\,\mathcal{B}\bigr]^{t_0}$ and~$\mathcal{A}^{t_0} \mathcal{B}$ coincide
\begin{align*}
	&
	\bigl[\mathcal{A}\,\mathcal{B}\bigr]^{t_0}(z_0,z_1|w_0,w_1) 
	=
	\bigl[\mathcal{A}\,\mathcal{B}\bigr](w_0,z_1|z_0,w_1)
	= \int \mathrm{d}^2 w_1'\,\mathcal{A}(w_0,z_1|z_0,w_1')\,\mathcal{B}(w_1',w_1) \\
	& \qquad
	= \int \mathrm{d}^2 w_1' \, \mathcal{A}^{t_0}(z_0,z_1|w_0,w_1')\,\mathcal{B}(w_1',w_1) =
	\bigl[\mathcal{A}^{t_0} \mathcal{B}\bigr](z_0,z_1|w_0,w_1) \,,
\end{align*}
whence follows the first equality in~\eqref{transp-triv}. The second equality can be checked in a similar fashion.

\section{Some useful properties of the $\mathbb{R}$-operator} 
\label{app:ucu}
The general $\mathbb{R}$-operator, satisfying the Yang Baxter equation
\begin{equation*}
	\mathbb{R}_{12}(u-v)\,\mathbb{R}_{13}(u)\,\mathbb{R}_{23}(v)=\mathbb{R}_{23}(v)\,\mathbb{R}_{13}(u)\,\mathbb{R}_{12}(u-v)
\end{equation*}
was constructed in~\cite{DM09}. It takes the form 
\begin{align*}
	\mathbb{R}_{12}(u) = \mathbb{P}_{12}\,\check{R}_{12}(u) = \mathbb{P}_{12}\,
	[z_{12}]^{u+1-s_1-s_2}\,
	[\hat{p}_1]^{u+s_2-s_1}\,[\hat{p}_2]^{u+s_1-s_2}\,
	[z_{12}]^{u+s_1+s_2-1} \,,
\end{align*}	
The first property of this operator is \textit{unitarity}
\begin{equation}\nonumber
	\mathbb{R}_{12}(u)\,\mathbb{R}_{12}(-u) = \II \,.
\end{equation}	
To prove this we write
\begin{align*}
	\mathbb{R}_{12}(u)\,\mathbb{R}_{12}(-u) =  \mathbb{P}_{12}\,
	[z_{12}]^{u+1-s_1-s_2}\,
	[\hat{p}_1]^{u+s_2-s_1}\,[\hat{p}_2]^{u+s_1-s_2}\,
	[z_{12}]^{u+s_1+s_2-1} \mathbb{P}_{12}\,
	[z_{12}]^{u+1-s_1-s_2}\\
	[\hat{p}_1]^{u+s_2-s_1}\,[\hat{p}_2]^{u+s_1-s_2}\,
	[z_{12}]^{u+s_1+s_2-1} .
\end{align*}
Moving the $\mathbb{P}_{12}$-operator to the left we see that all factors cancel.
The next property we use is the so-called \textit{crossing-unitarity}
\begin{equation}\label{Rcuni}
	\mathbb{R}_{12}^{t_1}(u)\,\mathbb{R}_{12}^{t_1}(-u-2) =
	\mathbb{R}_{12}^{t_2}(u)\,\mathbb{R}_{12}^{t_2}(-u-2) =
	\rho_{s_1,s_2}(u)\,\II \,,
\end{equation}	
where the partial transposition of an integral operator is defined in Section~\ref{sect:generalT}, see formulas~\eqref{transpDef1},~\eqref{transpDef2}, and the coefficient reads
\begin{align*}
	\rho_{s_1,s_2}(u) & = \frac{c(u+s_2-s_1)c(u+s_1-s_2)c(-u-2+s_2-s_1)c(-u-2+s_1-s_2)}{c(u+s_1+s_2)c(-u-s_1-s_2)c(-u-2+s_1+s_2)c(u+2-s_1-s_2)} \\
	& =
	(-1-u\pm(s_1-s_2))^{-1}(1+u\pm(1-s_1-s_2))^{-1} \\
	& \quad \times
	(-1-\bar{u}\pm(\bar{s}_1-\bar{s}_2))^{-1}(1+\bar{u}\pm(1-\bar{s}_1-\bar{s}_2))^{-1} \,.
\end{align*}
The operator $\mathbb{R}_{12}(u)$ has the kernel 
\begin{align*}
	R(z_1,z_2|w_1,w_2) = c(u+s_2-s_1)\,c(u+s_1-s_2)\,
	\frac{[z_{21}]^{u+1-s_1-s_2}\,[w_{12}]^{u+s_1+s_2-1}}{[z_2-w_1]^{u+s_2-s_1+1}\,[z_1-w_2]^{u+s_1-s_2+1}} \,.
\end{align*}	
Then the kernel of the operator $\mathbb{R}_{12}^{t_1}(u)$ equals to 
	\begin{align*}
	R(w_1,z_2|z_1,w_2) = c(u+s_2-s_1)\,c(u+s_1-s_2)\,
	\frac{[z_2-w_1]^{u+1-s_1-s_2}\,[z_1-w_2]^{u+s_1+s_2-1}}{[z_2-z_1]^{u+s_2-s_1+1}\,[w_1-w_2]^{u+s_1-s_2+1}} \,.
\end{align*}	
Moving back to the $[\hat{p}]^\alpha$ and $[z]^\alpha$ operators we obtain
\begin{align*}
	\mathbb{R}_{12}^{t_1}(u) = \frac{c(u+s_2-s_1)\,c(u+s_1-s_2)}{c(-u+s_1+s_2-2)\,c(-u-s_1-s_2)}\,\mathbb{P}_{12}\,
	[z_{12}]^{-u+s_1-s_2-1} \\ [p_1]^{-u+s_1+s_2-2}\,[p_2]^{-u-s_1-s_2}\,[z_{12}]^{-u+s_2-s_1-1} \,.
\end{align*}	
Then multiplying $\mathbb{R}_{12}^{t_1}(u)$ and $\mathbb{R}_{12}^{t_1}(-u-2)$ we obtain \eqref{Rcuni}.
The similar equality for $\mathbb{R}_{12}^{t_2}$ can be checked in the same way.

In section~\ref{sect:Qfact} we used the following formulae for traces of $\mathcal{R}$-operator products
\begin{align}
	& \label{trRR1app}
	\tr_0\Bigl[\mathcal{R}^{(1)}_{10}(x)\,\mathcal{R}^{(1)}_{02}(y)\Bigr] =
	-\frac{[\imath]^{3s-x}\,\mathbf{\Gamma}(2s, x-s+1)}{(x+s)(\bar{x}+\bar{s})} \, \mathcal{R}^{(1)}_{12}(y) \,, \\
	& \label{trRR2app}
	\tr_0\Bigl[\mathcal{R}^{(2)}_{10}(x)\,\mathcal{R}^{(2)}_{02}(y)\Bigr] =
	-\frac{[\imath]^{-3s-y}\,\mathbf{\Gamma}(2-2s, y+s)}{(y-s+1)(\bar{y}-\bar{s}+1)} \, \mathcal{R}^{(2)}_{12}(x) \,, \\
	& \nonumber
	\tr_0\Bigl[\mathcal{R}^{(1)}_{10}(x)\,\mathcal{R}^{(2)}_{01}(y)\,\mathcal{R}^{(1)}_{00'}(\alpha)\,\mathcal{R}^{(2)}_{0''0}(\beta)\Bigr] \\
	& \label{trRRRRapp}
	\qquad\qquad\qquad\qquad\;\;\;
	= -\frac{[-1]^{2s} [\imath]^{-x-y}\,\mathbf{\Gamma}(x-s+1, y+s)}{(x+y)(\bar{x}+\bar{y})}
	\,\mathcal{R}^{(1)}_{10'}(\alpha)\,\mathcal{R}^{(2)}_{0''1}(\beta) \,.
\end{align}
All of them are, actually, consequences of one simple relation 
\begin{align}
	\nonumber
	\tr_0\Bigl([z_{10}]^\alpha[\hat{p}_0]^{\alpha+\beta}[z_{10}]^\beta
	\Phi(z_0)\Bigr) & =
	\tr_0\Bigl([\hat{p}_0]^{\beta} [z_{10}]^{\alpha+\beta} [\hat{p}_0]^{\alpha}
	\Phi(z_0)\Bigr) \\
	\label{trzpz}
	& = -\frac{[\imath]^{\beta-\alpha} \mathbf{\Gamma}(\alpha+1,\beta+1)}{(\alpha+\beta+1)(\bar{\alpha}+\bar{\beta}+1)} \, \Phi(z_1) \,,
\end{align}
where two traces are equal due to the star-triangle relation~\eqref{star-tr},  the result of taking the trace is the operator acting on functions of variable $z_1$, and by $\Phi(z_0)$ we denote the operator of multiplication by arbitrary function~$\Phi$.
The identity~\eqref{trzpz} follows directly from the definition of integral operator $[\hat{p}_0]^{\alpha}$~\eqref{d}.
Indeed, since the kernel of $[\hat{p}_0]^{\alpha}$ has the form
\begin{equation}
	[\hat{p}_0]^{\alpha}(z_0|w_0) = \frac{[\imath]^\alpha\,\mathbf{\Gamma}(\alpha+1)}{\pi}\,[z_0-w_0]^{-\alpha-1}
\end{equation}
and the only operator in the argument of the second trace in~\eqref{trzpz} acting nontrivially on functions of $z_1$ is the operator of multiplication by function $[z_{10}]^{\alpha+\beta}$,
one expresses the kernel of the integral operator $\hat{O}=\tr_0\Bigl([\hat{p}_0]^{\beta} [z_{10}]^{\alpha+\beta} [\hat{p}_0]^{\alpha}
\Phi(z_0)\Bigr)$ as follows
\begin{align}
	\nonumber
	\hat{O}(z_1|w_1) &
	= \delta^2(z_1-w_1) \, \frac{[\imath]^{\alpha+\beta}\,\mathbf{\Gamma}(\alpha+1,\beta+1)}{\pi^2}
	\int \mathrm{d}^2z_0\,\mathrm{d}^2z_{0'}\, \frac{[z_{10'}]^{\alpha+\beta}\,\Phi(z_0)}{[z_{00'}]^{\beta+1}\,[z_{0'0}]^{\alpha+1}} \\
	& \nonumber
	= \delta^2(z_1-w_1) \, \frac{[\imath]^{\alpha-\beta}\,\mathbf{\Gamma}(\alpha+1,\beta+1)}{\pi^2}
	\int \mathrm{d}^2z_0\,\mathrm{d}^2z_{0'}\, \frac{\Phi(z_0)}{[z_{10'}]^{-\alpha-\beta}\,[z_{0'0}]^{\alpha+\beta+2}} \\
	& \nonumber
	= - \delta^2(z_1-w_1) \, \frac{[\imath]^{\beta-\alpha}\,\mathbf{\Gamma}(\alpha+1,\beta+1)}{(\alpha+\beta+1)(\bar{\alpha}+\bar{\beta}+1)}
	\int \mathrm{d}^2z_0\,\delta^2(z_{10})\,\Phi(z_0) \\
	& \nonumber
	= - \delta^2(z_1-w_1) \, \frac{[\imath]^{\beta-\alpha}\,\mathbf{\Gamma}(\alpha+1,\beta+1)}{(\alpha+\beta+1)(\bar{\alpha}+\bar{\beta}+1)} \, \Phi(z_1) \,,
\end{align}
where the integral over $z_{0'}$ gave the delta function according to~\eqref{delta2}.
This way we obtain precisely the kernel of the resulting operator in~\eqref{trzpz}.

To deduce~\eqref{trRR1app} from~\eqref{trzpz} we note that the operator~$\mathcal{R}^{(1)}_{02}(y)$ acts in the $0$-th space as the operator of multiplication by function.
Namely, by definition of $\mathcal{R}$-operators~\eqref{R12def} we have
\begin{align}
	& \nonumber
	\tr_0\Bigl[\mathcal{R}^{(1)}_{10}(x)\,\mathcal{R}^{(1)}_{02}(y)\Bigr] =
	\tr_0\Bigl[[\hat{p}_0]^{2s-1}\,[z_{10}]^{x-1+s}\,[\hat{p}_0]^{x-s}\,[\hat{p}_2]^{2s-1}\,[z_{02}]^{y-1+s}\,[\hat{p}_2]^{y-s}\Bigr] \\
	& \nonumber
	=
	[\hat{p}_2]^{2s-1}\,
	\tr_0\Bigl[[\hat{p}_0]^{2s-1}\,[z_{10}]^{x-1+s}\,[\hat{p}_0]^{x-s}\,[z_{02}]^{y-1+s}\Bigr] \, [\hat{p}_2]^{y-s} \\
	& \nonumber
	=
	-\frac{[\imath]^{3s-x}\,\mathbf{\Gamma}(2s, x-s+1)}{(x+s)(\bar{x}+\bar{s})} \,
	[\hat{p}_2]^{2s-1}\,[z_{12}]^{y-1+s}\,[\hat{p}_2]^{y-s}
	= -\frac{[\imath]^{3s-x}\,\mathbf{\Gamma}(2s, x-s+1)}{(x+s)(\bar{x}+\bar{s})} \,
	\mathcal{R}^{(1)}_{12}(y) \,.
\end{align}
The derivation of~\eqref{trRR2app} is similar.
And the check of~\eqref{trRRRRapp} also resembles the above calculation since operators $\mathcal{R}^{(1)}_{00'}(\alpha)$ and $\mathcal{R}^{(2)}_{0''0}(\beta)$ also act in the $0$-th space as operators of multiplication by function
\begin{align}
	& \nonumber
	\tr_0\Bigl[\mathcal{R}^{(1)}_{10}(x)\,\mathcal{R}^{(2)}_{01}(y)\,\mathcal{R}^{(1)}_{00'}(\alpha)\,\mathcal{R}^{(2)}_{0''0}(\beta)\Bigr] \\
	& \nonumber
	=
	[-1]^{y-s}\,[\hat{p}_{0'}]^{2s-1}\,[\hat{p}_{0''}]^{\beta+s-1}
	\tr_0\Bigl[[z_{10}]^{x-s}\,[\hat{p}_0]^{x+y-1}\,[z_{10}]^{y+s-1}\,[z_{00'}]^{\alpha+s-1}[z_{0''0}]^{\beta-s}\Bigr] \\
	& \nonumber
	\quad \times [\hat{p}_{0'}]^{\alpha-s}\,[\hat{p}_{0''}]^{1-2s} \\
	& \nonumber
	=
	-\frac{[-1]^{2s} [\imath]^{-x-y}\,\mathbf{\Gamma}(x-s+1, y+s)}{(x+y)(\bar{x}+\bar{y})}
	\,[\hat{p}_{0'}]^{2s-1}[\hat{p}_{0''}]^{\beta+s-1} \,
	[z_{10'}]^{\alpha+s-1}\,[z_{0''1}]^{\beta-s} \\
	& \nonumber
	\quad \times [\hat{p}_{0'}]^{\alpha-s}\,[\hat{p}_{0''}]^{1-2s} \\
	& \nonumber
	= -\frac{[-1]^{2s} [\imath]^{-x-y}\,\mathbf{\Gamma}(x-s+1, y+s)}{(x+y)(\bar{x}+\bar{y})} \,
	\mathcal{R}^{(1)}_{10'}(\alpha)\,\mathcal{R}^{(2)}_{0''1}(\beta) \,.
\end{align}

In section~\ref{sect:Qfact} we also use the relation
\begin{multline} \label{RpR}
	\mathcal{R}^{(2)}_{k,k-1}(x)\dots\mathcal{R}^{(2)}_{32}(x)\,\mathcal{R}^{(2)}_{21}(x)\,[\hat{p}_1]^{x+y-1}\,\mathcal{R}^{(1)}_{12}(y)\,\mathcal{R}^{(1)}_{23}(y)\dots
	\mathcal{R}^{(1)}_{k-1,k}(y) \\
	=
	[\hat{p}_1]^{s+y-1}\,\mathcal{R}^{(1)}_{12}(y)\,\mathcal{R}^{(1)}_{23}(y)\dots
	\mathcal{R}^{(1)}_{k-1,k}(y) \,
	\mathcal{R}^{(2)}_{k,k-1}(x)\dots\mathcal{R}^{(2)}_{32}(x)\,\mathcal{R}^{(2)}_{21}(x)\,[\hat{p}_1]^{x-s} \,,
\end{multline}
which can be proven by induction with respect to~$k$.
To derive the basic formula for $k=2$ one employs the star-triangle relation~\eqref{star-tr} two times
\begin{align}
	& \nonumber
	\mathcal{R}^{(2)}_{21}(x)\,[\hat{p}_1]^{x+y-1}\,\mathcal{R}^{(1)}_{12}(y) \\
	& \nonumber
	= [-1]^{x-s}\,[\hat{p}_2]^{x+s-1}\,[z_{12}]^{x-s}\,[\hat{p}_2]^{1-2s}\,
	[\hat{p}_1]^{x+y-1}\,[\hat{p}_2]^{2s-1}\,[z_{12}]^{y+s-1}\,[\hat{p}_2]^{y-s} \\
	& \nonumber
	= [-1]^{x-s}\,[\hat{p}_2]^{x+s-1}\,[\hat{p}_1]^{y+s-1}\,[z_{12}]^{x+y-1}\,
	\,[\hat{p}_1]^{x-s}\,[\hat{p}_2]^{y-s}
	\\
	& \nonumber
	= [-1]^{x-s}\,[\hat{p}_1]^{y+s-1}\,[z_{12}]^{y-s}\,[\hat{p}_2]^{x+y-1}\,[z_{12}]^{x+s-1}\,[\hat{p}_1]^{x-s} \\
	& \nonumber
	=
	[-1]^{x-s}\,[\hat{p}_1]^{y+s-1}\,[z_{12}]^{y-s}\,[\hat{p}_2]^{y+s-1}\,[z_{12}]^{2s-1}\,[z_{12}]^{1-2s}\,[\hat{p}_2]^{x-s}\,[z_{12}]^{x+s-1}\,[\hat{p}_1]^{x-s} \\
	& \nonumber
	= [\hat{p}_1]^{y+s-1}\,\mathcal{R}^{(1)}_{12}(y)\,\mathcal{R}^{(2)}_{21}(x)\,[\hat{p}_1]^{x-s} \,.
\end{align}
To prove the induction step $k\to k+1$ we first use the checked $k=2$ identity with operators $\mathcal{R}^{(1)}_{12}(y)$ and $\mathcal{R}^{(2)}_{21}(x)$ rewritten in the explicit form in the right hand side
\begin{equation} \nonumber
	\mathcal{R}^{(2)}_{21}(x)\,[\hat{p}_1]^{x+y-1}\,\mathcal{R}^{(1)}_{12}(y) =
	[-1]^{x-s}\,[\hat{p}_1]^{y+s-1}\,[z_{12}]^{y-s}\,[\hat{p}_2]^{x+y-1}\,[z_{12}]^{x+s-1}\,[\hat{p}_1]^{x-s} \,.
\end{equation}
Thus we arrive at the expression for which it is possible to apply the induction proposition
\begin{align}
	& \nonumber
	\mathcal{R}^{(2)}_{k+1,k}(x)\dots\mathcal{R}^{(2)}_{32}(x)\,\mathcal{R}^{(2)}_{21}(x)\,[\hat{p}_1]^{x+y-1}\,\mathcal{R}^{(1)}_{12}(y)\,\mathcal{R}^{(1)}_{23}(y)\dots
	\mathcal{R}^{(1)}_{k,k+1}(y) \\
	& \nonumber
	= [-1]^{x-s}\,[\hat{p}_1]^{y+s-1}\,[z_{12}]^{y-s} \,
	\mathcal{R}^{(2)}_{k+1,k}(x)\dots\mathcal{R}^{(2)}_{32}(x) \,
	[\hat{p}_2]^{x+y-1}
	\mathcal{R}^{(1)}_{23}(y)\dots\mathcal{R}^{(1)}_{k,k+1}(y) \\
	& \nonumber
	\quad \times
	[z_{12}]^{x+s-1}\,[\hat{p}_1]^{x-s} \\
	& \nonumber
	= [-1]^{x-s}\,[\hat{p}_1]^{y+s-1}\,[z_{12}]^{y-s} \,
	[\hat{p}_2]^{y+s-1}\,\mathcal{R}^{(1)}_{23}(y)\dots\mathcal{R}^{(1)}_{k,k+1}(y)\,\mathcal{R}^{(2)}_{k+1,k}(x)\dots\mathcal{R}^{(2)}_{32}(x)\,[\hat{p}_2]^{x-s} \\
	& \nonumber
	\quad \times
	[z_{12}]^{x+s-1}\,[\hat{p}_1]^{x-s} \,.
\end{align}
Inserting into the last expression two mutually inverse factors $[z_{12}]^{2s-1}$ and $[z_{12}]^{1-2s}$, which commute with $\mathcal{R}^{(1)}_{23}(y)$ and $\mathcal{R}^{(2)}_{32}(x)$ (since the latter act in the $2$-nd space as operators of multiplication by function),
we obtain the needed operators $\mathcal{R}^{(1)}_{12}(y)$, $\mathcal{R}^{(2)}_{21}(x)$ and finish the proof of the induction step
\begin{align}
	& \nonumber
	\mathcal{R}^{(2)}_{k+1,k}(x)\dots\mathcal{R}^{(2)}_{32}(x)\,\mathcal{R}^{(2)}_{21}(x)\,[\hat{p}_1]^{x+y-1}\,\mathcal{R}^{(1)}_{12}(y)\,\mathcal{R}^{(1)}_{23}(y)\dots
	\mathcal{R}^{(1)}_{k,k+1}(y) \\
	& \nonumber
	= [-1]^{x-s}\,[\hat{p}_1]^{y+s-1}\,[z_{12}]^{y-s} \,
	[\hat{p}_2]^{y+s-1}\,[z_{12}]^{2s-1}\,\mathcal{R}^{(1)}_{23}(y)\dots\mathcal{R}^{(1)}_{k,k+1}(y)\,\mathcal{R}^{(2)}_{k+1,k}(x)\dots\mathcal{R}^{(2)}_{32}(x) \\
	& \nonumber
	\quad \times
	[z_{12}]^{1-2s}\,[\hat{p}_2]^{x-s}\,[z_{12}]^{x+s-1}\,[\hat{p}_1]^{x-s} \\
	& \nonumber
	= [\hat{p}_1]^{y+s-1}\,\mathcal{R}^{(1)}_{12}(y)\,\mathcal{R}^{(1)}_{23}(y)\dots \mathcal{R}^{(1)}_{k,k+1}(y) \,
	\mathcal{R}^{(2)}_{k+1,k}(x)\dots\mathcal{R}^{(2)}_{32}(x)\,\mathcal{R}^{(2)}_{21}(x) \, [\hat{p}_1]^{x-s} \,.
\end{align}

At last let us derive one more important relation used in Section~\ref{sect:Qfact}
\begin{equation} \label{trppApp}
	\tr_0\Bigl[[\hat{p}_0]^\alpha\,\Phi(z_0)\,[\hat{p}_0]^{-\alpha-2}\,\Psi(z_0)\Bigr] =
	-\frac{1}{\pi^2\,(\alpha+1)(\bar{\alpha}+1)} \,
	\int \mathrm{d}^2z \, \Phi(z) \, \int \mathrm{d}^2z \, \Psi(z) \,,
\end{equation}
where $\Phi$ and $\Psi$ are arbitrary functions.
The kernels of integral operators $\hat{\Phi}=[\hat{p}_0]^\alpha\,\Phi(z_0)$ and $\hat{\Psi}=[\hat{p}_0]^{-\alpha-2}\,\Psi(z_0)$ read
\begin{equation} \nonumber
	\hat{\Phi}(z_0|w_0) =
	\frac{[\imath]^\alpha\,\mathbf{\Gamma}(\alpha+1)}{\pi}\,\frac{\Phi(w_0)}{[z_0-w_0]^{\alpha+1}} \,, \quad
	\hat{\Psi}(z_0|w_0) =
	\frac{[\imath]^{-\alpha}\,\mathbf{\Gamma}(-\alpha-1)}{\pi}\,\frac{\Psi(w_0)}{[z_0-w_0]^{-\alpha-1}} \,,
\end{equation}
and we have
\begin{align}
	& \nonumber
	\tr_0\Bigl[[\hat{p}_0]^\alpha\,\Phi(z_0)\,[\hat{p}_0]^{-\alpha-2}\,\Psi(z_0)\Bigr] =
	\tr_0\Bigl[\hat{\Phi}\,\hat{\Psi}\Bigr] =
	\int \mathrm{d}^2z_0\,\mathrm{d}^2w_0 \,
	\hat{\Phi}(z_0|w_0)\,\hat{\Psi}(w_0|z_0) \\
	& \nonumber
	= \frac{\mathbf{\Gamma}(\alpha+1,-\alpha-1)}{\pi^2}
	\int \mathrm{d}^2z_0\,\mathrm{d}^2w_0 \,
	[-1]^{\alpha}\,\Phi(w_0)\,\Psi(z_0) \\
	& \nonumber
	= -\frac{1}{\pi^2\,(\alpha+1)(\bar{\alpha}+1)} \,
	\int \mathrm{d}^2w_0 \, \Phi(w_0) \, \int \mathrm{d}^2z_0 \, \Psi(z_0) \,.
\end{align}

\end{document}